\documentclass[aps,prb,twocolumn,amsmath,amssymb,superscriptaddress,scrartcl,eqsecnum,longbibliography,nofootinbib]{revtex4-2}
\usepackage[T1]{fontenc}
\usepackage[latin9]{inputenc}
\usepackage{color}
\usepackage{verbatim}
\usepackage{amsmath}
\usepackage{amssymb}
\usepackage{graphicx}
\usepackage{bm}
\usepackage{tikz}
\usepackage{diagbox}
\usepackage{array}
\usepackage{dsfont}
\usepackage{capt-of}
\usepackage{braket}
\usepackage{makecell}
\usepackage{multirow}
\usepackage{placeins}
\usepackage[percent]{overpic}
\usepackage[normalem]{ulem}

\usepackage[most]{tcolorbox}
\makeatletter

\PassOptionsToPackage{caption=false}{subfig} 
\usepackage{hyperref}
\hypersetup{
breaklinks=true,
colorlinks=true,
citecolor=blue,
linkcolor=black,
filecolor=black,
urlcolor=blue
}
\IfFileExists{lmodern.sty}{\usepackage{lmodern}}{}

\renewcommand{\bar}{\overline}
\renewcommand{\tilde}{\widetilde}
\renewcommand{\hat}{\widehat}
\renewcommand{\leq}{\leqslant}

\renewcommand{\Im}{\operatorname{Im}}

\definecolor{darkred}{rgb}{0.5,0.,0.}
\makeatletter

\newcommand*{\wideboxed}[1]{\setlength{\fboxsep}{1ex}%
  \fbox{\m@th$\displaystyle#1$}}
\makeatother
\def\be{\begin{equation}}
\def\ee{\end{equation}}

\makeatother

\begin{document}
\title{Universal Suppression of Dissipation across Conformal Interface \\ in Open Quantum Critical Systems}

\author{Ruhanshi Barad}

\affiliation{School of Physics, Georgia Institute of Technology, Atlanta, GA 30332, USA}

\author{Tian Wu}

\affiliation{School of Physics, Georgia Institute of Technology, Atlanta, GA 30332, USA}

\author{Qicheng Tang}

\affiliation{School of Physics, Georgia Institute of Technology, Atlanta, GA 30332, USA}

\author{Xueda Wen}

\affiliation{School of Physics, Georgia Institute of Technology, Atlanta, GA 30332, USA}

\begin{abstract}

Conformal interfaces provide an important setting for studying universal transmission phenomena in one-dimensional quantum critical systems. While energy and information transmission across such interfaces are characterized by universal quantities in closed systems, the corresponding role of conformal interfaces in dissipative dynamics is less understood. In this work, we study relaxation in locally dissipative quantum critical chains with a conformal interface and show that a universal characterization emerges at the level of individual relaxation modes.
We first show that the relaxation coefficient defined in the recent work \cite{Barad_2025} from the Liouvillian gap can become non-universal for certain boundary conditions because the mode determining the smallest decay rate can change as the interface transmission is varied. To resolve this ambiguity, we introduce a mode-resolved relaxation coefficient $c_{\rm relax}$ by continuously tracking the same Liouvillian rapidity mode as a function of the interface transmission. Using analytical and numerical calculations for a critical harmonic chain and a critical free-fermion chain, we find that, in the weak-dissipation regime, $c_{\rm relax}$ follows the same universal dependence on the interface transmission in all cases considered, independent of microscopic details such as the boundary conditions, dissipation strength, and location of the local dissipation. For boundary dissipation, this universal behavior persists even at finite dissipation strength. 
Our results establish a universal mode-resolved characterization of relaxation across conformal interfaces in open quantum critical systems.

\end{abstract}

\maketitle

\tableofcontents

\begin{figure*}[!t]
\noindent\hspace*{-8cm}
\begin{tikzpicture}[x=1pt,y=0.75pt,yscale=-1,xscale=1]

\begin{scope}[xshift=0pt]

\draw [color={rgb, 255:red, 128; green, 128; blue, 128}, draw opacity=1, line width=2.25] (350,171) -- (571,170.5);

\draw [color={rgb, 255:red, 74; green, 144; blue, 226}, draw opacity=1, line width=3] (461,120) -- (461,220);

\draw [dash pattern={on 4.5pt off 4.5pt}] (461,189) -- (480.57,199.1);
\draw [shift={(488,205.5)}, rotate=205.31] [fill={rgb, 255:red, 0; green, 0; blue, 0}] [line width=0.08] [draw opacity=0] (12,-3) -- (0,0) -- (12,3) -- cycle;

\draw [color={rgb, 255:red, 208; green, 2; blue, 27}, draw opacity=1, line width=1.5]
(340.89,131.53) .. controls (340.7,134.06) and (340.52,136.46) .. (341.94,137.57)
.. controls (343.37,138.69) and (346.01,138.2) .. (348.78,137.68)
.. controls (351.55,137.17) and (354.19,136.68) .. (355.62,137.79)
.. controls (357.04,138.9) and (356.86,141.31) .. (356.67,143.83)
.. controls (356.47,146.35) and (356.29,148.76) .. (357.72,149.87)
.. controls (359.14,150.98) and (361.78,150.5) .. (364.55,149.98)
.. controls (367.32,149.46) and (369.96,148.98) .. (371.39,150.09)
.. controls (372.82,151.2) and (372.63,153.61) .. (372.44,156.13)
.. controls (372.24,158.65) and (372.06,161.06) .. (373.49,162.17)
.. controls (374.92,163.28) and (377.56,162.79) .. (380.33,162.28)
.. controls (383.09,161.76) and (385.73,161.27) .. (387.16,162.39)
.. controls (388.59,163.5) and (388.41,165.9) .. (388.21,168.42)
.. controls (388.16,169.07) and (388.11,169.72) .. (388.09,170.33);


\draw (470,210) node [anchor=north west][inner sep=0.75pt] [color={rgb, 255:red, 74; green, 144; blue, 226}, opacity=1]
{$\textcolor[rgb]{0.29,0.4,0.89}{\text{conformal interface}}$};

\draw (470,225) node [anchor=north west][inner sep=0.75pt] [color={rgb, 255:red, 74; green, 144; blue, 226}, opacity=1]
{$\textcolor[rgb]{0.29,0.4,0.89}{\text{with total reflection, $t=0$}}$};

\draw (560,80) node {\large\textbf{(a)}};
\draw (388.1,172.5) node {$\textcolor{red}{\bullet}$};
\draw (395,180) node {$x=n_d$};
\draw (145,180) node {};

\draw (415,105) node {Relaxing modes};
\draw (515,105) node {Oscillating modes};

\draw (340,105) node {\textcolor{red}{Local}};
\draw (345,122) node {\textcolor{red}{dissipation}};

\draw (340+250,105) node {\textcolor{red}{Local}};
\draw (345+250,122) node {\textcolor{red}{dissipation}};

\draw[
    rounded corners=6pt,
    dash dot,
    thick
] (470,165) rectangle (560,125);
\foreach \i in {1,...,20}{
    \pgfmathsetmacro{\randomx}{475 + (80)*rnd}
    \pgfmathsetmacro{\randomy}{130 + (30)*rnd}
    \fill[teal!80!white] (\randomx,\randomy) circle[radius=2pt];
}

\foreach \i in {1,...,20}{
    \pgfmathsetmacro{\randomx}{380 + (70)*rnd}
    \pgfmathsetmacro{\randomy}{130 + (30)*rnd}
    \fill[orange!80!black] (\randomx,\randomy) circle[radius=2pt];
}

\end{scope}

\begin{scope}[xshift=250pt]
\draw [color={rgb, 255:red, 128; green, 128; blue, 128}, draw opacity=1, line width=2.25] (350,171) -- (571,170.5);

\draw [color={rgb, 255:red, 74; green, 144; blue, 226}, draw opacity=0.15, line width=3] (461,120) -- (461,220);

\draw [dash pattern={on 4.5pt off 4.5pt}] (461,189) -- (480.57,199.1);
\draw [shift={(488,205.5)}, rotate=205.31] [fill={rgb, 255:red, 0; green, 0; blue, 0}] [line width=0.08] [draw opacity=0] (12,-3) -- (0,0) -- (12,3) -- cycle;

\draw [color={rgb, 255:red, 208; green, 2; blue, 27}, draw opacity=1, line width=1.5]
(340.89,131.53) .. controls (340.7,134.06) and (340.52,136.46) .. (341.94,137.57)
.. controls (343.37,138.69) and (346.01,138.2) .. (348.78,137.68)
.. controls (351.55,137.17) and (354.19,136.68) .. (355.62,137.79)
.. controls (357.04,138.9) and (356.86,141.31) .. (356.67,143.83)
.. controls (356.47,146.35) and (356.29,148.76) .. (357.72,149.87)
.. controls (359.14,150.98) and (361.78,150.5) .. (364.55,149.98)
.. controls (367.32,149.46) and (369.96,148.98) .. (371.39,150.09)
.. controls (372.82,151.2) and (372.63,153.61) .. (372.44,156.13)
.. controls (372.24,158.65) and (372.06,161.06) .. (373.49,162.17)
.. controls (374.92,163.28) and (377.56,162.79) .. (380.33,162.28)
.. controls (383.09,161.76) and (385.73,161.27) .. (387.16,162.39)
.. controls (388.59,163.5) and (388.41,165.9) .. (388.21,168.42)
.. controls (388.16,169.07) and (388.11,169.72) .. (388.09,170.33);

\draw (440,210) node [anchor=north west][inner sep=0.75pt] [color={rgb, 255:red, 74; green, 144; blue, 226}, opacity=1]
{$\textcolor[rgb]{0.29,0.4,0.89}{\text{conformal interface}}$};

\draw (430,225) node [anchor=north west][inner sep=0.75pt] [color={rgb, 255:red, 74; green, 144; blue, 226}, opacity=1]
{$\textcolor[rgb]{0.29,0.4,0.89}{\text{with partial transmission, $0< t \le1$}}$};

\draw (560,80) node {\large\textbf{(b)}};
\draw (388.1,172.5) node {$\textcolor{red}{\bullet}$};
\draw (395,180) node {$x=n_d$};
\draw (145,180) node {};

\foreach \i in {1,...,20}{
    \pgfmathsetmacro{\randomx}{380 + (160)*rnd}
    \pgfmathsetmacro{\randomy}{130 + (30)*rnd}
    \fill[teal!80!white] (\randomx,\randomy) circle[radius=2pt];
}

\foreach \i in {1,...,20}{
    \pgfmathsetmacro{\randomx}{380 + (160)*rnd}
    \pgfmathsetmacro{\randomy}{130 + (30)*rnd}
    \fill[orange!80!black] (\randomx,\randomy) circle[radius=2pt];
}

\end{scope}
\end{tikzpicture}

\captionof{figure}{
Schematic illustration of the effect of a conformal interface on slowly relaxing modes.
Local dissipation is introduced in the left half  at $x=n_d$, while a conformal interface is placed at the center of the system.
In the perfectly reflecting limit, $t=0$, half of the modes are localized in the nondissipative chain (green dots) and remain purely oscillating, while the other half are localized in the dissipative chain (orange dots) and decay with finite relaxation rates. For $0<t\le 1$, the formerly oscillatory modes extend into the dissipative chain and therefore acquire finite relaxation rates. In this work, we focus on modes whose relaxation times scale as $\mathcal{O}(L^3)$, which we refer to as slowly relaxing modes.
It is also useful to view the effect of the conformal interface from the following limiting cases. In the absence of an interface, local dissipation can generally drive the entire system toward a steady state. By contrast, when a perfectly reflecting conformal interface is inserted and completely decouples the two halves, the right half in (a), which is isolated from the dissipative region, does not relax. This provides an intuitive picture of how a conformal interface can suppress relaxation.
}
\label{fig:physical picture}
\end{figure*}
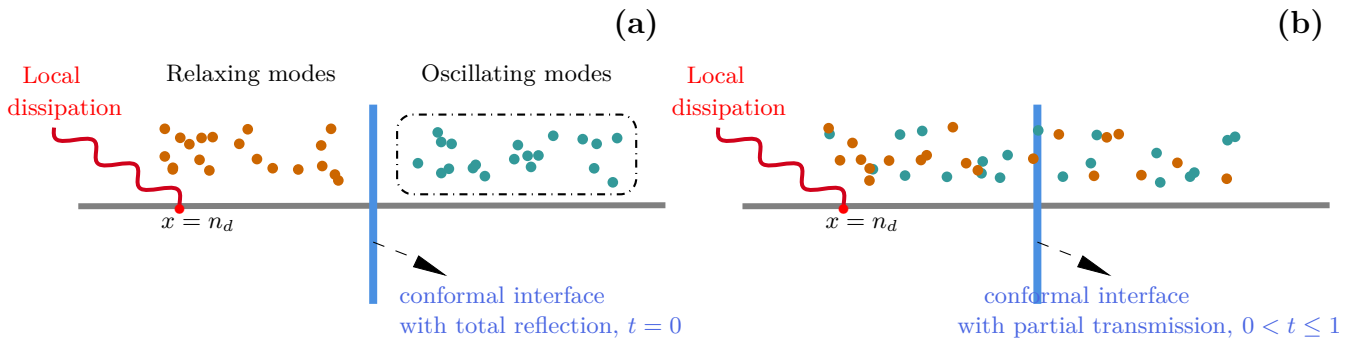

\section{Introduction}\label{sec:1}

Interfaces and impurities naturally arise in physical systems and can strongly affect both equilibrium and non-equilibrium properties. In one-dimensional quantum critical systems, interfaces provide a useful way to probe how excitations, energy, and information propagate between different parts of the system.
A particularly important class is formed by conformal interfaces \cite{Oshikawa_1996, Oshikawa_1997,Bachas_2002, Wong_1994, Chamon_2003, Oshikawa_2006}, which preserve the maximal amount of conformal symmetry and give rise to universal transmission properties \cite{Affleck_1992, Wong_1994, Affleck_1995, Quella_2002, Sakai_2008, Quella_2007, Brehm_2015, Eisler_2010, Frohlich_2004, Bill_2016, Capizzi_2022}. 

In closed systems, conformal interfaces have been studied extensively from both lattice and continuum field-theory perspectives. Their transmission properties are characterized by universal quantities such as the effective central charge $c_{\rm eff}$ \cite{Sakai_2008,Eisler_2010, Eisler_2012,Calabrese_2012,Brehm_2015,Gutperle_2016,Wen_2018, Capizzi_2023,Tang_2024,Karch_2024}, which characterizes the transmission of quantum information across the interface, and $c_{\rm LR}$ \cite{Quella_2007, Bill_2016, Brunner_2016, Meineri_2020,Bachas_2020, Bachas_2023}, which characterizes energy transmission. 
Conformal interfaces in closed quantum systems have also been studied extensively using holographic methods, where defects and interfaces are realized through dual gravitational geometries ~\cite{Bachas_2002, Bachas_2023, Bachas_2020, Karch_2023, KarchMianqi_2023, baig_2024, Karch_2024, Liu_2025}. This provides a complementary perspective on interface entanglement, boundary entropy, and other universal properties associated with defect and interface degrees of freedom.

The study of conformal interfaces has also been extended beyond conventional unitary closed systems.  More recently, non-unitary conformal interfaces and boundaries have provided another setting in which transmission and reflection properties can be generalized beyond the standard unitary framework \cite{Tang_2026,Furuta_2026,2026_Li,2025_Ye,2026_Takayanagi,2512_Zhu}. 
Conformal interfaces have also found recent applications in quantum measurement problems.
In particular, the weak measurement can be viewed as an interface separating the pre- and post-measurement states~\cite{Rajabpour2015_measurement_cft, Watanabe2016_project_holography, Swingle2022_holo_measure, Popov2022_MIPT_holo, Swingle2022_holo_measure_2, Swingle2023_holo_measure, 
Alicea2023_measure_ising, JianCM2023_measure_ising, JianSK_2023_holo_weak_measure, JianSK2023_measure_LL, Wei2023_MIPT_holo, Alicea2024_teleport_critical,2025_Tang}, and is related to an interface CFT after a spacetime rotation~\cite{JianCM2023_measure_ising, JianSK_2023_holo_weak_measure}. 
These developments further demonstrate the broad role of conformal interfaces as universal structures in quantum critical systems.

In contrast, the role of conformal interfaces in open quantum systems is much less understood. In this setting, local dissipation introduces a new kind of transmission problem: dissipation applied on one side of the interface can influence the relaxation of the full system toward a non-equilibrium steady state (See Fig.\ref{fig:physical picture}). Recent work introduced the quantity $c^{\rm global}_{\rm relax}$ \cite{Barad_2025}, defined in terms of the globally smallest nonzero decay rate, or equivalently the Liouvillian gap $g$, as a measure characterizing the transmission of dissipation, i.e., 
\footnote{It is noted that in Ref.~\cite{Barad_2025}, the symbol of $c_{\rm relax}$ was used in the definition in Eq.\eqref{eq:crelax old}. To distinguish it from the quantity introduced in the present work, we will denote the definition of Ref.~\cite{Barad_2025} by $c^{\rm global}_{\rm relax}$ throughout the remainder of this paper.  }
\begin{align}
\label{eq:crelax old}
 \frac{c^{\rm global}_{\rm relax}}{c} := \dfrac{g \text{ for partial transmission}}{g \text{ for total transmission}},
\end{align}
where $c$ is the central charge of the underlying CFT.
Physically, $c^{\rm global}_{\rm relax}$ quantifies how strongly the relaxation of the full system is suppressed by the conformal interface. Ref.~\cite{Barad_2025} further showed that, for the setup considered in Fig.\ref{fig:physical picture}, $c^{\rm global}_{\rm relax}$ is insensitive to microscopic details such as the dissipation strength and the location where we introdcue the local dissipation.

\smallskip

Motivated by the lattice results of Ref.~\cite{Barad_2025}, Karch and Wang recently investigated dissipation across conformal interfaces in a strongly coupled holographic setting~\cite{Karch_2026}. They considered a two-dimensional holographic interface CFT with only one side of the system coupled to a thermal bath. By analyzing the resulting damping of bulk modes, they identified a projected ultraviolet quantity $c_{\rm UV}$ that characterizes the suppression of dissipation induced by the interface. Remarkably, in the large-frequency limit they found
$
c_{\rm UV}/c=c_{\rm eff}/c,
$
with the result being independent of microscopic details of the system-bath coupling and the bath temperature. The precise relation between the holographic quantity $c_{\rm UV}$ and the relaxation coefficients $c^{\rm global}_{\rm relax}$ and $c_{\rm relax}$ considered in lattice open-system dynamics remains an open question. Nevertheless, the results in Ref.~\cite{Karch_2026} provide complementary evidence that the suppression of dissipation by a conformal interface can exhibit universal behavior governed by the interface itself.

\begin{figure}
    \centering
    \begin{tikzpicture}

     \node[inner sep=0pt] (russell) at (0pt, 0pt)
    {\includegraphics[width=0.5\textwidth]{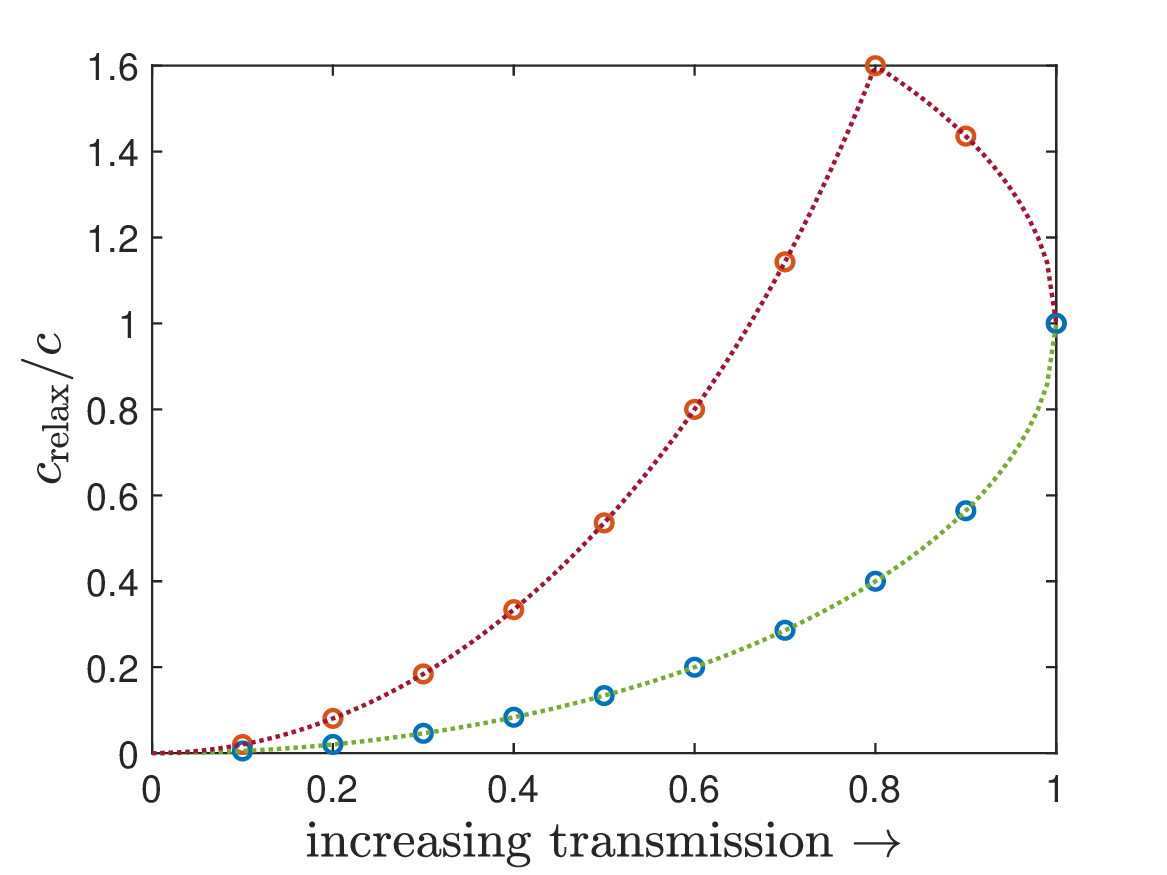}}; 

    \draw (-10pt,-12pt) node [anchor=north west][inner sep=0.75pt]  [rotate=-305.61]  
    {\textcolor{red}{\textbf{Non-universal}}};

    \draw (15pt,-42pt) node [anchor=north west][inner sep=0.75pt]  [rotate=-335.61]  
    {\textcolor[rgb]{0.25,0.46,0.02}{\textbf{Universal}}};
    
 \end{tikzpicture}
    \caption{$c_{\rm relax}$ as a function of the interface transmission for the critical harmonic chain and the critical fermion chain. For $c^{\rm global}_{\rm relax}$, defined in terms of the Liouvillian gap according to Eq.~\eqref{eq:crelax old}, the result may coincide with $c_{\rm relax}$ for some boundary conditions (green), but can become nonuniversal for others (red). See Sec.~\ref{sec:6} and Fig.~\ref{fig: Harmonic chain free left and right} for details. In contrast, the quantity $c_{\rm relax}$ introduced in this work is universal (green). 
    } 
\label{fig:relaxation coefficient}
\end{figure}

\subsection{Motivations}

Very recently, we learned from Karch, Sanyal, and Wang that the quantity $c^{\rm global}_{\rm relax}$ studied in Ref.~\cite{Barad_2025} is actually not universal for certain choices of boundary conditions in  quantum critical chains (See Sec.\ref{sec:6} for details)
\footnote{We thank Andreas Karch, Ainesh Sanyal, and Mianqi Wang for helpful discussions and personal communications regarding this subtle feature in lattice systems.}.
As illustrated in Fig.~\ref{fig:relaxation coefficient}, $c^{\rm global}_{\rm relax}$ can exhibit qualitatively different dependence on the interface transmission for different choices of boundary conditions. This observation motivates the central question of the present work:

\begin{tcolorbox}[
    colback=blue!3,
    colframe=blue!30,
    boxrule=0.5pt,
    arc=1pt
]
In a locally dissipative quantum critical system, can one define a universal quantity $c_{\rm relax}$ that characterizes the transmission of dissipation across a conformal interface?
\end{tcolorbox}

In this work, we answer this question affirmatively by resolving the relaxation dynamics mode by mode.
We refer to this framework as \emph{mode-resolved relaxation} in the presence of a conformal interface. By tracking the relaxation rate of individual modes as the interface transmission is varied, we define a universal quantity $c_{\rm relax}$ that characterizes the suppression of relaxation induced by the conformal interface.

\begin{figure*}[!t]
    \centering
    \begin{tikzpicture}
            \node[inner sep=0pt] (russell) at (-250pt, 0pt)
    {\includegraphics[width=0.5\textwidth]{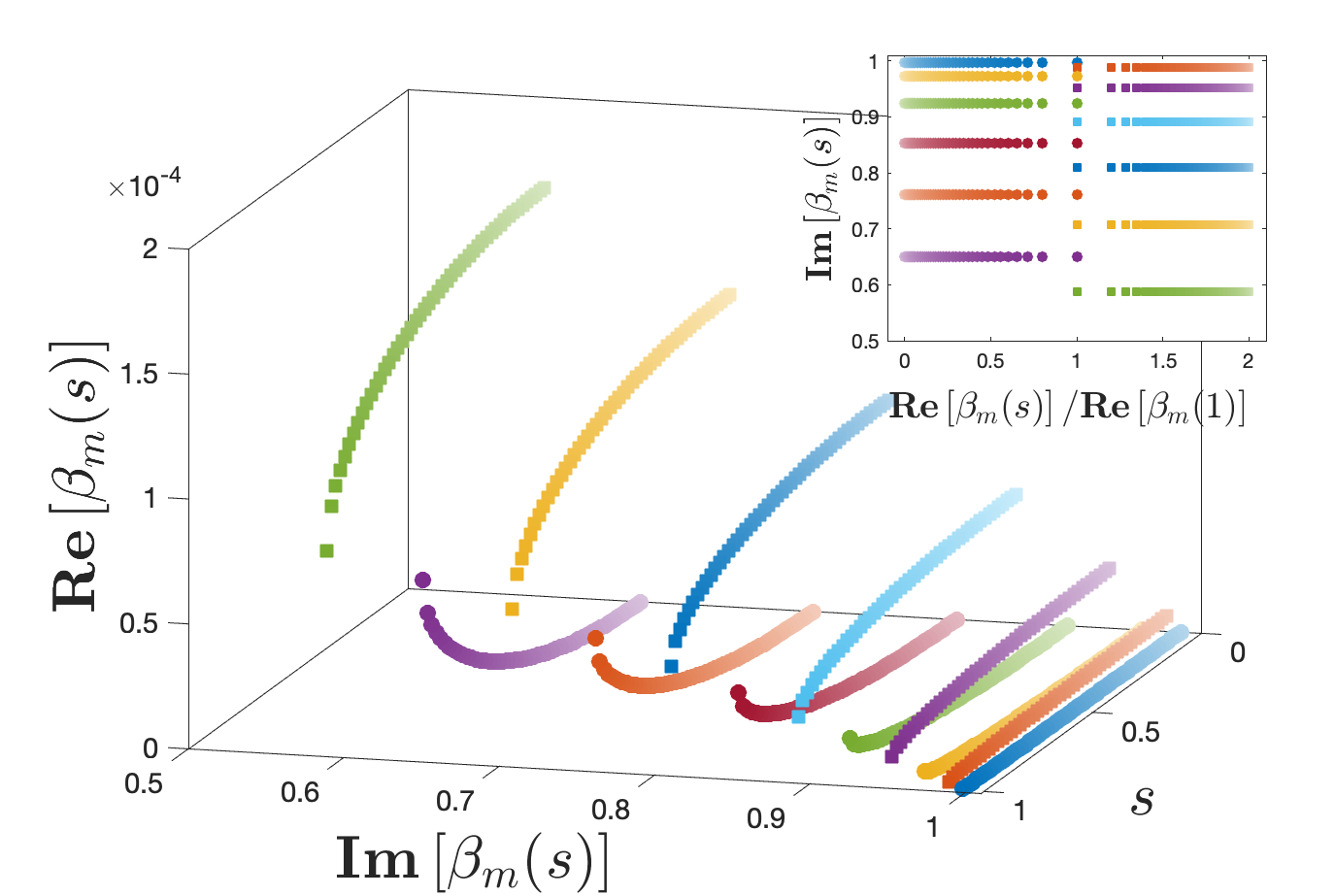}};
            \node[inner sep=0pt] (russell) at (0pt, 0pt)
    {\includegraphics[width=0.5\textwidth]{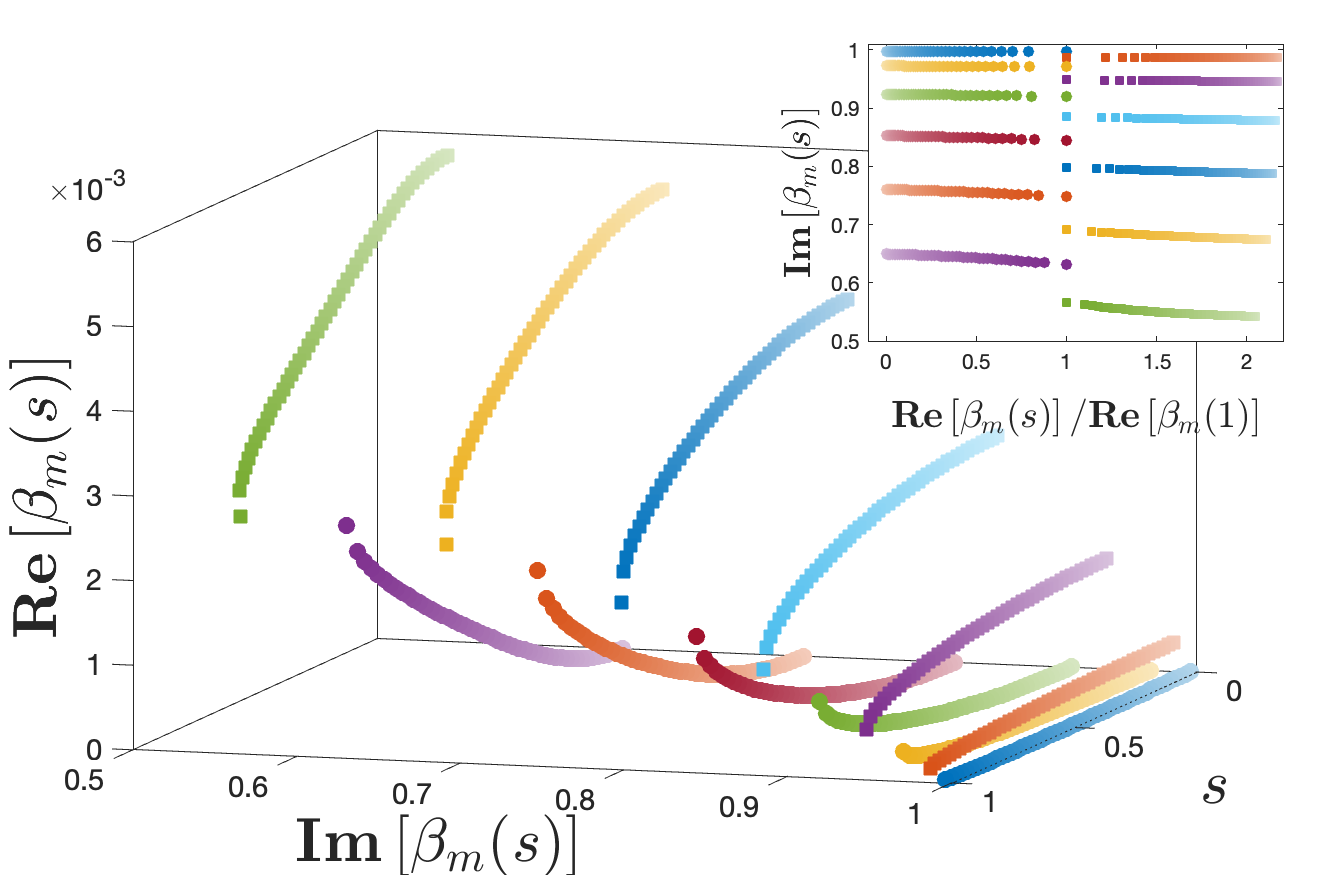}}; 
   
    \node at (-9,2) {\textbf{(a)}};
    \node at (0,2) {\textbf{(b)}};
    
    \end{tikzpicture}
    \caption{Complex single-particle eigenspectrum of the critical harmonic chain with a conformal interface coupled to a local environment. For each mode $m$, the corresponding complex eigenvalue $\beta_{m}(s)$ is shown as the transmission parameter $s$ is varied. The real part $\text{Re}\left[\beta_{m}(s)\right]>0$ characterizes the relaxation rate, while the imaginary part $\text{Im}\left[\beta_{m}(s)\right]$ determines the coherent part of the mode evolution. Here, different colors label different modes, while different shades of the same color show how each mode evolves as $s$ is varied from $1$ to $0$, with lighter shades corresponding to smaller values of $s$. Note, the curves separate into two distinct branches: the circle-marked curves, which bend downward, correspond to modes localized in the non-dissipative chain at $s=0$, while the square-marked curves, which bend upward, correspond to modes localized in the dissipative chain at $s=0$. Panel (a) shows the perturbative regime, $\gamma=0.005$, while panel (b) shows the finite-dissipation regime, $\gamma=4$. Here, we identify $t=s$ as the conformal interface parameter for the critical harmonic chain, which we introduce in  Sec.~\ref{sec:2}. The total system size is $2L=20$. }
    
    \label{fig: Mode resolved rapidities}
\end{figure*}

\subsection{Physical picture}
\label{sec:physical picture}
Before introducing the refined definition of $c_{\rm relax}$, we first describe the physical picture underlying our analysis in this work. In particular, we explain why $c^{\rm global}_{\rm relax}$ can become non-universal and why a mode-resolved description of relaxation is needed.
This discussion is also intended to provide an intuitive introduction to our main ideas for readers who are less familiar with open quantum critical systems.

\smallskip
We begin by illustrating how a conformal interface can suppress relaxation. In a one-dimensional quantum critical system without an interface, local dissipation can generally drive the entire system toward a steady state. By contrast, as shown in Fig.~\ref{fig:physical picture}(a), a perfectly reflecting conformal interface completely decouples the two halves of the system. The right half, being isolated from the dissipative region, therefore does not relax. This limiting case provides a simple physical picture of how a conformal interface can suppress relaxation.

As the interface parameters are varied away from the perfectly reflecting limit, the two halves become increasingly coupled. One therefore expects the relaxation induced on the nondissipative side to become stronger as the transmission coefficient increases. Equivalently, the suppression of relaxation by the conformal interface is expected to weaken monotonically with increasing interface transmission.

Second, we provide an intuitive picture for why the quantity $c^{\rm global}_{\rm relax}$ defined in Eq.\eqref{eq:crelax old} is not universal for general choices of boundary conditions. As discussed in detail in Sec.~\ref{sec:6} and illustrated in Fig.\ref{fig: Harmonic chain free left and right}, for certain boundary conditions, the slowest decay rate can be associated with different modes as the transmission of the conformal interface is varied. This mode switching leads to non-smooth behavior in $c^{\rm global}_{\rm relax}$, as shown in Fig.\ref{fig:relaxation coefficient}.
By contrast, when no mode switching occurs in the slowest decay rate, $c^{\rm global}_{\rm relax}$ remains a smooth function of the interface transmission, as also shown in Fig.\ref{fig:relaxation coefficient}, and exhibits the same universal behavior as the refined quantity $c_{\rm relax}$ introduced below.

To resolve the non-universality of $c^{\rm global}_{\rm relax}$, we move beyond the Liouvillian gap and instead study relaxation in a mode-resolved manner. Specifically, we continuously track a given mode as the interface transmission is varied and examine the corresponding decay rate as a function of interface transmission parameter $t$, based on which one can define the universal $c_{\rm relax}$. This procedure avoids the mode-switching ambiguity inherent in defining relaxation solely through the smallest decay rate.

\smallskip
The mode-resolved picture also provides a simple physical interpretation of how a conformal interface suppresses relaxation. Since the dissipation is local, the decay rate of a given mode is controlled by how much weight that mode has in the dissipative region. 
The effect of the conformal interface is  to redistribute the spatial profile of the modes. For a mode predominantly supported on the nondissipative side (see Fig.\ref{fig:physical picture}), decreasing the interface transmission suppresses its penetration into the dissipative region and hence reduces its weight at the dissipative site. As a result, the corresponding decay rate decreases and the relaxation time becomes longer. In the perfectly reflecting limit, the two sides are completely decoupled, and modes localized on the nondissipative side have vanishing overlap with the local dissipation and therefore do not relax.

This picture also clarifies why a mode-resolved description is natural. By continuously following the same mode as the interface transmission is varied, one directly tracks how the conformal interface modifies the mode weight seen by the local dissipation. The resulting change in the decay rate then provides a direct measure of the suppression of relaxation induced by the interface.

\subsection{Main results}
\label{sec:main results}
In this work, we address the question raised in the motivation
using two one-dimensional quantum critical systems with a conformal interface: a critical free-fermion chain and a critical harmonic chain, each studied with different choices of boundary conditions.

For later comparison, we recall that $c^{\rm global}_{\rm relax}$ was defined as the ratio between the relaxation rate of the full system in the presence of a partially transmissive interface, $g(t)$, and that for a fully transmissive interface, $g(1)$:
\begin{align}
\label{eq:old crelax formula}
\dfrac{c^{\text{global}}_{\rm relax}}{c} := \dfrac{g(t)}{g(1)},\quad 0\le t\le 1,
\end{align}
where $t$ denotes the transmission parameter of the conformal interface. Here, $t=0$ corresponds to the perfectly reflecting limit, while $t=1$ corresponds to the fully transmissive limit.

For the setup considered in Ref.~\cite{Barad_2025} (See also Fig.\ref{fig:physical picture}), this quantity was found to exhibit universal behavior in the weak-dissipation regime, in the sense that it is independent of both the dissipation site $n_d$ and the dissipation strength $\gamma$. In particular,
\begin{align}
\label{c_global_intro}
    \dfrac{c^{\rm global}_{\rm relax}}{c}  \approx 1-\sqrt{1-t^{2}}. 
\end{align}
For boundary dissipation, this relation continues to hold even at finite dissipation strength $\gamma$.

As discussed in the motivation and analyzed in detail in Sec.~\ref{sec:6}, however, changing the boundary conditions of the critical chains can lead to non-universal behavior of $c^{\rm global}_{\rm relax}$, as illustrated in Fig.~\ref{fig:relaxation coefficient}.

\smallskip

Following the physical picture discussed in Sec.\ref{sec:physical picture}, we first analyze relaxation at the level of individual single-particle modes.
For the free systems considered here, the dissipative dynamics can be reduced to an effective non-Hermitian single-particle problem. Its complex eigenspectrum allows the modes to be resolved individually, as shown in Fig.~\ref{fig: Mode resolved rapidities}. The imaginary parts of the eigenvalues determine the oscillation frequencies, while the real parts determine the relaxation rates. Because the oscillation frequencies remain well separated as the interface transmission is varied, the modes can be tracked continuously without crossings. This makes it possible to define a mode-resolved relaxation coefficient for each individual mode.

\begin{figure*}[!t]
    \centering
    \begin{tikzpicture}
            \node[inner sep=0pt] (russell) at (-250pt, 0pt)
    {\includegraphics[width=0.5\textwidth]{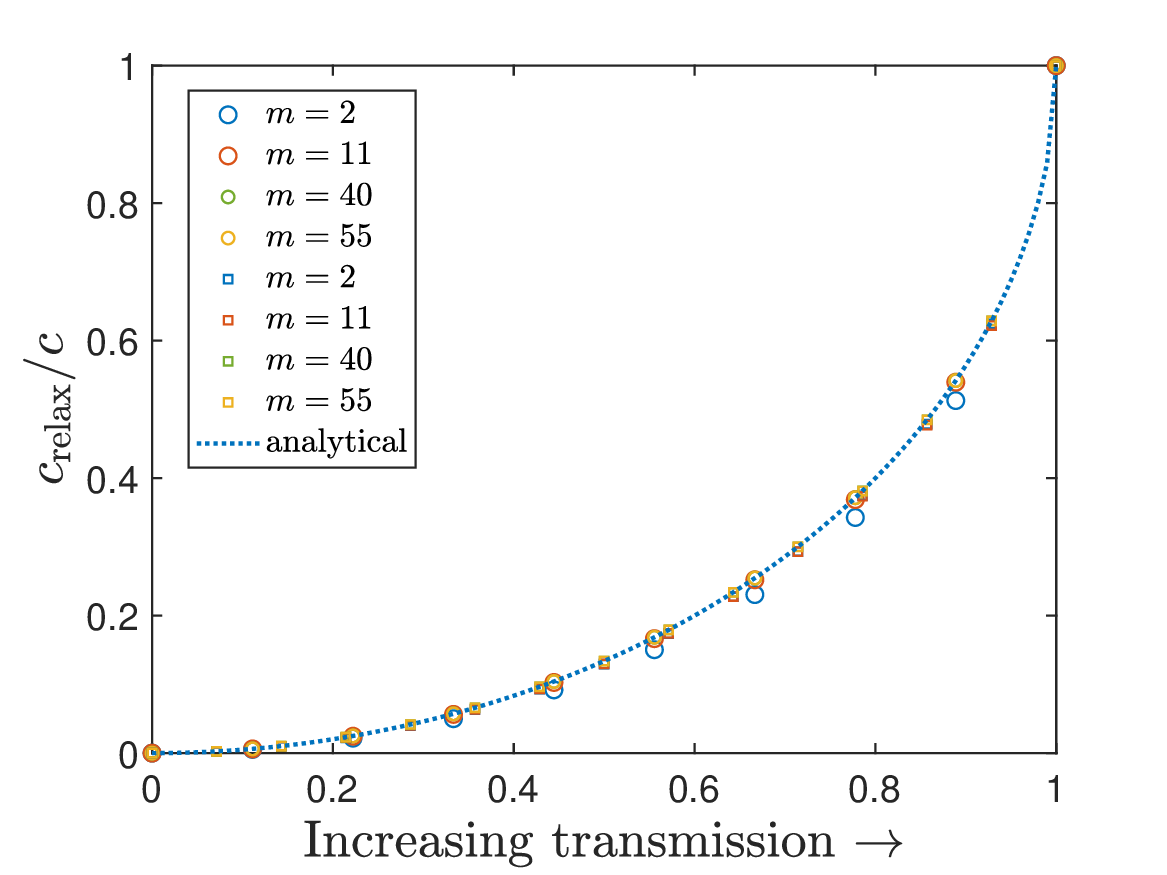}};
            \node[inner sep=0pt] (russell) at (0pt, 0pt)
    {\includegraphics[width=0.5\textwidth]{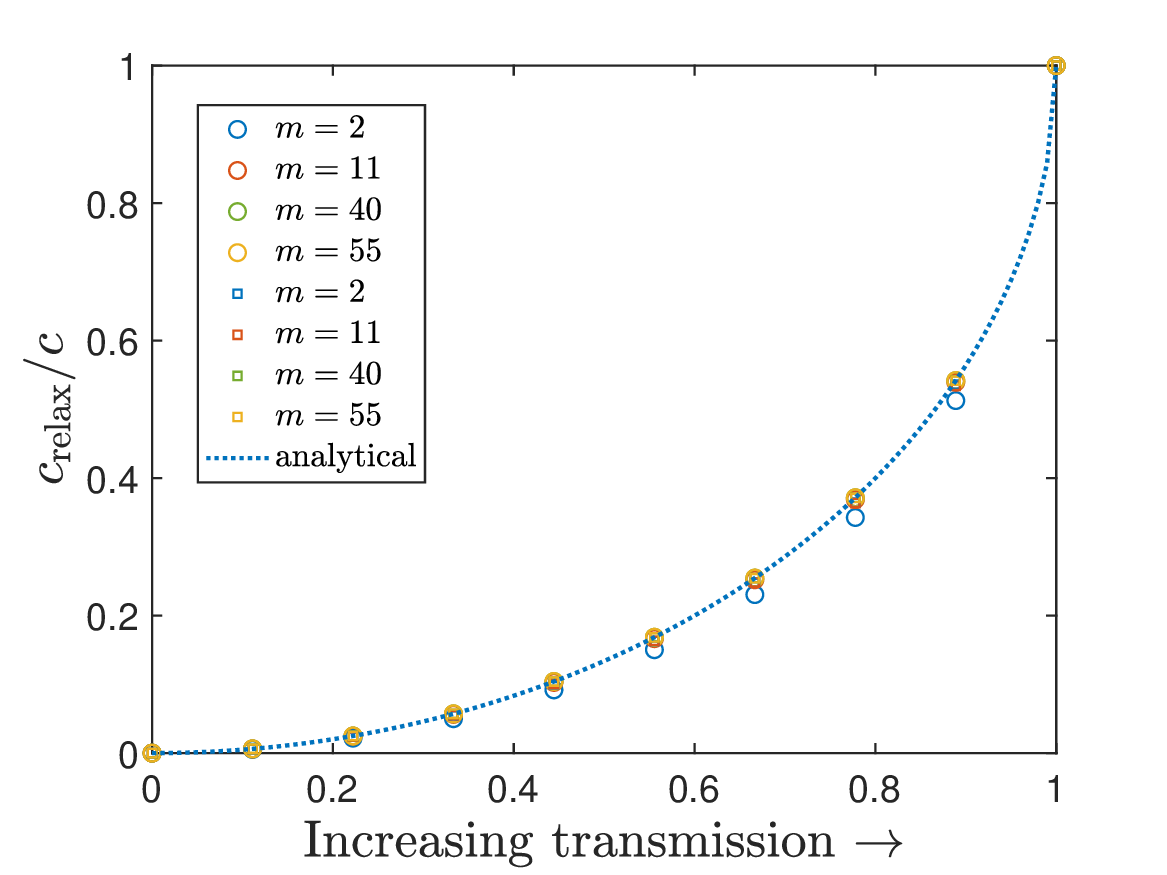}}; 
   
    \node at (-8,2) {\textbf{(a)}};
    \node at (1,2) {\textbf{(b)}};
    
    \end{tikzpicture}
    \caption{Mode-resolved relaxation coefficient $c_{\rm relax}$ in Eq.\eqref{eq: new crelax} for different modes in the weak-dissipation regime. Panel (a) shows free boundary conditions, while panel (b) shows fixed boundary conditions. Here, $m$ denotes the mode index. For each panel, we consider two choices of boundary dissipation: circles correspond to local dissipation site $n_d$ in the left chain, while squares correspond to local dissipation site $n_d$ in the right chain.}
    \label{fig: figure new crelax}
\end{figure*}

Based on the relaxation rate of the $m$-th mode 
in the presence of a conformal interface, we define $c_{\rm relax}$ as follows:
\begin{align}
\label{eq: new crelax}
    \dfrac{c_{\rm relax}}{c}  := \dfrac{g_m(t)-g_m(0)}{g_m(1)-g_m(0)},
\end{align}
which characterizes how the conformal interface modifies the relaxation of individual modes relative to the perfectly reflective limit.

Note, for the modes $m$ localized in the non-dissipative region at $t=0$ (See Fig.\ref{fig:physical picture}), these modes are purely oscillating, so $g_{m}(0) =0$. Therefore the above definition of $c_{\rm relax}$ in \eqref{eq: new crelax} reduces to 
\begin{align}
\label{eq: new crelax_01}
    \dfrac{c_{\rm relax}}{c}  := \dfrac{g_m(t)}{g_m(1)},
\end{align}
which has the same form as \eqref{eq:old crelax formula}, but is now applied to a fixed mode $m$ rather than to the Liouvillian gap.
More generally,
\eqref{eq: new crelax} also applies to modes localized in the dissipative region at $t=0$, for which the relaxation rate is already finite, $g_m(0)>0$. In this case, $c_{\rm relax}$ characterizes the change in the relaxation rate of the same mode relative to the perfectly reflecting limit.

By studying both the critical harmonic chain and the critical free-fermion chain, we find that, in the weak-dissipation regime,
\begin{align}
\label{c_relax_value}
     \frac{c_{\rm relax}}{c} \approx 1-\sqrt{1-t^{2}}.
\end{align}
We note that the expression for $c_{\rm relax}$ in Eq.~\eqref{c_relax_value} has the same functional form as that previously obtained for $c^{\rm global}_{\rm relax}$ in Eq.~\eqref{c_global_intro}. The crucial difference, as discussed in Secs.~\ref{sec:6} and~\ref{sec:7}, is that Eq.~\eqref{c_global_intro} can fail for certain choices of boundary conditions, whereas Eq.~\eqref{c_relax_value} remains universal across all microscopic variations considered in this work.


As shown in Fig.\ref{fig: figure new crelax},  the refined mode-resolved relaxation coefficient $c_{\rm relax}$ defined in Eq.\eqref{eq: new crelax} exhibits universal behavior in the weak-dissipation regime: it is independent of the choice of boundary conditions, the dissipation strength, and the position of the local dissipation. 
This constitutes the main result of this work. For the special case of boundary dissipation, we further obtain analytical expressions for the relaxation rates of the slowly decaying modes, which relax on a time scale of order $\mathcal{O}(L^{3})$ \cite{2021_Yamanaka,2020_Katsura}.

In addition, for both solvable lattice models considered in this work, the inequalities relating $c_{\rm
relax}$, $c_{\rm LR}$ and $c_{\rm eff}$ remain the same as in our previous work \cite{Barad_2025},  i.e., 
\begin{align}
   0 \le c_{\rm relax} \le c_{\rm LR} \le c_{\rm eff} \le c. 
\end{align}

\bigskip

The structure of the rest of this work is organized as follows. In Sec.~\ref{sec:2}, we review the two quantum critical systems studied in this work, namely the critical harmonic chain and critical free fermion chain, each with a conformal interface. For each quantum critical system, we consider two choices of boundary conditions. In Sec.~\ref{sec:3}, we introduce local dissipation in the presence of a conformal interface. We then analyze the spectral properties of the resulting open system for both setups using third quantization, formulated in terms of an effective non-Hermitian Hamiltonian. Building on this discussion, in Sec.~\ref{sec:4}, we calculate the relaxation rates of the slowly decaying modes for both setups in the perturbative regime. These modes relax on a time scale of order $\mathcal{O}(L^3)$, where $2L$ is the system size. We consider local dissipation applied at the boundary on one side of the interface, and then repeat the analysis for a boundary on the other side. In Sec.~\ref{sec:5}, we extend the above analysis to finite dissipation strength. In Sec.~\ref{sec:6}, we discuss the Liouvillian gap of the two chains connected through a conformal interface and comment on the previous Liouvillian gap based definition of $c^{\rm global}_{\rm relax}$. Based on this, in Sec.~\ref{sec:7} we introduce the mode-resolved relaxation coefficient Eq.\eqref{eq: new crelax}, which is the main result of this work. In Sec.~\ref{sec:8}, for completeness, we extend the perturbative analysis to an arbitrary choice of the local dissipation site on both sides of the interface in the weak dissipation regime. For illustrating this more general case, we present numerical results for Harmonic chain with free boundary condition. In Sec.~\ref{sec:9}, we conclude with a discussion and mention several ongoing and future directions along this line. Additional details and technical derivations are provided in the appendices. In Appendix~\ref{section:Eigenvalues and Eigenfunctions}, we present the details of the single-particle eigen spectrum of both setups for the two choices of boundary conditions. Next, in Appendix~\ref{appendix:Equation of motion for the covariance matrix}, we provide an alternative approach to deriving the effective non-Hermitian Hamiltonian from the equation of motion for the two-point correlation matrix in the harmonic chain. The corresponding derivation for the free-fermion chain was given in our previous work. Based on this, in Appendix~\ref{appendix:Finite dissipation case}, we provide the details of the eigenvalue problem for the harmonic chain at finite dissipation strength. Finally, in Appendix~\ref{appendix:cLR and ceff Harmonic chain}, we show how to extract $c_{\rm LR}$ and $c_{\rm eff}$ for the critical harmonic chain with both choices of boundary conditions.

\section{Setup and model}\label{sec:2}
As discussed in the introduction, we consider two examples of quantum critical systems with a conformal interface: a harmonic chain and a free fermion chain. For each system, we study two choices of boundary conditions, namely free boundary conditions at both ends and fixed boundary conditions at both ends. In this section, we briefly introduce the two setups and their corresponding boundary conditions. Details of the single-particle eigenspectrum for both setups and both choices of boundary conditions are provided in Appendix~\ref{section:Eigenvalues and Eigenfunctions}.

\smallskip
Throughout this work, we use $t$ as a common transmission parameter to characterize the conformal interface in both setups, with $t=s$ for the harmonic chain and $t=\lambda$ for the free fermion chain. Further details are provided below.

\subsection{Harmonic chain}

In this subsection, we briefly recall the quantum harmonic chain introduced in Ref.~\cite{Eisler_2012}. The system consists of $2L$ harmonic oscillators coupled by springs, with a conformal interface located at the center of the chain. The Hamiltonian is given by
\begin{equation}
\begin{aligned}
H = &\sum_{n=1}^{2L}\left(-\frac{1}{2m_n}\frac{\partial^2}{\partial x^2_n} + \frac{1}{2}m_n\Omega_0^2x^2_n\right) + V(x_1) + V(x_{2L}) \\&+\frac{1}{2}\sum_{n=1}^{2L-1}D_n\left(x_n - x_{n+1}\right)^2,\label{eq:initial Hamiltonian}
\end{aligned}
\end{equation}
with $\Omega_0$ denoting the oscillator eigenfrequency. In this setup, the spring constants  $D_n$ and masses $m_n$ are chosen such that the ratio $D_n/m_n$ is fixed on both sides of the interface. In particular, away from the interface, their values are given by
\begin{align}
\displaystyle D_n = m_n= \begin{cases}K_1 = e^{\theta}\hspace{1mm} \quad \hspace{1mm}n<L,\\K_2 = e^{-\theta}\hspace{1mm}\quad\hspace{1mm}n>L, \end{cases}
\label{eq:conformal parameters}
\end{align}
while at the center of the chain the coupling constant and mass term across the interface are given by
\begin{align}
 D_{L} = m_{L} = {\displaystyle K_0 = \frac{2K_1K_2}{K_1 + K_2} = {\rm sech}(\theta)}\hspace{1mm}.  
\end{align}
In the continuum limit, this model describes two free boson CFTs connected by a conformal interface, as introduced in Ref.\cite{Sakai_2008}. In this work, we also include boundary potentials $V(x_1)$ and $V(x_{2L})$ to realize both free and fixed boundary conditions at the two ends of the harmonic chain with a conformal interface. When the boundary potentials vanish,
\be
V(x_1) = V(x_{2L}) = 0,
\ee
the system has free boundary conditions at both ends. Similarly, we also consider the fixed boundary case studied in Ref.\cite{Lu_2011}, where fixed boundary conditions at both ends are realized by choosing the boundary potentials as 
\begin{align}
V(x_1) = \dfrac{1}{2}x^{2}_{1} \quad \text{and}, \quad V(x_{2L}) = \dfrac{1}
{2}x^{2}_{2L}. 
\label{eq:potential terms}
\end{align}
As studied in Ref.\cite{Eisler_2012}, the conformal interface in (\ref{eq:initial Hamiltonian}) is characterized by the parameter $\theta$, which enters through the modified central spring constant $D_L = K_0 = \rm{sech}(\theta)$. In the large $|\theta|$ limit, $K_0\to 0$, so the two chains become nearly decoupled. This corresponds to a totally reflective limit of conformal interface. In contrast, for small $|\theta|$, $K_0\to 1$, so the two chains are homogeneously coupled. This corresponds to the totally transmissive limit of the interface. Following Ref.\cite{Eisler_2012}, we parametrize the Hamiltonian in Eq.\eqref{eq:initial Hamiltonian} by the transmission amplitude $s$ of the conformal interface,
\begin{equation}
    K_{0} = \operatorname{sech}(\theta) \equiv s\in [0,1].\label{eq:transmission amplitude s}
\end{equation}
Here, $s=0$ corresponds to the totally reflective limit, while $s=1$ corresponds to the totally transmissive limit. This parallels the free fermion chain, where the conformal interface is characterized by the transmission parameter $\lambda \in [0, 1]$  ~\cite{Eisler_2010}. By rescaling the coordinates, as shown in Ref.\cite{Eisler_2012} and summarized in Appendix~\ref{section:Eigenvalues and Eigenfunctions}, the Hamiltonian in Eq.\eqref{eq:initial Hamiltonian} takes the following form:
\begin{align}
\label{eq: initial Hamitonian in simplified form}
    H = \dfrac{1}{2}\sum_{i=1}^{2L}\pi^{2}_{i} + \dfrac{1}{2}\sum_{i,j=1}^{2L}u_i\left(\mathbf{K}\right)_{ij}u_j,
\end{align}
where $u_j= \sqrt{m_j}x_j$ and $\pi_{j} = -i\partial/\partial u_{j}$, and the
non-zero elements of $\mathbf{K}$ matrix are
\begin{align}
\label{eq:K matrix}
    &K_{j,j+1} = K_{j+1,j} = \begin{cases}
        -1,  \quad j\neq L \\
        -s,  \quad j = L  
    \end{cases} \\ \nonumber \\ \nonumber
    &K_{j,j}  - \Omega^{2}_{0} = 2, \quad j \neq \{1, L, L+1, 2L\} \\ \nonumber \\\nonumber
    &K_{j,j} - \Omega^{2}_{0} = \begin{cases}
        1 + V_{0}, \quad  j  = \{1,2L\}\\
        2-\sqrt{1-s^{2}},  \quad j  = L\\
        2+\sqrt{1-s^{2}},  \quad j  = L+1.
    \end{cases}      
\end{align}
Here, $V_{0} = 0$ corresponds to free boundary conditions at both ends, while $V_{0} = 1$  corresponds to fixed boundary conditions at both ends. For the harmonic chain, we first express the Hamiltonian \eqref{eq: initial Hamitonian in simplified form} in terms of bosonic operators $a_i$ and $a^{\dagger}_{j}$ (see Appendix~\ref{appendix:Equation of motion for the covariance matrix} for details):
\begin{align}
\label{eq: Bosonic Hamiltonian}
H = \sum_{i,j = 1}^{2L}\left[A_{ij}a_i^{\dagger}a_j + \frac{1}{2}\left(B_{ij}a_i^{\dagger}a_j^{\dagger} + B_{ij}^{\ast}a_ia_j\right)\right]
+\text{const}, 
\end{align}
where the local bosonic operators satisfy 
$[a_i,a_j^\dagger] = \delta_{ij}$ and $[a_i,a_j] = [a^\dagger_i,a^\dagger_j] = 0$. 

\subsection{Free fermion chain}
Next, we briefly discuss the critical free fermion chain with a conformal interface at the middle of the chain. This setup is similar to those considered in previous work  ~\cite{Eisler_2010, Eisler_2012, Wen_2018, Capizzi_2023, Barad_2025}. However, here we also study the effect of the boundary conditions imposed at both ends of the chain. The system is described by the Hamiltonian
\be
\label{eq:H1}
H = \frac{1}{2}\sum\limits^{2L}_{i,j=1}H_{i,j}c^{\dagger}_{i}c_{j},
\ee
where $c_i^\dag$ and $c_j$ are fermionic operators that satisfy $\{c_i^\dag,c_j\}=\delta_{ij}$ and $\{c_i,c_j\}=\{c_i^\dag,c_j^\dag\}=0$. The non-zero elements of the Hamiltonian matrix in \eqref{eq:H1} are 
\be
\begin{split}
\label{eq:H2}
H_{i,i+1} = H_{i+1,i} = \begin{cases}
    -1, \quad  i\neq L\\
    -\lambda, \quad i = L
\end{cases}\\
-H_{L,L} = H_{L+1,L+1} = \sqrt{1-\lambda^{2}}, \end{split}
\ee
with boundary potential at two ends 
\begin{align}
\label{eq: boundary free fermion}
 H_{1,1} = H_{2L,2L} = -V_{0}.    
\end{align}
Here, the parameter $\lambda\in[0,1]$ characterizes the transmission through the conformal interface. Similarly, $V_{0} = 0$ corresponds to free boundary conditions at both ends, while $V_{0} = 1$ corresponds to fixed boundary conditions at both ends.\footnote{Here we emphasize that, for the lattice free-fermion model, the terms free and fixed boundary conditions are used only for notational simplicity. As discussed in Appendix~\ref{Appendix:fermion_BC}, these two cases are more appropriately referred to as Dirichlet-type and Neumann-type boundary conditions, respectively. In the following, we will continue to use the terms free and fixed boundary conditions whenever no ambiguity arises.
}

For both the critical harmonic and the free fermion chain, the Hamiltonian with a partially transmissive conformal interface, $0\le t<1$, is unitarily related to the  Hamiltonian in the fully transmissive limit, $t=1$, where the interface becomes transparent. As a result, the spectrum is unchanged, while the eigenfunctions are modified by the interface. Here, we present only the modified eigenfunctions needed for the derivation in the next section. A more detailed discussion of the eigenvalues and eigenfunctions for both setups, with either free or fixed boundary conditions, is provided in Appendix~\ref{section:Eigenvalues and Eigenfunctions}. 

Let $\phi^{\rm free}_{n}(j)$ and $\phi^{\rm fixed}_{n}(j)$ denote the eigenfunctions of the homogeneous case $(t=1)$, with free and fixed boundary conditions, respectively. For the inhomogeneous case $(0\le t<1)$, the eigenvalues remain unchanged, while the eigenfunctions are modified by the interface. Following Refs.\cite{Eisler_2012, Eisler_2010}, the modified eigenfunctions $\tilde{\phi}^{\rm free}_{n}(j)$ and $\tilde{\phi}^{\rm fixed}_{n}(j)$ which leave the eigenvalues unchanged, are given by:
\begin{enumerate}
    \item Both ends free ($V_0 = 0$)
    \begin{equation}
\tilde{\phi}^{\text{free}}_n(j) = \begin{cases}\alpha_n\phi^{\text{free}}_n(j),\quad 1 \le j \le L,\\\beta_n\phi^{\text{free}}_n(j),\quad L+1 \le j \le 2L,\end{cases}\label{eq:modified single-particle eigenfunction free}
\end{equation}
    \item Both ends fixed ($V_0 = 1$)
 \begin{equation}
\tilde{\phi}^{\text{fixed}}_n(j) = \begin{cases}\beta_n\phi^{\text{fixed}}_n(j),\quad 1 \le j \le L,\\\alpha_n\phi^{\text{fixed}}_n(j),\quad L+1 \le j \le 2L,\end{cases}\label{eq:modified single-particle eigenfunction free}
\end{equation}
\end{enumerate}
with coefficients $\alpha_n$ and $\beta_n$ characterizing the two sides of the interface. For the harmonic chain, they satisfy
\begin{equation}
\alpha^2_n = 1 + (-1)^n\sqrt{1-t^2},\quad \beta^2_n = 1 - (-1)^n\sqrt{1-t^2}.\label{eq:alpha and beta in terms of s}
\end{equation}  
For the free fermion chain, the corresponding coefficients are given by
\begin{equation}
\alpha^2_n = 1 - (-1)^n\sqrt{1-t^2},\quad \beta^2_n = 1 + (-1)^n\sqrt{1-t^2}.\label{eq:alpha and beta in terms of s fermion}
\end{equation}
Next, we couple these setups to an environment, thereby introducing local dissipation into the system.

\section{Liouvillian spectrum}
\label{sec:3}
In this section, we study the dissipative dynamics generated by local dissipation in the quantum critical systems with a conformal interface introduced in the previous section. To describe the open system dynamics, we use the Lindblad master equation \cite{Petruccione_2007},
\begin{equation}
\label{eq:Lindblad}
\hat{\mathcal{L}}[\rho] = \frac{d\rho}{dt} = -i[H, \rho] + \sum_{\mu}\left(L_\mu\rho L_\mu^{\dagger} - \frac{1}{2}\{L_\mu^{\dagger}L_\mu, \rho\}\right),
\end{equation}
where $H$ is the Hamiltonian of the system with a conformal interface, and $L_\mu$ are jump operators describing coupling to the environment. Here, we introduce local dissipation into the harmonic or free fermion chain, corresponding to particle loss or gain at a single site $n_d \in [1, 2L]$. 
For local dissipation, we consider the following choices of jump operators $L_\mu$. First, for the harmonic chain, the jump operators are chosen as
\begin{align}
\label{eq:bosonic jump operators}
    \displaystyle L_{1} = \sqrt{\gamma_l}\,a_{n_d}, \quad \text{and} \quad L_{2} = \sqrt{\gamma_g}\,a^{\dagger}_{n_d}. 
\end{align}
For the free fermion chain, we choose
\begin{align}
\label{eq:fermionic jump operators}
    \displaystyle L_{1} = \sqrt{\gamma_l}\, c_{n_d}, \quad \text{and} \quad L_{2} = \sqrt{\gamma_g}\, c^{\dagger}_{n_d}. 
\end{align}
These jump operators describe local particle gain and loss at a single site $n_d$, with dissipation strengths $\gamma_g$ and $\gamma_{l}$, respectively. With the above choice of jump operators, the system remains quasi-free, since the Hamiltonian is quadratic and the jump operators are linear in the creation and annihilation operators. As a result, the eigenspectrum of the Liouvillian in Eq.(\ref{eq:Lindblad}) can be obtained exactly using the third quantization \cite{Prosen_2008, Prosen_2010, Seligman_2010, Barthel_2022}.

In our previous work \cite{Barad_2025}, we considered only one choice of boundary conditions for which $c^{\rm global}_{\rm relax}$, defined in Eq.\eqref{eq:old crelax formula}, is independent of the local dissipation site. In other words, $c^{\rm global}_{\rm relax}$ takes the same value whether dissipation is introduced on the left or right side of the interface. In this work, we show that, for the two choices of boundary conditions introduced in Sec.\ref{sec:2}, the previous Liouvillian gap based $c^{\rm global}_{\rm relax}$ can depend on the dissipation location. More concretely, we discuss how $c^{\rm global}_{\rm relax}$ depends on the transmission properties of the conformal interface when local dissipation is introduced on either side of the interface, namely in the left half of the chain, $n_d =1, \dots, L$, or in the right half $n_d = L+1,\dots, 2L$. We also show that this quantity is sensitive to the boundary conditions imposed at two ends of the chain. As shown in Fig.\ref{fig:relaxation coefficient}, changing the boundary conditions can lead to non-universal behavior.

In this section, we analyze the spectral properties of the quadratic Liouvillian in Eq.\eqref{eq:Lindblad} using third quantization. Since this formalism for quadratic open systems is well established, we review only the necessary details required for our spectral analysis and refer readers to Refs.\cite{Prosen_2008, Prosen_2010, Seligman_2010} for detailed derivations.
\begin{figure*}
\centering
    \begin{tikzpicture}
            \node[inner sep=0pt] (russell) at (-250pt, -85pt)
    {\includegraphics[width=0.25\textwidth]{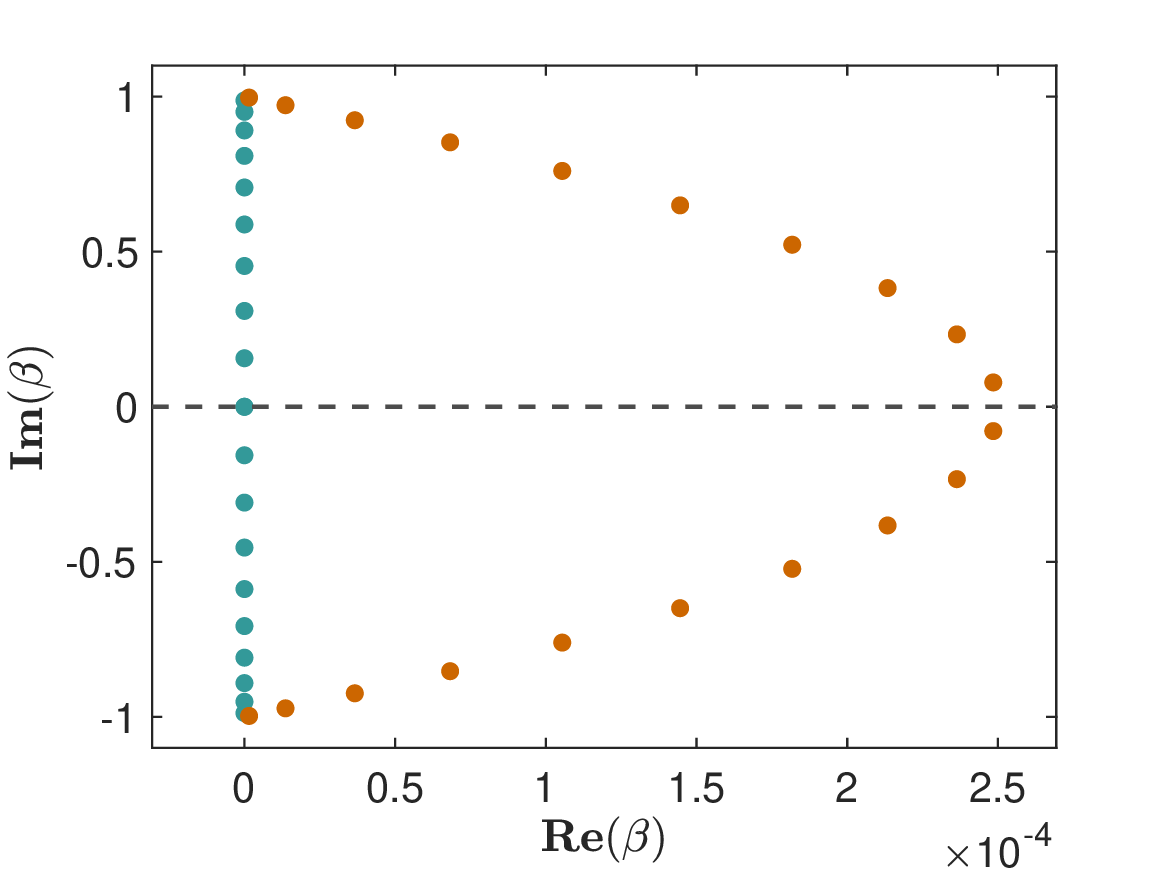}};
            \node[inner sep=0pt] (russell) at (-125pt, -85pt)
    {\includegraphics[width=0.25\textwidth]{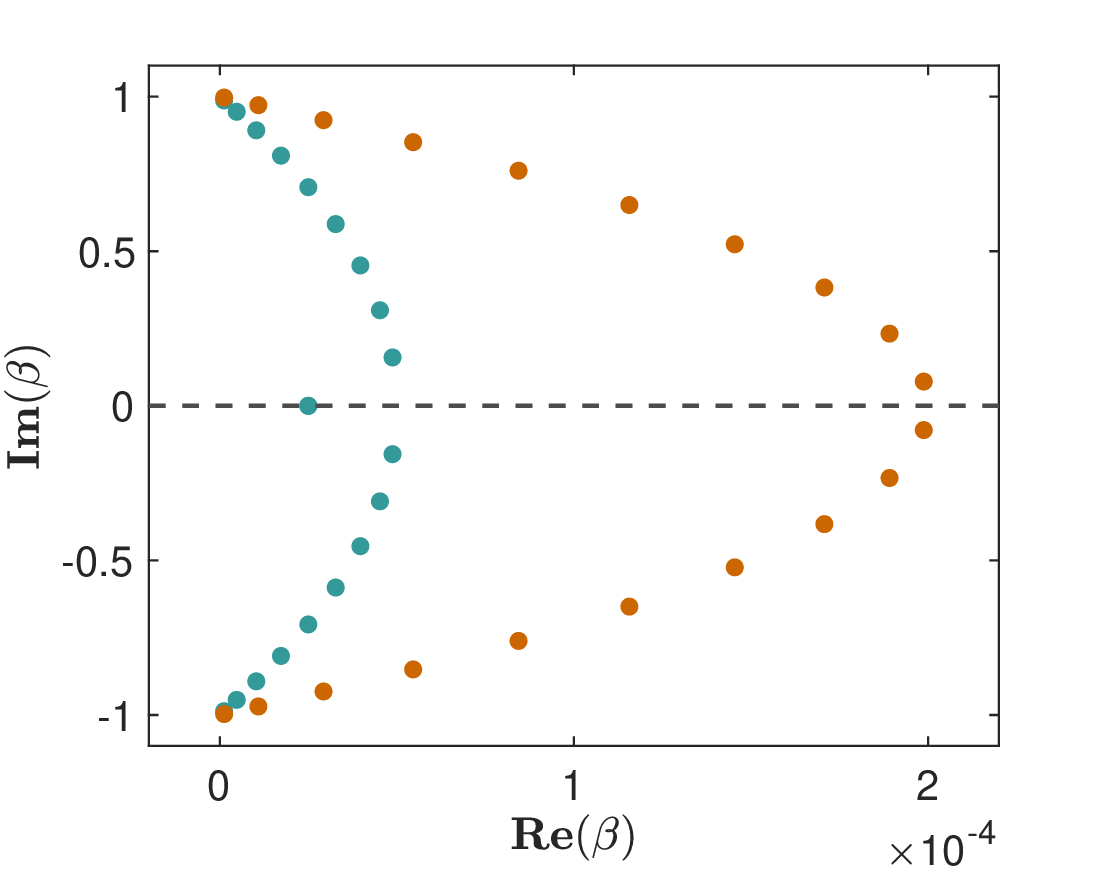}}; 
             \node[inner sep=0pt] (russell) at (0pt, -85pt)
    {\includegraphics[width=0.25\textwidth]{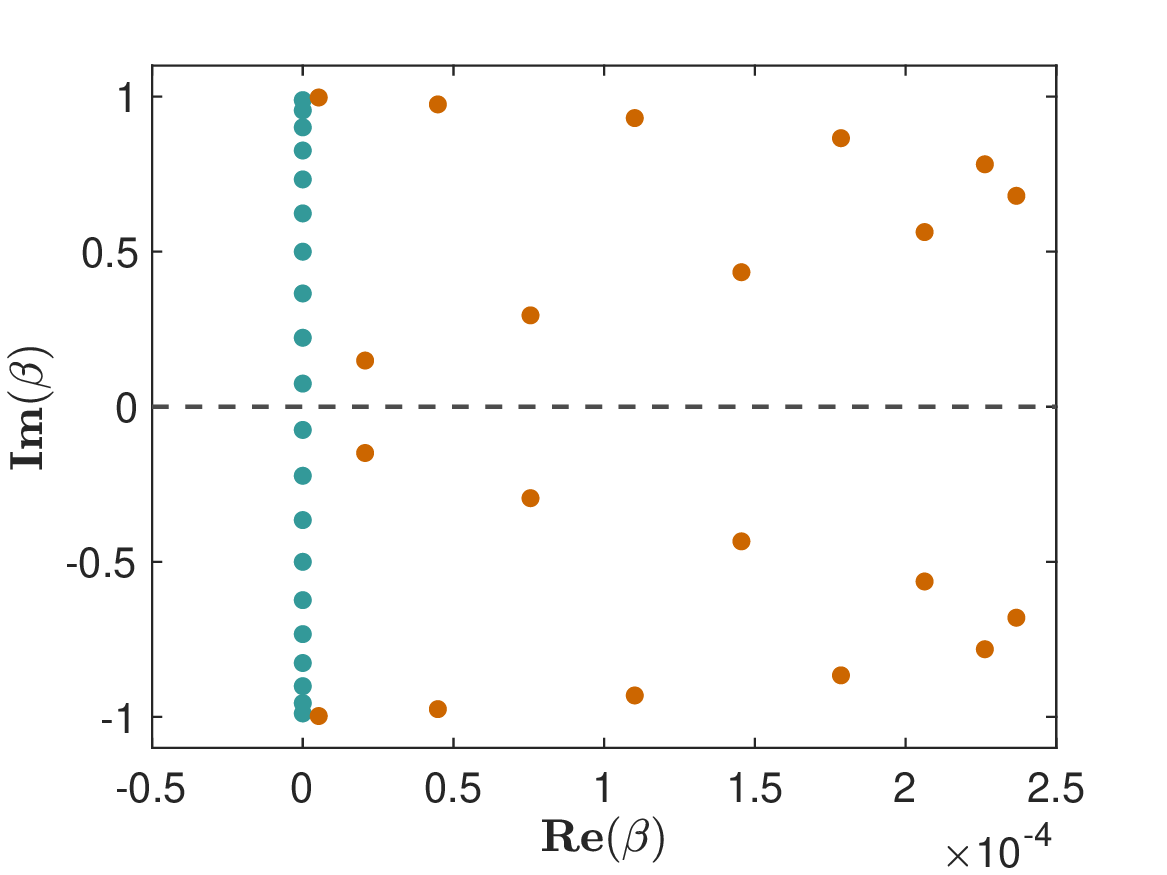}};
             \node[inner sep=0pt] (russell) at (125pt, -85pt)
    {\includegraphics[width=0.25\textwidth]{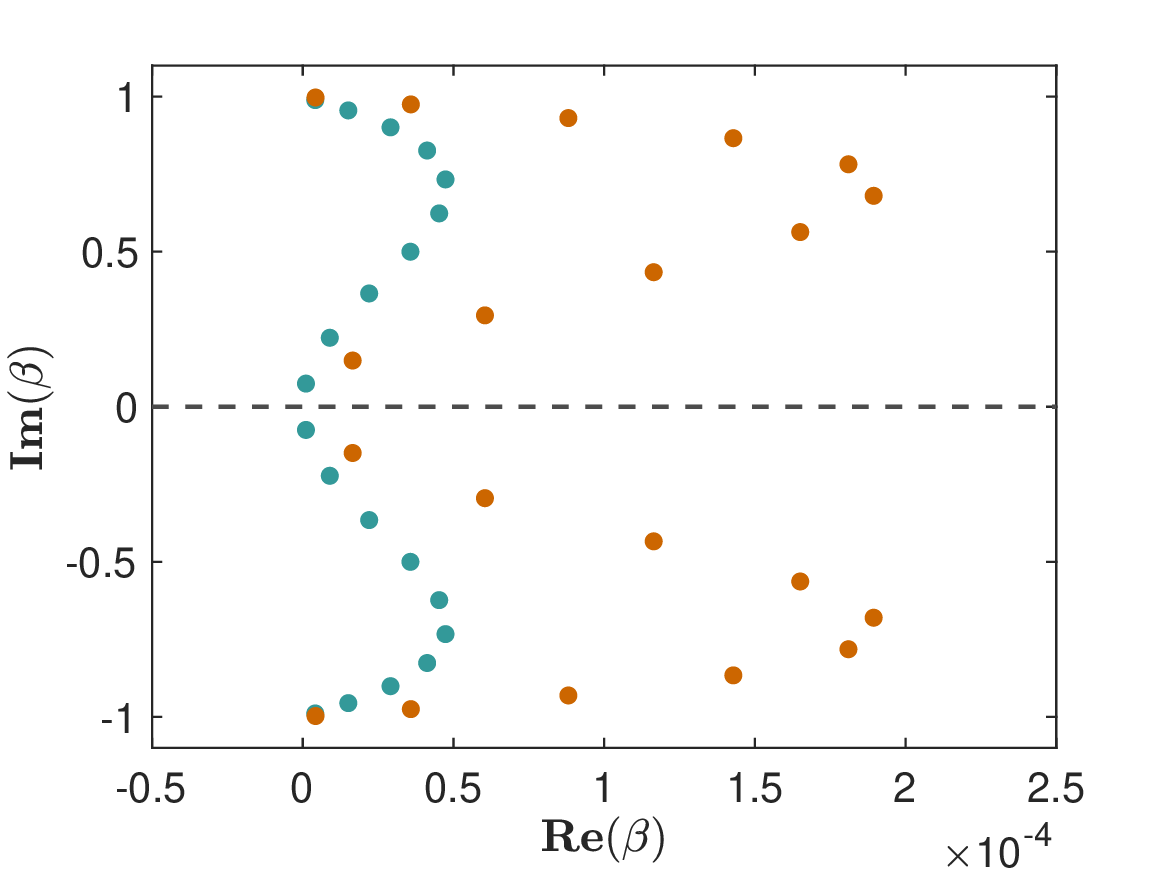}};

            \node[inner sep=0pt] (russell) at (-250pt, -185pt)
    {\includegraphics[width=0.25\textwidth]{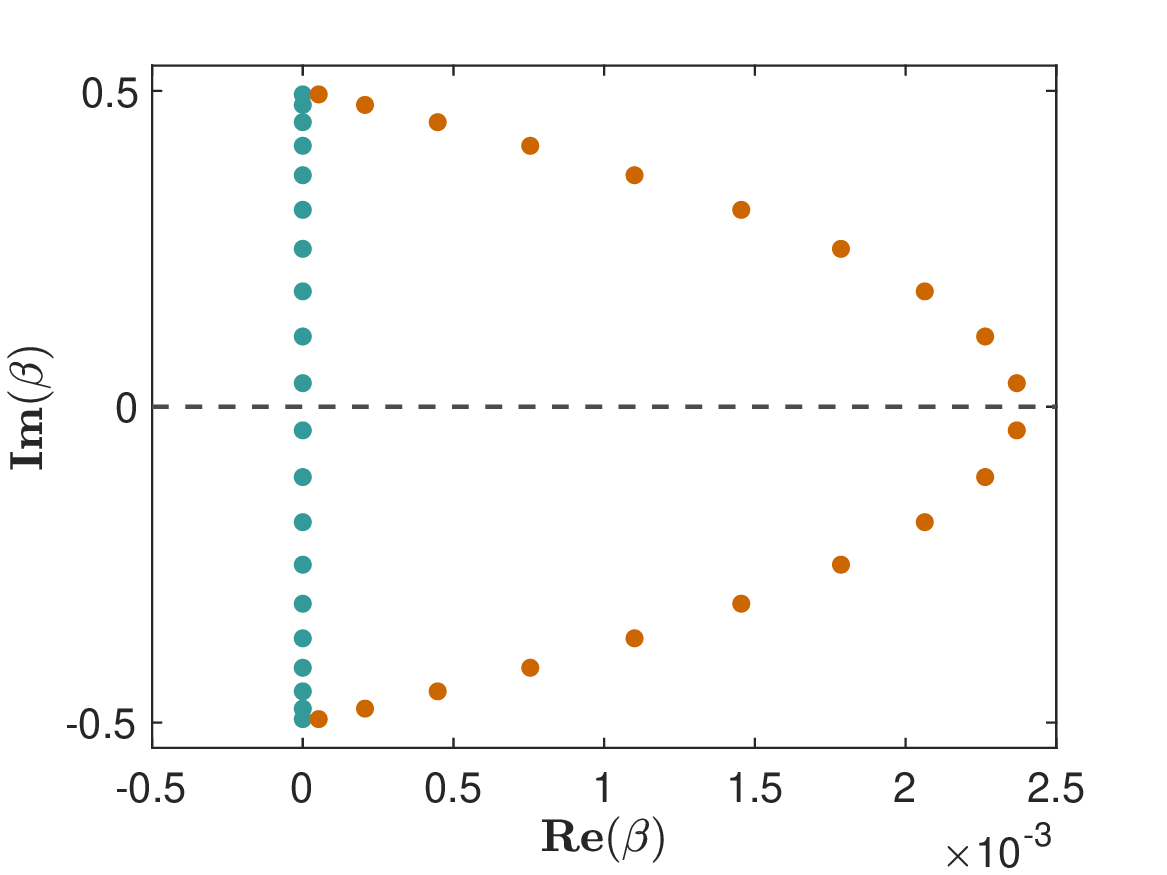}};
            \node[inner sep=0pt] (russell) at (-125pt, -185pt)
    {\includegraphics[width=0.25\textwidth]{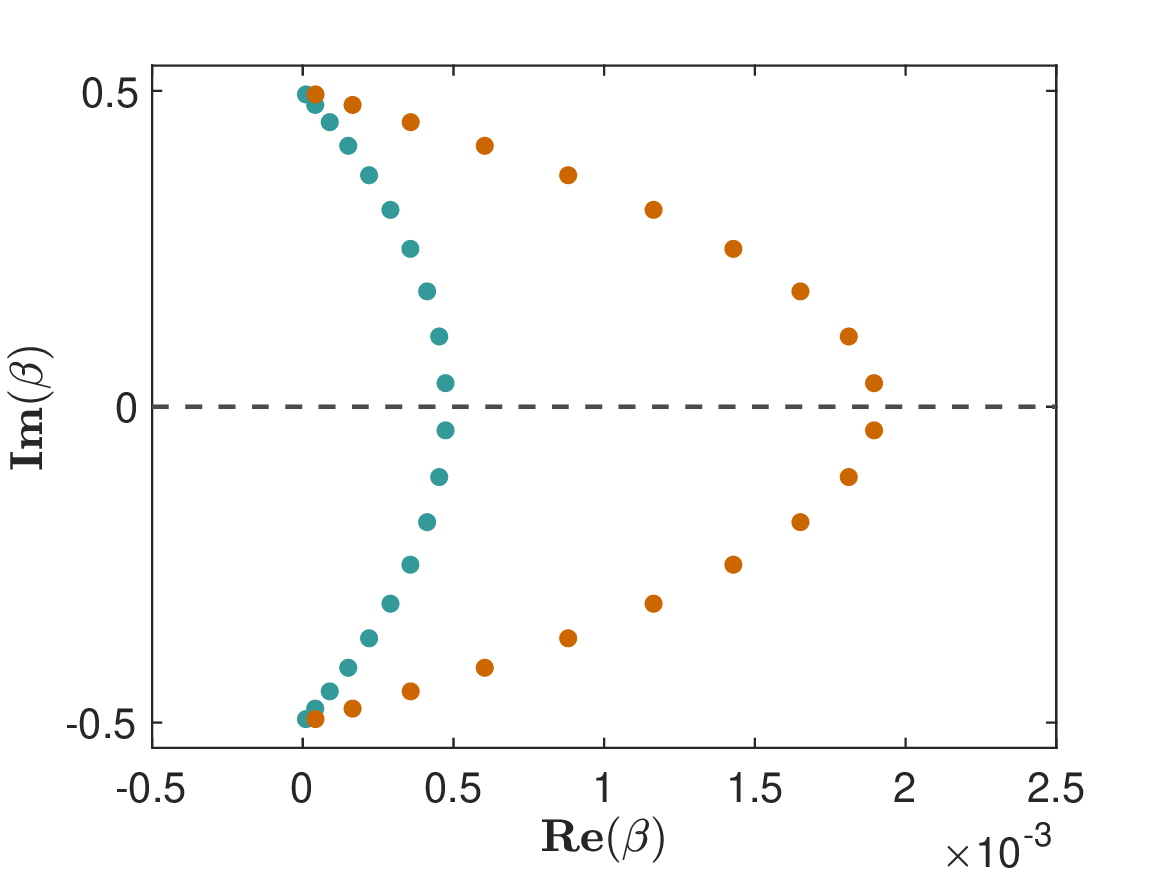}}; 
             \node[inner sep=0pt] (russell) at (0pt, -185pt)
    {\includegraphics[width=0.25\textwidth]{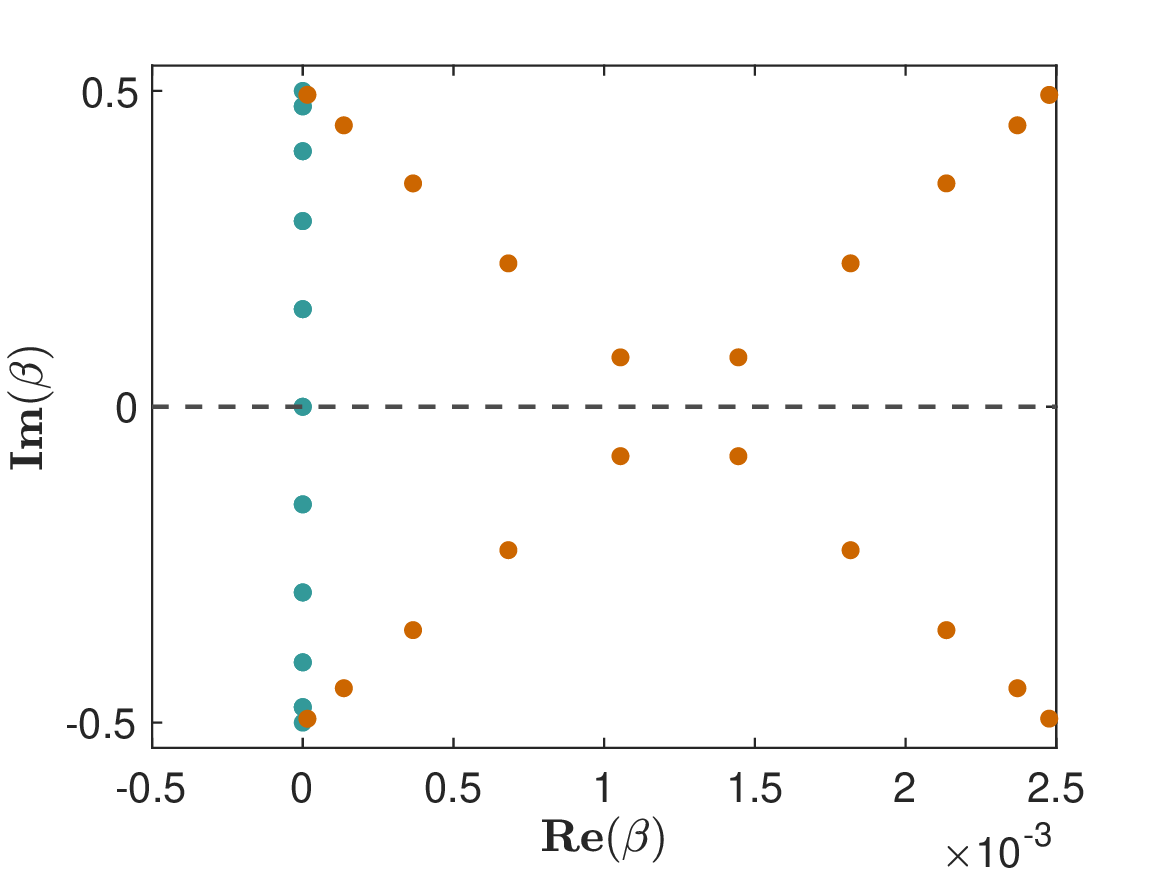}};
             \node[inner sep=0pt] (russell) at (125pt, -185pt)
    {\includegraphics[width=0.25\textwidth]{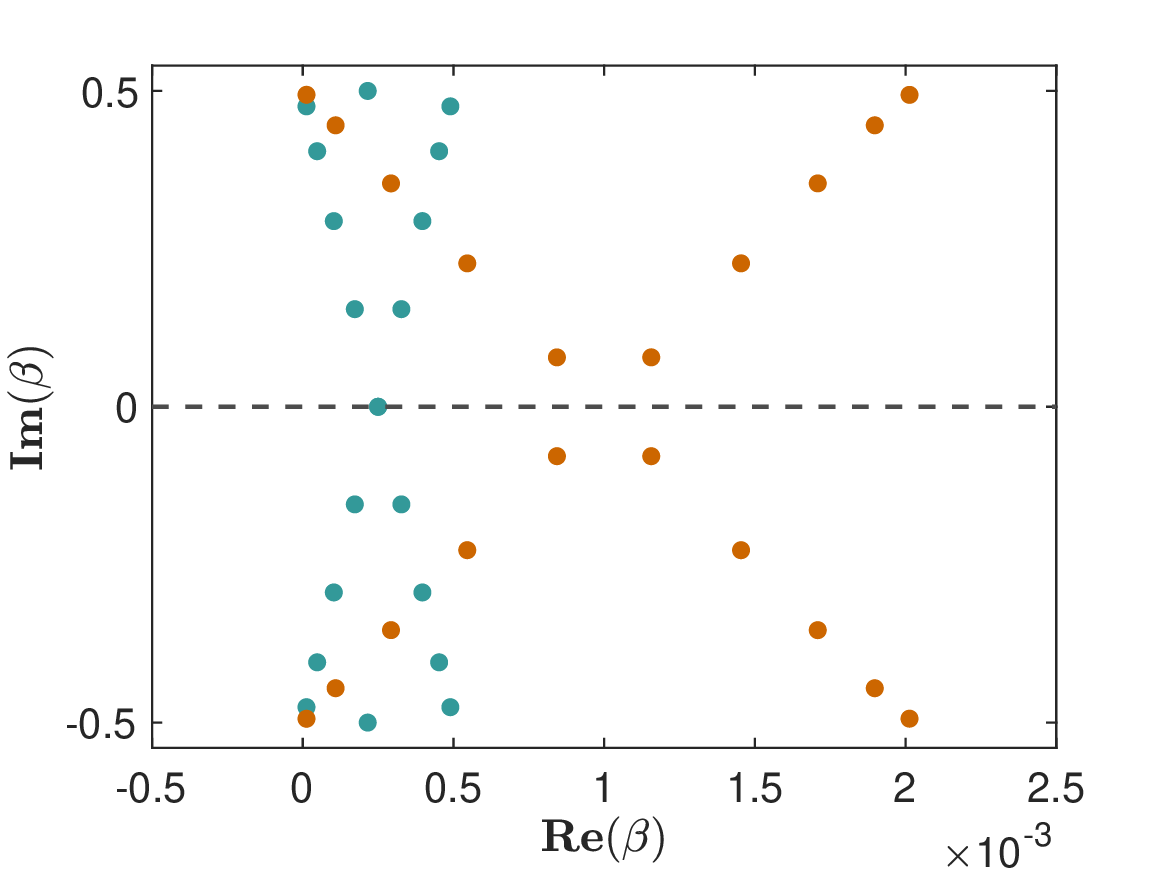}};

    \node at (-9,-2.5) {\bf (a)};
    \node at (-4.2,-2.5) {\bf (b)};
    \node at (0,-2.5) {\bf (c)};
    \node at (4.2,-2.5) {\bf (d)};
    \node at (-9,-6) {\bf (e)};
    \node at (-4.2,-6) {\bf (f)};
    \node at (0,-6) {\bf (g)};
    \node at (4.2,-6) {\bf (h)};
    \end{tikzpicture}
\caption{(a$-$d) Rapidity spectrum $\{\beta_{i}\}$ for a harmonic chain  of length $N = 2L = 20$. The local dissipation is introduced on the right boundary (i.e. $n_d =N$) with interface parameter $s$ and dissipation strengths ($\gamma_l$ and $\gamma_g$, are set as:  $\gamma_l=0.015$ and $\gamma_g=0.01$. We consider free boundary conditions in (a$-$b) and fixed boundary conditions in (c$-$d). In (a,\,\,c) we tune the conformal interface parameter $s=0$ such that green dots show purely oscillating modes that do not decay or localized in non-dissipative chain. In (b,\,\,d), we tune $s =0.8$ such that both modes have finite relaxation rate. (e$-$h) Rapidity spectrum $\{\beta_{i}\}$ for a fermionic chain  of length $N = 2L = 20$. The local dissipation is introduced on the right boundary (i.e. $n_d =N$) with interface parameter $\lambda$ and dissipation strengths ($\gamma_l$ and $\gamma_g$, are set as:  $\gamma_l=0.05$ and $\gamma_g=0$. We consider free boundary conditions in (a$-$b) and fixed boundary conditions $V_{0} = 1$ in (c$-$d). In (e,\,\,g) we tune the conformal interface parameter $\lambda=0$ such that green dots show purely oscillating modes that do not decay or localized in non-dissipative chain. In (f,\,\,h), we tune $\lambda =0.8$ such that both modes have finite relaxation rate.}
\label{fig:Rapidity spectrum}
\end{figure*}
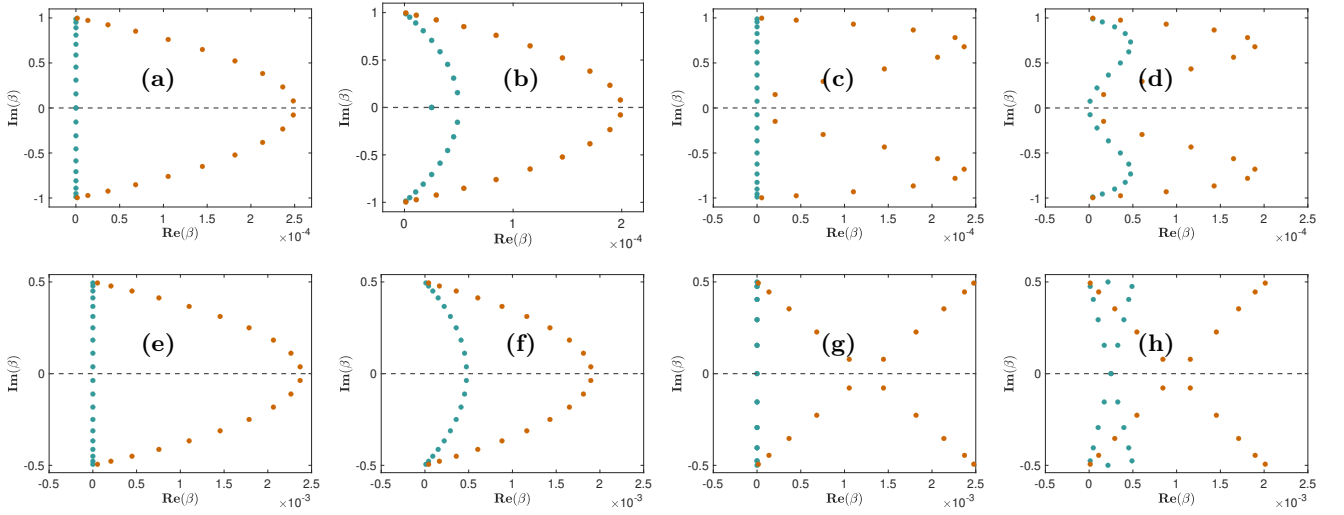

\subsection{Harmonic chain}
For our choice of jump operators in Eq.\eqref{eq:bosonic jump operators}, the Lindblad generator remains quadratic in the bosonic operators. Therefore, its spectrum can be obtained using the  third quantization formalism for bosons introduced in Ref.\cite{Seligman_2010}. In this approach, the Liouvillian eigenvalues are determined by the eigenvalues of an effective $4L\times 4L$ non-Hermitian matrix $\mathbf{X}$. The construction of this matrix follows Ref.~\cite{Seligman_2010}, to which we refer the reader for details. For a general quadratic Hamiltonian and linear jump operators, 
\begin{subequations}
\begin{align}
    &H = \mathbf{a}^{\dagger}\cdot  \mathbf{H} \mathbf{a} +  \mathbf{a}\cdot  \mathbf{K}  \mathbf{a} + \mathbf{a}^{\dagger}\cdot  \mathbf{\bar{K}}  \mathbf{a}^{\dagger}\\
    & L_{\mu} = \mathbf{l}_{\mu}\cdot  \mathbf{a} +  \mathbf{k}_{\mu}\cdot  \mathbf{a}^{\dagger}
\end{align}
\end{subequations}
the resulting $4L\times 4L$ non-Hermitian matrix $\mathbf{X}$ takes the form
\begin{align}
    \mathbf{X}  = \dfrac{1}{2}\begin{pmatrix}
        i\bar{\mathbf{H}} -\bar{\mathbf{N}} +\mathbf{M}  && -2i\mathbf{K} -\mathbf{L}+\mathbf{L}^{\rm T} \\
        2i\bar{\mathbf{K}} -\bar{\mathbf{L}}+\bar{\mathbf{L}}^{\rm T} &&  -i\mathbf{H} -\mathbf{N} +\bar{\mathbf{M}}
    \end{pmatrix},
\end{align}
with $2L\times2L$ bath matrices are defined by
\begin{align}
    &\mathbf{M} = \dfrac{1}{2}\sum_{\mu}\mathbf{l}_{\mu}\mathbf{l}^{\dagger}_{\mu},\quad 
    \mathbf{N} =\dfrac{1}{2}\sum_{\mu}\mathbf{k}_{\mu}\mathbf{k}^{\dagger}_{\mu},\quad 
    \mathbf{L} = \dfrac{1}{2}\sum_{\mu}\mathbf{l}_{\mu}\mathbf{k}^{\dagger}_{\mu}.
\end{align}
The same non-Hermitian matrix can be derived equivalently from the equations of motion for the two-point correlation functions. Since the system is quasi-free, these equations close at the level of the correlation matrix, which therefore fully determines the Gaussian dynamics (see Appendix~\ref{appendix:Equation of motion for the covariance matrix} for details). Assuming $\mathbf{X}$ is diagonalizable, one can write  
\begin{align}
    \mathbf{X} = \mathbf{P}\mathbf{\Delta}\mathbf{P}^{-1}, \quad\mathbf{\Delta} = \text{diag}\{\beta_{1},\beta_{2},\dots,\beta_{4L}\},
\end{align}
where the complex eigenvalues $\{\beta_{i}\}$ of the matrix $\mathbf{X}$ are known as \textit{rapidities} (see Fig.\ref{fig:Rapidity spectrum}). Its rapidity spectrum is then used to analyze the stability of the dynamics and to determine the Liouvillian gap.

For bosonic systems, the existence of a stable fixed point is not always guaranteed because the Hilbert space is infinite-dimensional and the occupation number can grow without bound. Stability requires the dissipative dynamics to provide sufficient damping; otherwise, gain processes can lead to unbounded growth and prevent the existence of a normalizable steady state. In other words, for the system to be stable, all rapidities should lie to the right of the imaginary axis, i.e. ${\rm Re}(\beta_{j})>0~\forall~j$. In the stable case, the Liouvillian gap is defined as
\begin{align}
\label{eq:rapidity}
    g = 2\underset{{\text{Re}[\beta_{j}]\neq0}}{\text{min}}\{\text{Re}[\beta_{j}]\}.
\end{align}
For all numerical calculations involving the harmonic chain in this work, we diagonalize the matrix $\mathbf{X}$ numerically to obtain the rapidity spectrum $\{\beta_{i}\}$. For our setup, the transpose of the non-Hermitian matrix $\mathbf{X}$ takes the following form
\begin{align}
\mathbf{X}^{T}  = \dfrac{1}{2}\left(i\Sigma_{z}\tilde{\mathbf{H}}_{\text{BdG}}(s) - \dfrac{\tilde{\mathbf{\Gamma}}}{2}\right)
\label{eq:X matrix after 3rd quantization}
\end{align}
where $\tilde{\mathbf{H}}_{\text{BdG}}(s)$ is the $4L \times 4L$ bosonic Bogoliubov--de Gennes Hamiltonian matrix defined in Eq.\eqref{eq: Bosonic Hamiltonian}, with  
\begin{align}
\label{eq:sigma_z and Gamma_tilde}
    \Sigma_{z} = \begin{pmatrix}
        \mathbb{I} & 0 \\
        0 & -\mathbb{I}
    \end{pmatrix}, \quad \text{and} \quad \tilde{\mathbf{\Gamma}} = \begin{pmatrix}
        \mathbf{\Gamma} & 0 \\
        0 & \mathbf{\Gamma}
    \end{pmatrix}.
\end{align}
Here, $\mathbf{\Gamma} = \mathbf{\Gamma}_{-} - \mathbf{\Gamma}_{+}$ is the dissipation matrix,  with  matrix elements
\begin{align}
\label{eq: gain and loss Gamma matrices}
    \left(\Gamma_{+}\right)_{i,j} = \gamma_{g}\delta_{i,n_d}\delta_{j,n_d}, \quad\quad\left(\Gamma_{-}\right)_{i,j} = \gamma_{l}\delta_{i,n_d}\delta_{j,n_d},
\end{align}
or equivalently,
\begin{align}
\label{eq: gain and loss Gamma matrices}
    \mathbf{\Gamma_{+}} = \gamma_{g}\,|n_d\rangle\langle n_d|, \quad  \quad\mathbf{\Gamma_{-}} = \gamma_{l}\,|n_d\rangle\langle n_d|.
\end{align}

In Figs.~\ref{fig:Rapidity spectrum}(a)$-$\ref{fig:Rapidity spectrum}(d), we show the complex rapidity spectrum of the harmonic chain for both choices of boundary conditions. Figures~\ref{fig:Rapidity spectrum}(a) and \ref{fig:Rapidity spectrum}(b) correspond to free boundary conditions at both ends, while Figs.~\ref{fig:Rapidity spectrum}(c) and \ref{fig:Rapidity spectrum}(d) correspond to fixed boundary conditions at both ends. For the harmonic chain, we obtain the spectrum by numerically diagonalizing the non-Hermitian matrix $\mathbf{X}$ defined in Eq.~\eqref{eq:X matrix after 3rd quantization}. In all cases shown, dissipation is applied to the right boundary, $n_d = 2L$. Here, we study the spectrum for two values of the conformal interface transmission parameter $s \in [0,1]$: the perfectly reflective limit $s=0$ and the partially transmissive case $s=0.8$. 

\subsection{Fermionic chain}
For quadratic fermionic systems, we again use the third quantization formulation \cite{Prosen_2008}. 
Instead of repeating the full construction, we start from the non-Hermitian single particle matrix $\mathbf{Z}$ whose eigenvalues generate the elementary fermionic Liouvillian rapidities (see Sec.III of Ref.\cite{Barad_2025}).
In the present work, we identify the matrix $\mathbf{Z}$ with the effective non-Hermitian Hamiltonian $\mathbf{H}_{\rm eff}(\lambda)$ as
\be\label{eq:effective_Ham}
\textbf{Z} = \dfrac{1}{2}\left( \tilde{\textbf{H}}(\lambda) - i\frac{\mathbf{\Gamma_{+}+\Gamma_{-}}}{2}\right) = \dfrac{1}{2}\mathbf{H}_{\rm eff}(\lambda),
\ee
where $\tilde{\textbf{H}}(\lambda)$ is the Hamiltonian matrix in the original complex Dirac fermion representation defined in Eq.\eqref{eq:H1}. The effective non-Hermitian matrix $\mathbf{H}_{\rm eff}(\lambda)$ contains both the Hamiltonian contribution and the local dissipative terms. The matrices $\mathbf{\Gamma_{+}}$ and $\mathbf{\Gamma_{-}}$ encode gain and loss at the dissipative site $n_d$, respectively. For the jump operators in Eq.\eqref{eq:fermionic jump operators}, their matrix elements are
\begin{align}
\label{eq: gain and loss Gamma matrices fermion}
\left(\Gamma_{+}\right)_{i,j} = \gamma_{g}\delta_{i,n_d}\delta_{j,n_d}, \quad  \quad\left(\Gamma_{-}\right)_{i,j} = \gamma_{l}\delta_{i,n_d}\delta_{j,n_d},
\end{align}
or equivalently,
\begin{align}
\label{eq: gain and loss Gamma matrices fermion}
    \mathbf{\Gamma_{+}} = \gamma_{g}|n_d\rangle\langle n_d|, \quad\quad\mathbf{\Gamma_{-}} = \gamma_{l}|n_d\rangle\langle n_d|.
\end{align}
The above matrices have nonzero entries only at the site $n_d$, where local dissipation is applied. In the fermionic case, the full Liouvillian eigenspectrum $\{\Lambda_{i}\}$ is constructed from the rapidities $\{ \beta_i \}$, defined as the eigenvalues of the non-Hermitian structure matrix $\mathbf{A}$~\cite{Prosen_2010}. The rapidities $\{ \beta_i \}$ are related to eigenvalues $\{E_i\}$ of the non-Hermitian effective Hamiltonian $\mathbf{H}_{\rm eff}(\lambda)$ by 
\be \label{eq:relation_rapidity_Z}
\beta_i = \dfrac{1}{2}\{- i E_i^*, i E_i\}.
\ee
Thus, one can express the Liouvillian gap in the following form 
\be \label{eq:gap_def_Z}
g =  -\max_{\Im[E_{i}] \neq 0} \left\{\Im[E_{i}] \right\}.
\ee 
That is, the problem is reduced to solving the eigenvalues $\{ E_i \}$ of the non-Hermitian effective Hamiltonian matrix $\mathbf{H}_{\rm eff}(\lambda)$ and identifying the largest imaginary part of the eigenvalues other than zero. 

In Figs.~\ref{fig:Rapidity spectrum}(e)--\ref{fig:Rapidity spectrum}(h), we show the rapidities of the free fermion chain in the complex plane for both choices of boundary conditions. Figures~\ref{fig:Rapidity spectrum}(e) and \ref{fig:Rapidity spectrum}(f) correspond to free boundary conditions at both ends, $V_0 = 0$, while  Figs.~\ref{fig:Rapidity spectrum}(g) and \ref{fig:Rapidity spectrum}(h) correspond to fixed boundary conditions at both ends, $V_0 = 1$. Here, the rapidities are obtained by diagonalizing the non-Hermitian structure matrix $\mathbf{A}$, defined in Eq.(21) of Ref.\cite{Barad_2025}. In all cases shown, dissipation is applied to the right boundary. We again study the spectrum for two values of the conformal interface transmission parameter $\lambda \in [0,1]$: the perfectly reflective limit $\lambda =0$ and the partially transmissive case $\lambda =0.8$.

In summary, in the perfectly reflective limit, both the harmonic chain and the free fermion chain separate into two decoupled chains: a dissipative chain and a non-dissipative chain. The modes localized in the non-dissipative chain have zero relaxation rate and appear as \textit{purely oscillating coherences}, shown by the green dots in dots in Fig.\ref{fig:Rapidity spectrum}. In contrast, the modes localized in the dissipative chain, which we choose to be the right chain, have finite relaxation rates and are indicated by the orange dots in Fig.\ref{fig:Rapidity spectrum}. In the partially transmissive case, these initially non-dissipative modes (i.e. green dots or modes localized in the non-dissipative chain at $t=0$) extend across the interface, spread over the full chain, and acquire finite relaxation rates. This behavior is visible for both choices of boundary conditions. We use these two boundary conditions to illustrate how the rapidity spectrum spreads in the complex plane as the interface transmission is increased. This provides a useful perspective on the global relaxation rate of the system, or equivalently the Liouvillian gap, as discussed in section \ref{sec:6}.

In the next two sections, we compute the relaxation rates for both classes of modes discussed above and study how the conformal interface distinguishes between them. We focus especially on the slowly decaying modes, whose relaxation times scale as $\mathcal{O}(L^3)$, or equivalently whose relaxation rates scale as $\mathcal{O}\left(1/L^3\right)$.

\section{Relaxation rate: Perturbative analysis}\label{sec:4}

In this section, our goal is twofold. First, using perturbation theory, we evaluate the relaxation rates of the slowly decaying modes, whose relaxation times scale as $\mathcal{O}(L^3)$. Second, we study how these relaxation rates depend on the transmission properties of the conformal interface. We perform this analysis for both the critical harmonic chain and the free fermion chain, considering the two boundary conditions introduced in the Sec.\ref{sec:2}. More concretely, we start from the effective non-Hermitian matrix $\mathbf{H}_{\rm eff}(t)$ and use first order perturbation theory in the weak dissipation regime, where the dissipation strength $\gamma$ is small compared with the characteristic single particle energy scale of each setup. This allows us to obtain the relaxation rates of the slowly decaying modes and determine their dependence on the conformal interface parameter $t$, the dissipation strength $\gamma$, the system size $L$ and the mode index, denoted by $m$ for the harmonic chain and $k$ for the free-fermion chain.
 
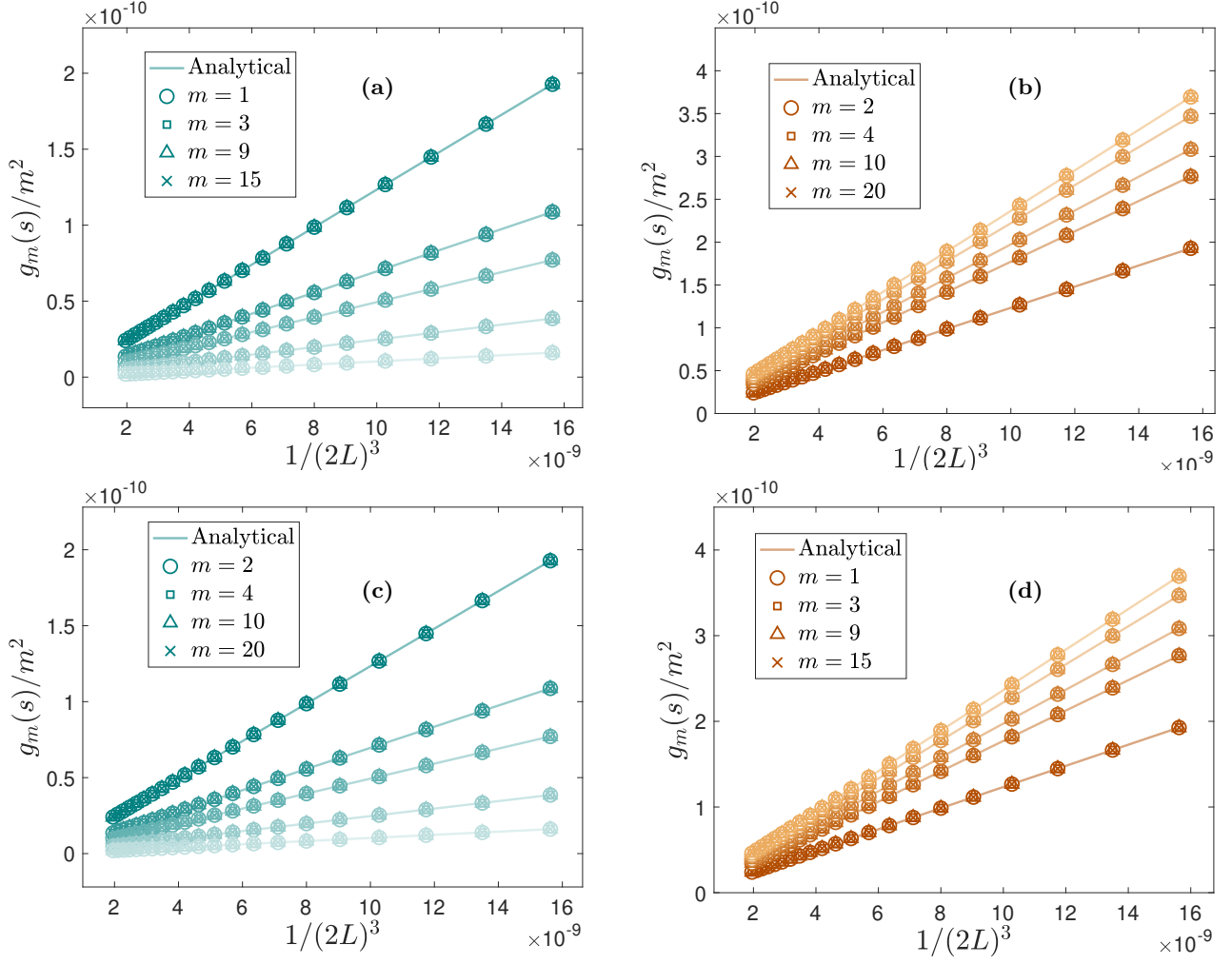
\begin{figure*}[t]
    \centering
    \begin{tikzpicture}
            \node[inner sep=0pt] (russell) at (-250pt, 0pt)
    {\includegraphics[width=0.5\textwidth]{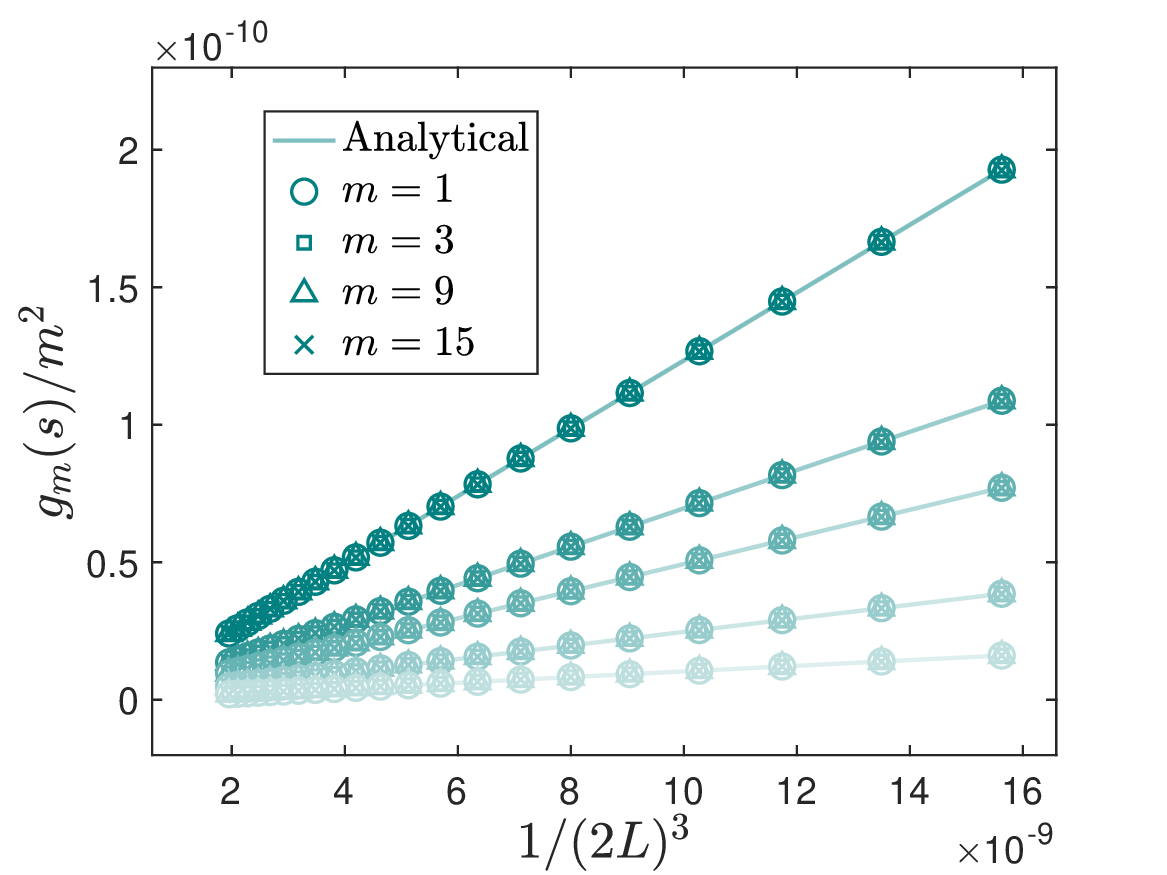}};
            \node[inner sep=0pt] (russell) at (0pt, 0pt)
    {\includegraphics[width=0.5\textwidth]{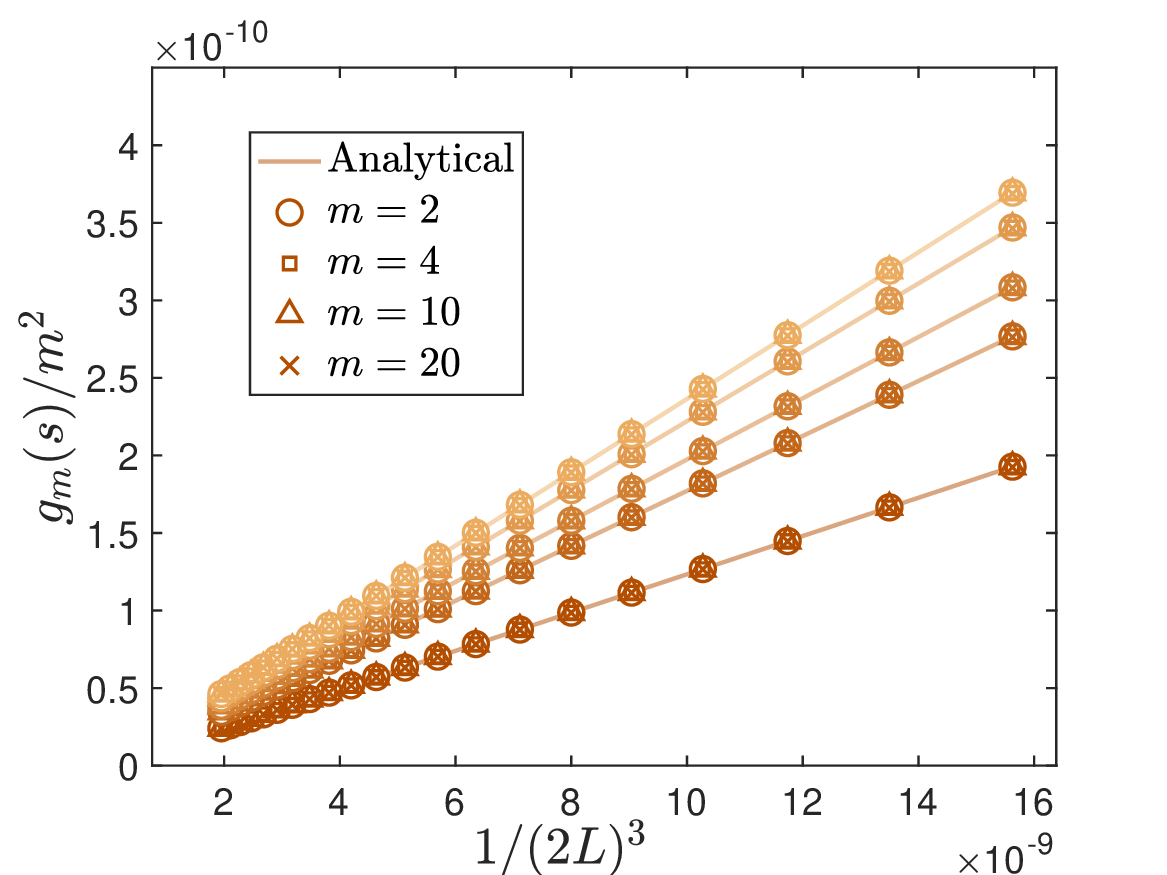}}; 
                 \node[inner sep=0pt] (russell) at (-250pt, -190pt)
    {\includegraphics[width=0.5\textwidth]{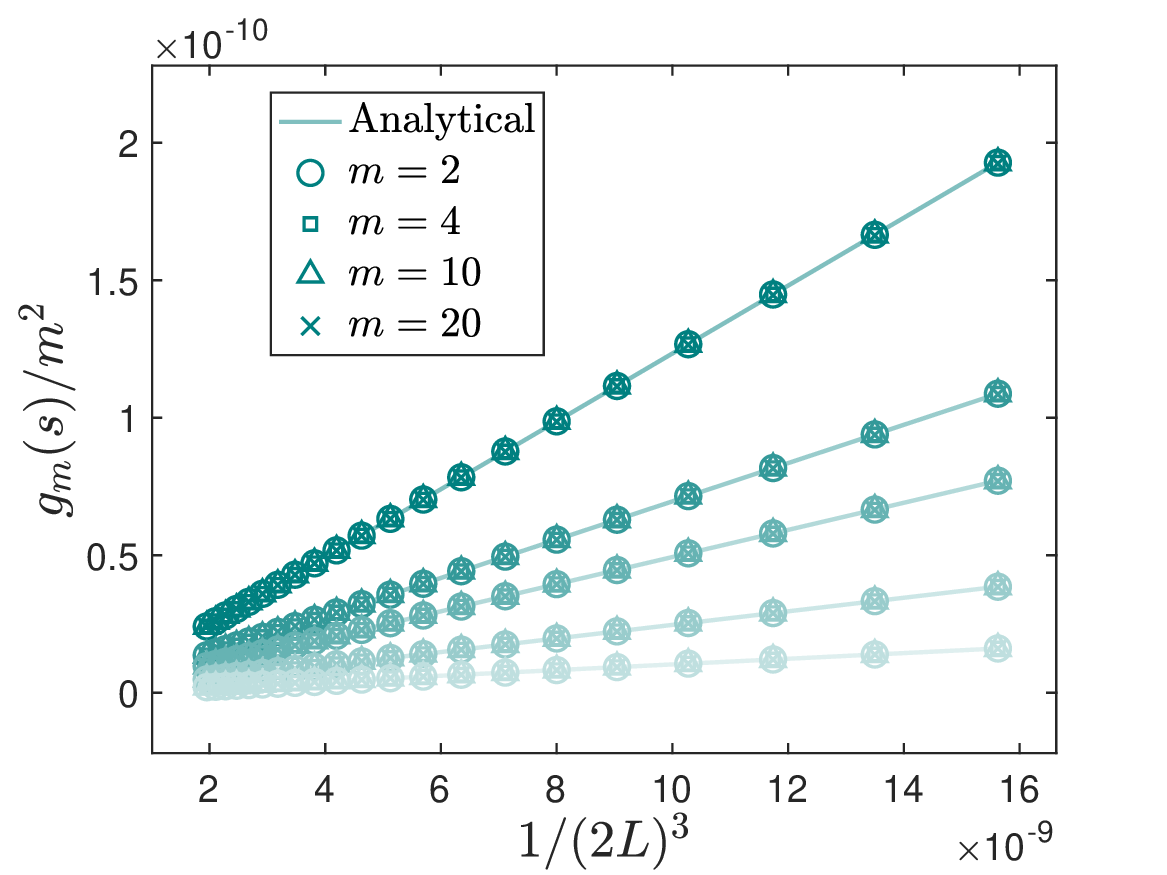}};
             \node[inner sep=0pt] (russell) at (0pt, -190pt)
    {\includegraphics[width=0.5\textwidth]{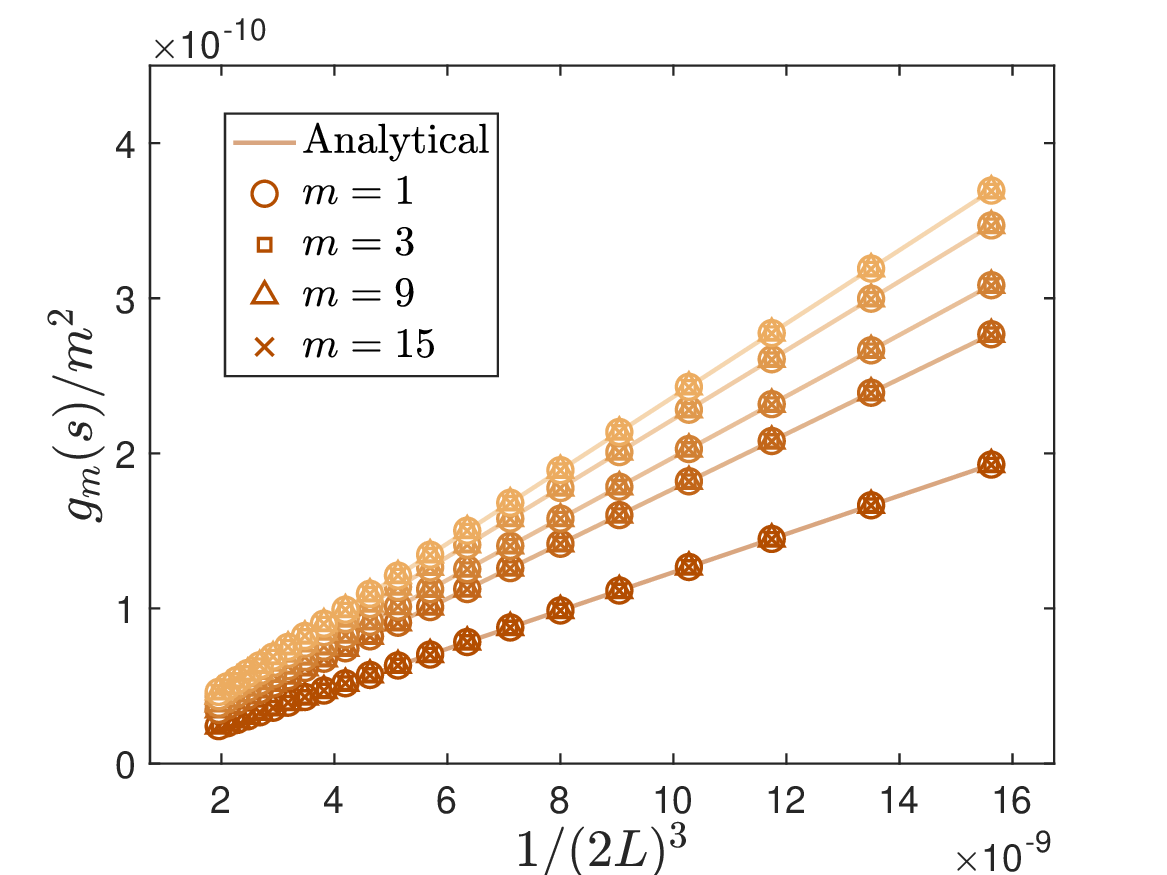}};    
    \node at (-8,2) {\textbf{(a)}};
    \node at (1,2) {\textbf{(b)}};
    \node at (-8,-5) {\textbf{(c)}};
    \node at (1,-5) {\textbf{(d)}};
    
    \end{tikzpicture}
    \caption{Rescaled relaxation rate $g_{m}(s)/m^{2}$ for the slowly decaying modes (i.e. $m\ll L$) in the harmonic chain for free boundary conditions as a function of the system size $2L$. Panels (a) and (b) correspond to local dissipation at the left boundary ($n_d = 1$), while panels (c) and (d) correspond to local dissipation at the right boundary ($n_d = 2L$). The color gradient indicates decreasing values of the interface parameter $s$, from darker to lighter shades, with $s = 1, 0.9, 0.8, 0.6,0.4$. For each fixed $s$, different markers denote different slowly decaying modes $m$; their collapse shows that the scaled relaxation rate is independent of $m$ for slowly decaying modes. There is an excellent agreement between the numerical results (markers) and the analytical result from perturbative analysis (lines). 
    Here the half-system size is chosen as $L = 200, 210, 220, \dots, 400$, $\Omega_0$ is taken as $0$ for the critical system, and the dissipation strength  $\gamma$ is set to be $0.005$. For notational simplicity, we suppress the explicit dependence on the dissipation site $n_d$ in the figures and throughout the remainder of the paper, writing $g_m(s)$ instead of $g_{m,n_d}(s)$ whenever $n_d$ is fixed and clear from the context.
    }
    \label{fig: Plot of relaxation ratefor boson}
\end{figure*}
\subsection{Harmonic chain}
For the harmonic chain with a conformal interface, the effective non-Hermitian Hamiltonian follows from the third quantization matrix in Eq.\eqref{eq:X matrix after 3rd quantization} and takes the form
\begin{align}
   i\mathbf{H}_{\rm eff}(s) = i\Sigma_{z}\tilde{\mathbf{H}}_{\rm BdG}(s) - \dfrac{\tilde{\mathbf{\Gamma}}}{2}.
\end{align}
In the \textit{closed system} setting, the bosonic commutation relations imply that the physical frequencies of the Hamiltonian in Eq.(\ref{eq:initial Hamiltonian}) are obtained from the eigenvalues of the dynamical matrix $\Sigma_z\tilde{\mathbf{H}}_{\rm BdG}(s)$, rather than from the eigenvalues of $\tilde{\mathbf{H}}_{\rm BdG}(s)$ itself. We start with the harmonic chain defined in Eq.\eqref{eq: initial Hamitonian in simplified form}. The corresponding normal mode frequencies $\Omega_{m}$ are obtained from the eigenvalue equation
\begin{align}
    \mathbf{K}(s)|\tilde{\Phi}_{m}\rangle = \Omega_{m}^{2}|\tilde{\Phi}_{m}\rangle,
\end{align}
where $\langle j |\tilde{\Phi}_{m}\rangle = \tilde{\Phi}_{m}(j)$ denotes the $j$-th component of the eigenvector $|\tilde{\Phi}_{m}\rangle$. Using this relation, the dynamical matrix $\Sigma_z\tilde{\mathbf{H}}_{\rm BdG}(s)$ can be diagonalized as 
\begin{align}
    \Sigma_z\tilde{\mathbf{H}}_{\rm BdG}(s)|\tilde{\Phi}^{\pm}_{m}\rangle = \pm\Omega_{m} |\tilde{\Phi}^{\pm}_{m}\rangle.
\end{align}
Here, $|\tilde{\Phi}^{\pm}_{m}\rangle$ denotes the eigenvector with eigenvalue $\pm\Omega_{m}$. In terms of the eigenvector $|\tilde{\Phi}_{m}\rangle$, it takes the form
\begin{align}
\label{eq: eigenvectors}
    |\tilde{\Phi}^{\pm}_{m}\rangle = \dfrac{1}{2\sqrt{\Omega_{m}}}\begin{pmatrix}
(\Omega_{m}\pm1)|\tilde{\Phi}_{m}\rangle \\
        -(\Omega_{m}\mp 1)|\tilde{\Phi}_{m}\rangle
    \end{pmatrix}.
\end{align}
The dynamical matrix $\Sigma_z\tilde{\mathbf{H}}_{\rm BdG}(s)$ is pseudo-Hermitian and has particle-hole symmetry. Therefore, its eigenvalues appear in pairs $\left(\Omega_{m}, -\Omega_{m}\right)$, and the corresponding eigenvectors are normalized as \begin{align}
\langle\tilde{\Phi}^{\sigma}_{m}|\Sigma_{z}|\tilde{\Phi}^{\sigma'}_{m}\rangle = \sigma\delta_{mn}\delta_{\sigma\sigma'} \quad\quad \sigma,\sigma' = \pm.
\end{align}

Next, we analyze the eigenspectrum of the matrix $i\mathbf{H}_{\rm eff}(s)$ analytically. In the \textit{open system} setting, we can express relaxation rate for mode $m$ in the following form
\begin{align}
    g_{m}(s) = - {\rm Re}[i\tilde{\Omega}^{\pm}_m]
\end{align}
where $i\tilde{\Omega}^{\pm}_{m}$ are complex eigenvalues of non-Hermitian matrix $i\mathbf{H}_{\rm eff}(s)$. Starting from $i\mathbf{H}_{\rm eff}(s)$, the eigenvalues up to first order in $\tilde{\mathbf{\Gamma}}$ are
\begin{align}
    i\tilde{\Omega}^{\pm}_{m} = \pm i\Omega_{m} - \dfrac{1}{2}\left\langle\tilde{\Phi}^{\pm}_{m}\right|\Sigma_{z}\tilde{\mathbf{\Gamma}}\left|\tilde{\Phi}^{\pm}_{m}\right\rangle + \mathcal{O}(\tilde{\mathbf{\Gamma}}^{2}).
\end{align}
Equivalently, defining $\left(\Delta\tilde{\Omega}^{\pm}_{m}\right) = \tilde{\Omega}^{\pm}_{m}\mp\Omega_{m}$, the first order shift satisfies 
\begin{align}
i\left(\Delta\tilde{\Omega}^{\pm}_{m}\right) = -\dfrac{1}{2}\left\langle\tilde{\Phi}^{\pm}_{m}\right|\Sigma_{z}\tilde{\mathbf{\Gamma}}\left|\tilde{\Phi}^{\pm}_{m}\right\rangle + \mathcal{O}(\tilde{\mathbf{\Gamma}}^{2}).
\end{align}
Using the definitions of $\tilde{\mathbf{\Gamma}}$ and $|\tilde{\Phi}^{\pm}_{m}\rangle$ given in Eq.(\ref{eq:sigma_z and Gamma_tilde}) and  Eq.(\ref{eq: eigenvectors}), respectively, we obtain
\begin{align}
i\left(\Delta\tilde{\Omega}^{\pm}_{m}\right) = -\dfrac{\gamma}{2} \bigg|\langle n_d|\tilde{\Phi}_{m}\rangle \bigg|^2 + \mathcal{O}(\tilde{\mathbf{\Gamma}}^{2}).
\end{align}
For our choice of local jump operators in Eq.(\ref{eq: gain and loss Gamma matrices}), this becomes
\begin{align}
i\left(\Delta\tilde{\Omega}^{\pm}_{m}\right) = -\dfrac{\gamma}{2} \Big|\tilde{\phi}_{m}(n_d)\Big|^{2} + \mathcal{O}(\gamma^{2}),
\end{align}
where $\gamma = \gamma_{l} -\gamma_{g}$. We assume $\gamma\ge0$, which is the stability condition for the open harmonic chain. Therefore, to first order in the dissipation strength, the spectrum of $i\mathbf{H}_{\rm eff}(s)$ is
\begin{align}
 i\tilde{\Omega}^{\pm}_{m} \approx \pm i\Omega_{m}  -\dfrac{\gamma}{2} \Big|\tilde{\phi}_{m}(n_d)\Big|^{2}.
\end{align}
Here, $n_{d}$ denotes the site at which local dissipation is applied. The  corresponding perturbative expression for the relaxation rate for mode $m$ is 
\begin{align}
\label{eq:Liouvillian gap pert}
    g_{m,n_d}(s) \approx \dfrac{\gamma}{2}\Big|\tilde{\phi}_{m}(n_d)\Big|^{2}.
\end{align}
This expression is valid up to first order in the dissipation strength $\gamma$.

Next, we evaluate the relaxation rate $g_{m,n_d}(s)$ in Eq.(\ref{eq:Liouvillian gap pert}) for several cases. We first fix the boundary conditions of the harmonic chain introduced in Sec.\ref{sec:2}, and derive analytical expressions for two choices of the local dissipation site: the left boundary of the left chain and the right boundary of the right chain. In other words, our goal is to study how the relaxation rate of each mode $m$ depends on the transmission parameter $s$ when dissipation is applied to one of the two chains, and to determine whether this dependence is sensitive to the choice of boundary conditions.

\subsubsection{Free boundary condition} 
We first consider the harmonic chain in Eq.\eqref{eq:initial Hamiltonian} with free boundary conditions. Combining the modified single-particle eigenfunctions in Eq.\eqref{eq:modified single-particle eigenfunction free} with the perturbative expression in Eq.\eqref{eq:Liouvillian gap pert}, the relaxation rate for an arbitrary dissipation site $n_d$ is 
\begin{align}
\label{eq: g for free}
    g^{\rm free}_{m, n_d}(s) \approx \dfrac{\gamma}{2} \frac{\eta^{\rm free}_{n_d}(m)}{L}\cos^{2}\left(\frac{\left(2n_d-1\right)\theta_m}{2}\right).
\end{align}
Here, $g^{\rm free}_{m, n_d}(s)$ denotes the relaxation rate of mode $m$ as a function of the transmission parameter $s$ for free boundary conditions, with local dissipation applied at the fixed site $n_d$. The coefficient $\eta^{\rm free}_{n_d}(m)$ depends on the choice of dissipation site $n_d$, specifically whether it lies on the left or right side of the conformal interface. Explicitly, 
\begin{align}
    \eta^{\rm free}_{n_d}(m) = \begin{cases}
        \alpha^{2}_{m}, \quad  1\le n_d \le L,\\
        \beta^{2}_{m}, \quad   L+1\le n_d \le 2L.
    \end{cases}
\end{align}
All transmission properties of the conformal interface are encoded in the coefficients given in Eq.\eqref{eq:alpha and beta in terms of s}. In general, obtaining a closed-form expression for $g^{\rm free}_{m, n_d}(s)$ in Eq.(\ref{eq: g for free}) is difficult. However, for specific choices of the local dissipation site $n_d$, the relaxation rates take simple explicit forms. We first consider dissipation applied at the left boundary, $n_d =1 $, and write the quantized wave number $\theta^{\rm free}_m$ as 
\begin{align}
    \theta^{\rm free}_{m} = \pi - \dfrac{m\pi}{2L},
\end{align}
where $m = 1, 2, \dots 2L$ is the mode index labeling the normal modes. The perturbative expression for the relaxation rate for mode $m$ then becomes
\begin{align}
    g^{\rm free}_{m,1}(s) \approx \dfrac{\gamma}{2} \frac{\alpha^{2}_{m}}{L}\sin^{2}\left(\frac{m\pi}{4L}\right).
\end{align}
At long times, the relaxation dynamics is dominated by the slowly decaying modes with $m \ll L$. These modes have relaxation times that scales as $\mathcal{O}(L^3)$.  Thus, in the large  $L$ limit, the corresponding relaxation rate simplifies to 
\begin{align}
    g^{\rm free}_{m,1}(s) \approx \left(1+(-1)^{m}\sqrt{1-s^2}\right)\frac{\pi^2m^{2}\gamma }{32L^3}\label{eq:relax rate for weak boundary dissipation left}.
\end{align}
Modes with odd mode indices $m$ are localized in the non-dissipative chain at $s=0$, as indicated by their vanishing relaxation rates in the perfectly reflective limit; see Fig.\ref{fig: Plot of relaxation ratefor boson}(a). When the transmission is finite, $0<s\le 1$, these modes extend into the dissipative chain and acquire finite relaxation rates. By contrast, modes with even mode indices $m$ are localized in the dissipative chain at $s=0$ and therefore have finite relaxation rates even in the perfectly reflective limit; see Fig.\ref{fig: Plot of relaxation ratefor boson}(b).
\smallskip

Next, we consider the case where dissipation is applied at the right boundary, $n_d = 2L$. For a given quantized wave number $\theta^{\rm free}_m$, the relaxation rate becomes 
\begin{align}
    g^{\rm free}_{m,2L}(s) \approx \dfrac{\gamma}{2} \frac{\beta^{2}_{m}}{L}\cos^{2}\left(\frac{(4L-1)}{2}\left(\pi -\frac{m\pi}{2L}\right)\right).
\end{align}
For slowly decaying modes with $m\ll L$, this expression in the large $L$ limit reduces to
\begin{align}
    g^{\rm free}_{m,2L}(s) \approx \left(1-(-1)^{m}\sqrt{1-s^2}\right)\frac{\pi^2m^{2}\gamma}{32L^3}
    \label{eq:relax rate for weak boundary dissipation right}.
\end{align}
For right boundary dissipation, the mode-index parity is reversed, but the physical picture remains unchanged. Modes initially localized in the non-dissipative chain acquire finite relaxation rates once the interface transmission becomes finite; see Fig.\ref{fig: Plot of relaxation ratefor boson}(c) and Fig.\ref{fig: Plot of relaxation ratefor boson}(d). 

\subsubsection{Fixed boundary condition} 
Next, we impose fixed boundary conditions on the harmonic chain. In this case, the perturbative expression for the relaxation rate becomes
\begin{align}
    g^{\rm fixed}_{m, n_d}(s) \approx \gamma \frac{\eta^{\rm fixed}_{n_d}(m)}{2L+1}\sin^{2}\left(n_d\theta_m\right). 
\end{align}
Here, $\eta^{\rm fixed}_{nd}(m)$ is again a site dependent coefficient determined by the location of the local dissipation site $n_d$. For fixed boundary conditions, the two coefficients on the left and right sides of the interface are exchanged, so that 
\begin{align}
    \eta^{\rm fixed}_{n_d}(m) = \begin{cases}
        \beta_m^{2}, \quad  1\le n_d \le L,\\
        \alpha_m^{2}, \quad   L+1\le n_d \le 2L.
    \end{cases}
\end{align}
Following the free boundary case, we now repeat the calculation for fixed boundary conditions. We derive closed-form perturbative expressions for the relaxation rate when dissipation is applied to either side of the interface. First, we consider dissipation at the left boundary, $n_d = 1$, and choose the quantized wave number as 
\begin{align}
    \theta^{\rm fixed}_{m} = \pi -\dfrac{m\pi }{2L+1}.
\end{align}
where $m = 1, 2, \dots 2L$ is the mode index labeling the normal modes. The relaxation rate for mode with index $m$ is
\begin{align}
    g^{\rm fixed}_{m,1}(s) \approx  \frac{\gamma\beta^{2}_{2L+1-m}}{2L+1}\sin^{2}\left(\frac{m\pi}{2L+1}\right),
\end{align} 
or equivalently,
\begin{align}
    g^{\rm fixed}_{m,1}(s) \approx  \frac{\gamma\alpha^{2}_m}{2L+1}\sin^{2}\left(\frac{m\pi}{2L+1}\right). 
\end{align}
Again, we focus on the long time relaxation dynamics, which is dominated by modes with $m \ll L$. Thus, in the large $L$ limit, the relaxation rate simplifies to
\begin{align}
\label{eq: weak result boson left fixed}
    g^{\rm fixed}_{m,1}(s) \approx \left(1+(-1)^{m}\sqrt{1-s^2}\right)\frac{\pi^2 m^{2}\gamma}{(2L+1)^3}.
\end{align}
Similarly, when dissipation is applied at the right boundary, $n_d =2L$, the relaxation rate in the large $L$ limit of modes with $m\ll L$ is given by
\begin{align}
\label{eq: weak result boson left fixed}
    g^{\rm fixed}_{m, 2L}(s) \approx \left(1-(-1)^{m}\sqrt{1-s^2}\right)\frac{\pi^2 m^{2}\gamma}{(2L+1)^3}.
\end{align}

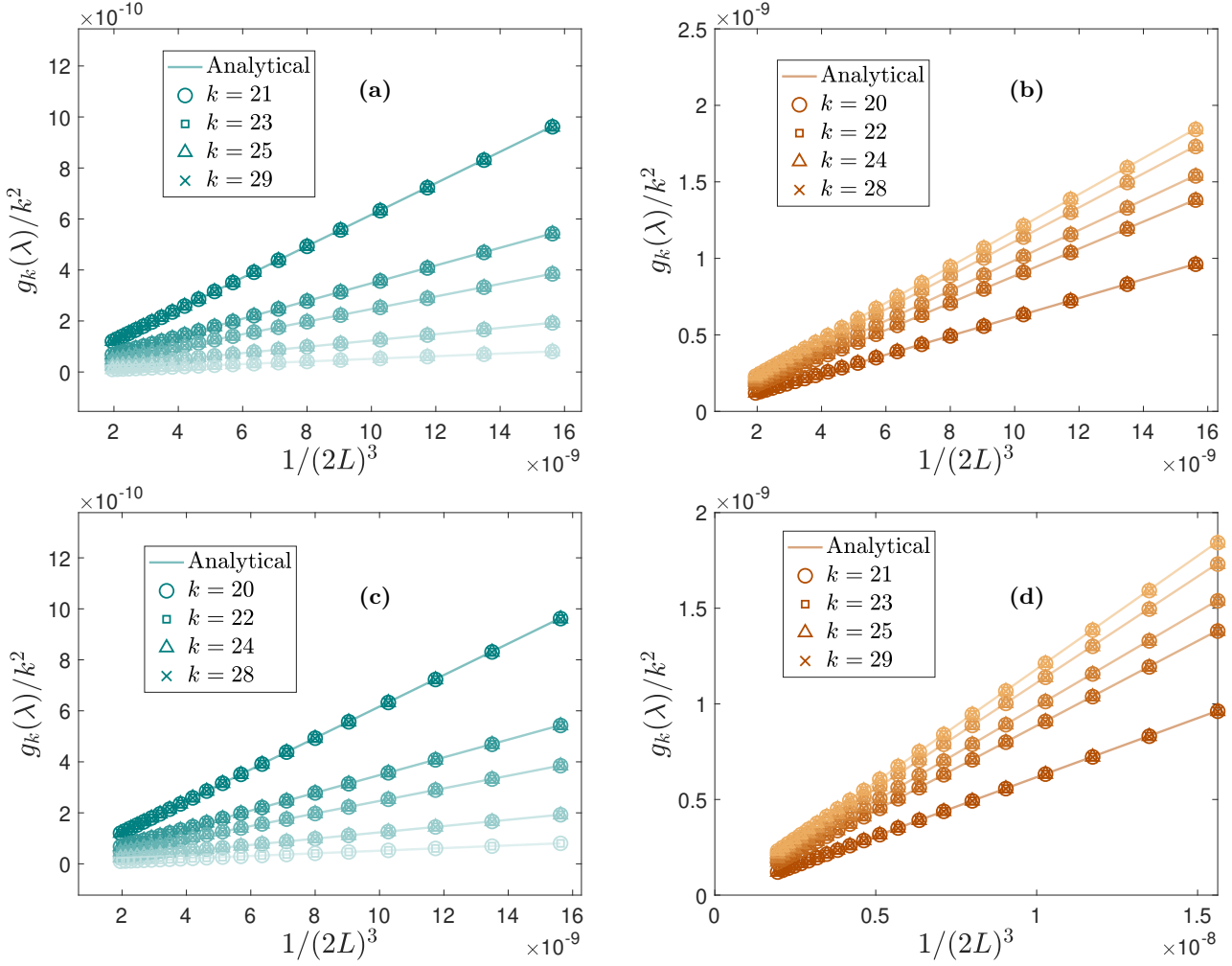
\begin{figure*}[t]
    \centering
    \begin{tikzpicture}
            \node[inner sep=0pt] (russell) at (-250pt, 0pt)
    {\includegraphics[width=0.5\textwidth]{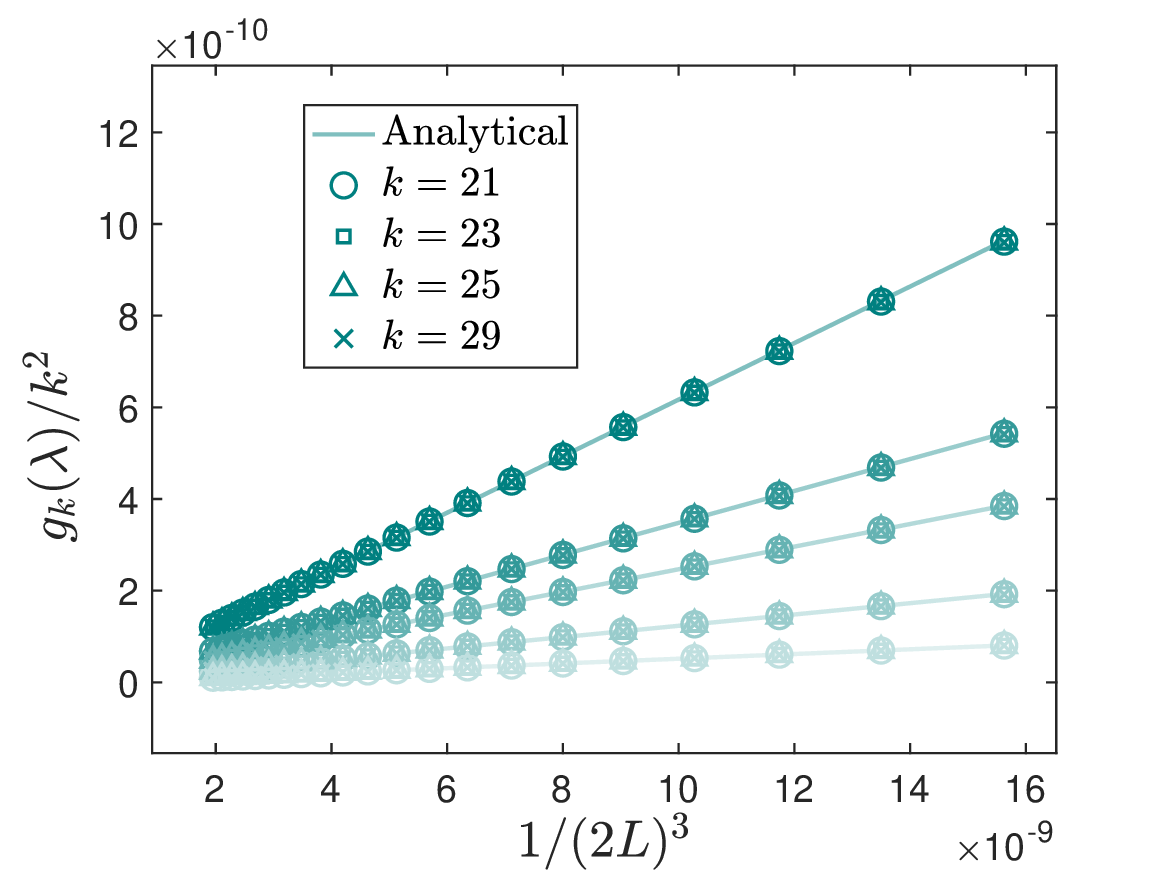}};
            \node[inner sep=0pt] (russell) at (0pt, 0pt)
    {\includegraphics[width=0.5\textwidth]{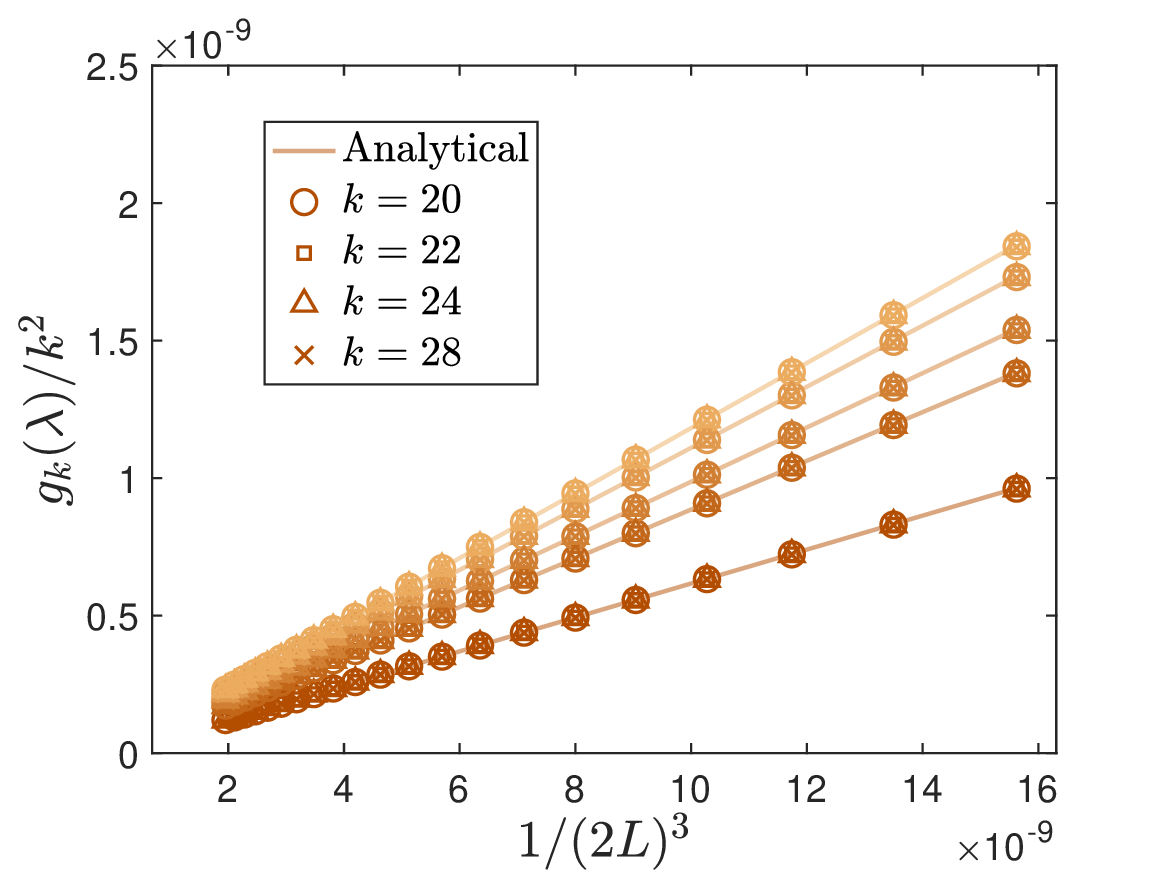}}; 
                 \node[inner sep=0pt] (russell) at (-250pt, -190pt)
    {\includegraphics[width=0.5\textwidth]{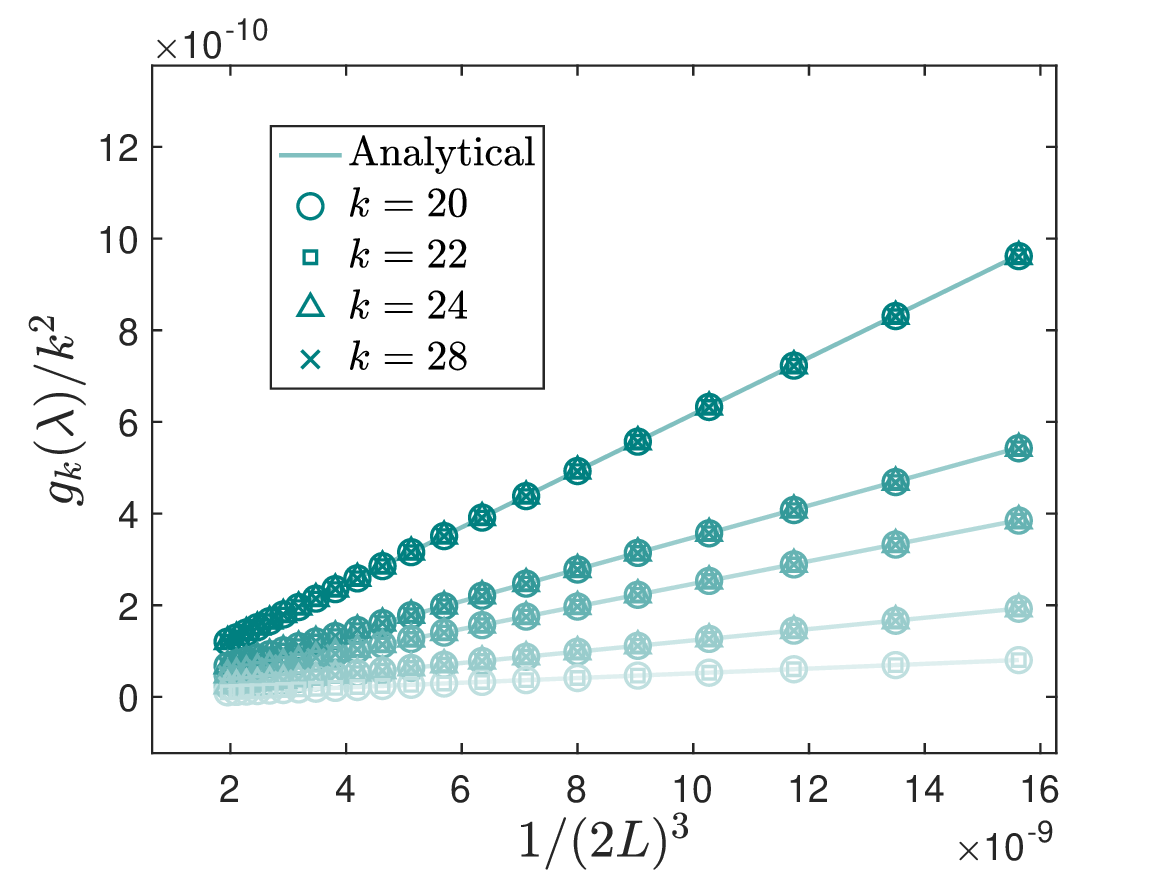}};
             \node[inner sep=0pt] (russell) at (0pt, -190pt)
    {\includegraphics[width=0.5\textwidth]{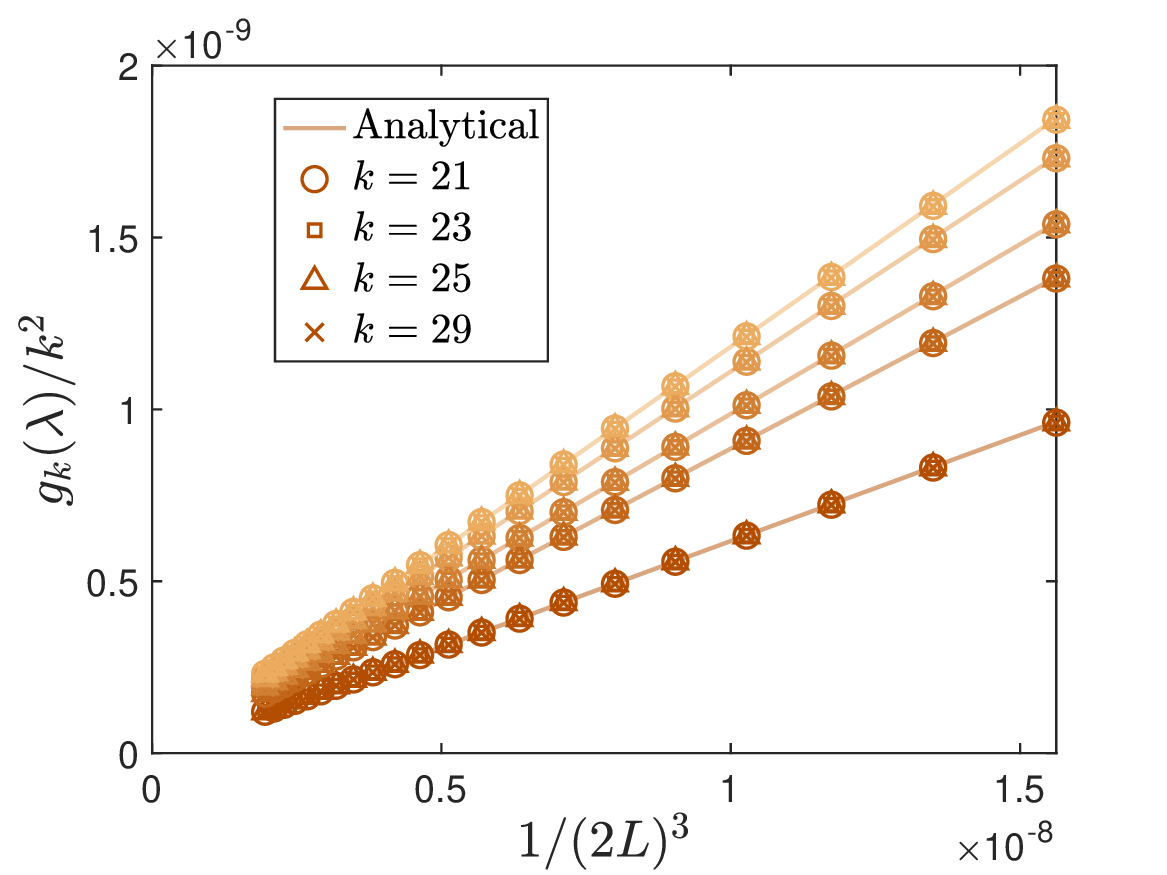}};    
    \node at (-8,2) {\textbf{(a)}};
    \node at (1,2) {\textbf{(b)}};
    \node at (-8,-5) {\textbf{(c)}};
    \node at (1,-5) {\textbf{(d)}};
    
    \end{tikzpicture}
    \caption{Rescaled relaxation rate $g_{k}(\lambda)/k^{2}$ for the slowly decaying modes (i.e. $k\ll L$) in the free fermion chain for fixed boundary condition as a function of the system size $2L$. Panels (a) and (b) correspond to local dissipation at the left boundary ($n_d = 1$), while panels (c) and (d) correspond to local dissipation at the right boundary ($n_d = 2L$). The color gradient indicates decreasing values of the interface parameter $\lambda$, from darker to lighter shades, with $\lambda = 1, 0.9, 0.8, 0.6,0.4$. For each fixed $\lambda$, different markers denote different slowly decaying modes $k$; their collapse shows that the scaled relaxation rate is independent of $k$ for slowly decaying modes. There is an excellent agreement between the numerical results (markers) and the analytical result from perturbative analysis (lines). 
    Here the half-system size is chosen as $L = 200, 210, 220, \dots, 400$, and the dissipation strength  $\gamma$ is set to be $0.05$. For notational simplicity, we suppress the explicit dependence on the dissipation site $n_d$ in the figures and throughout the remainder of the paper, writing $g_k(\lambda)$ instead of $g_{k,n_d}(\lambda)$ whenever $n_d$ is fixed and clear from the context.
    }
    \label{fig: Plot of relaxation ratefor fermion}
\end{figure*}

\subsection{Free fermion chain}
Next, we consider the critical free-fermion chain, for which the effective non-Hermitian Hamiltonian takes the form
\begin{align}
     \mathbf{H}_{\rm eff}(\lambda) = \tilde{\textbf{H}}(\lambda) - i\frac{\mathbf{\Gamma_{+}+\Gamma_{-}}}{2}.
\end{align}
where $\tilde{\textbf{H}}(\lambda)$ is the Hamiltonian matrix of the free fermion chain with a conformal interface of transmission $\lambda$, defined by 
\be
\widetilde{\mathbf{H}}(\lambda) = \sum_{i, j=1}^{2L} \widetilde{H}_{i,j}(\lambda) c_i^\dagger c_j.
\ee
For any transmission $\lambda$, the Hamiltonian can be diagonalized exactly as
\be
\widetilde{H}(\lambda) \, |\widetilde{\phi}_k\rangle = E_k \, |\widetilde{\phi}_k\rangle.
\ee
Following Eq.(32) of \cite{Barad_2025}, the first-order energy shift of the non-Hermitian effective Hamiltonian $\mathbf{H}_{\rm eff}(\lambda)$, relative to the dissipation free Hamiltonian $\widetilde{\mathbf{H}}(\lambda)$, is given by 
\be
\begin{aligned}
\Delta \omega_k & = E'_k - E_k
\\ & 
= - i \frac{\gamma}{2} \widetilde{\phi}^*_k(n_d) \widetilde{\phi}_k(n_d) + \mathcal{O}(\gamma^2), 
\end{aligned}
\ee
where $E'_k$ denotes the perturbed single-particle eigenvalue of $\mathbf{H}_{\rm eff}(\lambda)$, while $|\widetilde{\phi}_k\rangle$ is the eigenstate of the dissipation free Hamiltonian $\widetilde{\mathbf{H}}(\lambda)$. For dissipation applied at site $n_d$, the the relaxation rate of mode $k$ is then given by
\be
\begin{aligned}
g_{k, n_d}(\lambda) & = -   \Im[E'_{k} ]
\\ & 
= \dfrac{\gamma}{2}  \left| \widetilde{\phi}_k(n_d) \right|^{2} + \mathcal{O}(\gamma^2). 
\end{aligned}
\label{eq:Liouvillian gap pert fermion}
\ee 

Next, we evaluate the relaxation rate  $g_{k, n_d}(\lambda)$ in Eq.\eqref{eq:Liouvillian gap pert fermion} for the slowly decaying modes of the free fermion chain. We study how this rate depends on the transmission parameter $\lambda$ when dissipation is applied to one of the two chains, and whether this dependence is sensitive to the choice of boundary conditions. Since the derivation is similar to that for the harmonic chain, we do not repeat it here and instead state the final results directly. The only difference is the correspondence between boundary conditions: free boundary conditions in the harmonic chain correspond to fixed boundary conditions in the free fermion chain, while fixed boundary conditions in the harmonic chain correspond to free boundary conditions in the free fermion chain. Thus, the results for the free fermion chain can be obtained from the harmonic chain expressions by exchanging the corresponding boundary condition labels.

\subsubsection{Free boundary conditions}
When dissipation is applied at the left boundary, $n_d=1$, we write the quantized wave number $\theta^{\rm free}_{k}$ for the mode with index $k$ as
\begin{align}
    \theta^{\rm free}_{k} = \pi -\dfrac{k\pi }{2L+1}.
\end{align}
At long times, the relaxation dynamics is dominated by modes with $k\ll L$. Thus, in the large $L$ limit, the corresponding relaxation rate simplifies to
\begin{align}
\label{eq: weak result fermion left free}
    g^{\rm free}_{k,1}(\lambda) \approx \left(1+(-1)^{k}\sqrt{1-\lambda^2}\right)\frac{\pi^2 k^{2}\gamma}{(2L+1)^3}.
\end{align}
Similarly, when dissipation is applied at the right boundary, $n_d =2L$, the relaxation rate is
\begin{align}
\label{eq: weak result fermion left free}
    g^{\rm free}_{k, n_d}(\lambda) \approx \left(1-(-1)^{k}\sqrt{1-\lambda^2}\right)\frac{\pi^2 k^{2}\gamma}{(2L+1)^3}.
\end{align}

\subsubsection{Fixed boundary conditions}
We now evaluate the relaxation rate for fixed boundary conditions, $V_{0} = \pm1$. We first consider the case $V_{0}= +1$, for which we choose the quantized wave number as  
\begin{align}
    \theta^{\rm fixed,+}_{k} = \pi - \dfrac{k\pi}{2L}
\end{align}
such that the relaxation rate of the slowly decaying modes with mode index $k$, for dissipation applied at the left boundary, is given by
\begin{align}
\label{eq: weak result fermion left free}
    g^{\rm fixed,+}_{k,1}(\lambda) \approx \left(1+(-1)^{k}\sqrt{1-s^2}\right)\frac{\pi^2 k^{2}\gamma}{32L^3}.
\end{align}
Similarly, for $n_d =2L$, the relaxation rate is given by
\begin{align}
\label{eq: weak result fermion right free}
    g^{\rm fixed,+}_{k, 2L}(\lambda) \approx \left(1-(-1)^{k}\sqrt{1-s^2}\right)\frac{\pi^2 k^{2}\gamma}{32L^3}.
\end{align}
For the second case, with boundary potential $V_{0}= -1$ boundary potential, we choose the quantized wave number as 
\begin{align}
    \theta^{\rm fixed,-}_{k} =  \dfrac{k\pi}{2L}.
\end{align}
The relaxation rate for modes $k\ll L$ is then evaluated separately for the two choices of dissipation site
For left-boundary dissipation, $n_d =1$, the relaxation rate is given by 
\begin{align}
    g^{\rm fixed,-}_{k, 1}(\lambda) \approx \left(1+(-1)^{k}\sqrt{1-s^2}\right)\frac{\pi^2 k^{2}\gamma}{32L^3}
    \label{eq: weak result fermion left fixed}
\end{align}
while for right-boundary dissipation, $n_d =2L$, it becomes
\begin{align}
    g^{\rm fixed,-}_{k, 2L}(\lambda) \approx \left(1-(-1)^{k}\sqrt{1-s^2}\right)\frac{\pi^2 k^{2}\gamma}{32L^3}.
    \label{eq: weak result fermion left fixed}
\end{align}

In Fig.\ref{fig: Plot of relaxation ratefor fermion}, we show the relaxation rates of the free fermion chain with fixed boundary conditions, $V_0 = +1$. For left boundary dissipation, $n_d=1$, modes with odd mode indices $k$ are localized in the non-dissipative chain at $\lambda=0$, as indicated by their vanishing relaxation rates in the perfectly reflective limit; see Fig.\ref{fig: Plot of relaxation ratefor fermion}(a). For finite transmission, $0<\lambda\le 1$,  these modes extend into the dissipative chain and acquire finite relaxation rates. By contrast, modes with even mode indices $k$ are localized in the dissipative chain at $\lambda=0$ and therefore retain finite relaxation rates even in the perfectly reflective limit; see Fig.\ref{fig: Plot of relaxation ratefor fermion}(b). Figures~\ref{fig: Plot of relaxation ratefor fermion}(c) and~\ref{fig: Plot of relaxation ratefor fermion}(d) show the corresponding results when the dissipation site $n_d$ is moved from the left side of the interface to the right side. In this case, the parity assignment of the two classes of modes is reversed.

\section{Relaxation rate: finite dissipation strength}\label{sec:5}
In the previous section, we considered the weak dissipation regime, where the dissipation strength $\gamma$ was assumed to be small compared with the characteristic energy scale of each setup. We then derived analytical expressions for the relaxation rates of the slowly decaying modes in the harmonic and free fermion chains. Here, we extend the analysis to finite dissipation strength and obtain exact expressions for these relaxation rates when dissipation is applied at either the left or right boundary.

\subsection{Harmonic chain}
For the harmonic chain, we begin with the following eigenvalue equation, derived in Appendix~\ref{appendix:Finite dissipation case}:
\begin{align}
\label{eq:finite dissipation case matrix}
    \left(\mathbf{K}(s) + \dfrac{\mathbf{\Gamma}^{2}}{4} -i E_{m}\mathbf{\Gamma} - E^{2}_{m}\mathbb{I}\right)|\boldsymbol\psi_m\rangle = 0.
\end{align}
Here, the matrices $\mathbf{K}(s)$ and $\mathbf{\Gamma} = \mathbf{\Gamma}_{-}-\mathbf{\Gamma}_{+}$ are defined in Eq.\eqref{eq:K matrix} and Eq.\eqref{eq: gain and loss Gamma matrices}, respectively. For an arbitrary dissipation site $n_d$, obtaining an exact expression for the relaxation rate is generally difficult. However, exact results can be derived for special cases such as boundary dissipation. In this section, we obtain the large $L$ solution for both free and fixed boundary conditions, considering two choices of the dissipation site: (a) dissipation at the left boundary, $n_d =1$, and (b) dissipation at the right boundary, $n_d =2L$. From the eigenvalue equation in Eq.\eqref{eq:finite dissipation case matrix}, we obtain the following conditions for the eigenfunction $\psi_{j}$ in the bulk and at the two boundaries:
(i) In the bulk, away from the boundary and the conformal interface, the eigenfunction $\psi_{j}$ satisfies
\begin{align}
-\psi_{j-1}+2\psi_j-\psi_{j+1} = E^2_{m}\psi_j.
\end{align}
(ii) At the left and right boundaries, the eigenfunction $\psi_{j}$ satisfies
\begin{subequations}
\label{eq: boundary conditions}
\begin{align}
\left(V_{0} + 1\right)\psi_1  + \delta_{n_d,1}\left(\frac{\gamma^2}{4} - iE_{m}\gamma\right)\psi_1 -  \psi_2 =  E^2_{m}\psi_1 \quad\quad\quad\\
    \left(V_{0} + 1\right)\psi_{2L}  + \delta_{n_d,2L}\left(\frac{\gamma^2}{4} - iE_{m}\gamma\right)\psi_{2L} -  \psi_{2L-1} =  E^2_{m}\psi_{2L}. 
\end{align}
\end{subequations}
Using the bulk eigenvalue equation for $\psi_j$, we introduce the plane-wave ansatz
\begin{align}
\label{eq:planewave initial}
    \psi_{j} \sim Az^{j}_{m} + Bz^{-j}_{m}, \quad z_{m} \in \mathbb{C},
\end{align}
which gives the dispersion relation
\begin{align}
\label{eq:eigenvalue E^2}
    E^{2}_{m} \sim 2- \left(z_{m} + z^{-1}_{m}\right).
\end{align}
However, the conformal interface at the center of the harmonic chain modifies the eigenfunction, leading to the following additional matching conditions:
\begin{subequations}
\begin{align}
&-\psi_{L-1}+(2-\xi)\psi_L-s\psi_{L+1}=E^2_{m}\psi_L,\label{eq:eigenvalye equation to the left of the defect-site}&\\
&-s\psi_L+(2+\xi)\psi_{L+1}-\psi_{L+2} = E^2_{m}\psi_{L+1}.\label{eq:eigenvalye equation to the right of the defect-site}
\end{align}
\end{subequations}
Here, $\xi = \sqrt{1-s^2}$ is the onsite potential. For the eigenfunction $\psi_{j}(m)$ of mode $m$ to satisfy the bulk, boundary, and interface conditions, we modify the plane-wave ansatz in Eq.\eqref{eq:planewave initial}. In particular, we introduce a prefactor that depends on whether the site $j$ lies to the left or right of the conformal interface, following a construction similar to that used in Ref.\cite{Barad_2025}
\begin{align}
\psi_j(m) = \begin{cases}\displaystyle \alpha_m\left(Az^j_{m} + Bz^{-j}_{m}\right),\quad 1 \le j \le L,\\\displaystyle \beta_m\left(Cz^j_{m} + Dz^{-j}_{m}\right),\quad L+1 \le j \le 2L,\end{cases}\label{eq:plane-wave ansatz equations with conformal defect}
\end{align}
where $\alpha_{m}$ and $\beta_{m}$ denote the prefactors on the left and right sides of the conformal interface, respectively, as defined in Eq.\eqref{eq:alpha and beta in terms of s}. We choose $z_m = \text{exp}\left(i\theta_{m}\right)$ in the plane-wave ansatz, where $\theta_{m}$ is in general complex. Substituting the ansatz in Eq.\eqref{eq:plane-wave ansatz equations with conformal defect} into the interface and boundary conditions, we obtain the following set of equations:
\begin{align}
\label{eq:ABCD equation 1} 
&A\left(\tilde{\gamma}_1z_{m}  + 1\right) + B\left(\tilde{\gamma}_1z^{-1}_{m}  + 1\right) =0, \\ \nonumber
&A\left(\alpha_mz^{L+1}_{m} - \alpha_m\xi z^L_{m}\right) + B\left(\alpha_mz^{-L-1}_{m} - \alpha_m\xi z^{-L}_{m}\right) +\\ \nonumber
&C\left(-\beta_mz^{L+1}s\right)+D\left(-\beta_mz^{-L-1}s\right)  = 0, \\ \nonumber
&A\left(\alpha_mz^{L}_{m}s\right)+B\left(\alpha_mz^{-L}_{m}s\right) + C\left(-\beta_mz^{L}_{m}- \beta_m\zeta z^{L+1}_{m}\right) \\ \nonumber
&+ D\left(-\beta_mz^{-L}_{m} - \beta_m\zeta z^{-L-1}_{m}\right) = 0, \\ \nonumber 
&C\left(\tilde{\gamma}_{2L}z^{2L}_{m}  + z^{2L+1}_{m}\right) + D\left(\tilde{\gamma}_{2L}z^{-2L}_{m}  + z^{-2L-1}_{m}\right) =0,
\end{align}  
where we have defined
\begin{align}
\label{eq: tilde gamma}
    \tilde{\gamma}_j = V_{0} -1 + \delta_{n_d,j}\left(\dfrac{\gamma^{2}}{4} - iE_{m}\gamma\right).
\end{align} 
with $E_{m} = 2\,\text{sin}\left(\theta_{m}/2\right)$, as follows from Eq.\eqref{eq:eigenvalue E^2}. A nontrivial solution exists only if $\theta_{m}$ satisfies the following quantization condition:
\begin{equation}
\label{eq:quantization}
\begin{split}
      \text{sin}[\left(2L+1\right)\theta_{m}] +  (\tilde{\gamma}_1+\tilde{\gamma}_{2L})\text{sin}[\left(2L\right)\theta_{m}] + \quad\quad\quad \\
\tilde{\gamma}_1\tilde{\gamma}_{2L}\text{sin}[\left(2L-1\right)\theta_{m}] + \sqrt{1-s^2}(\tilde{\gamma}_1-\tilde{\gamma}_{2L})\text{sin}[\theta_{m}] =0. 
\end{split}
\end{equation}
In the remainder of this subsection, we use this quantization condition to derive analytical expressions for the relaxation rates at finite dissipation strength $\gamma$ under both free and fixed boundary conditions.
\begin{figure*}
\centering
    \begin{tikzpicture}
            \node[inner sep=0pt] (russell) at (-250pt, -85pt)
    {\includegraphics[width=0.25\textwidth]{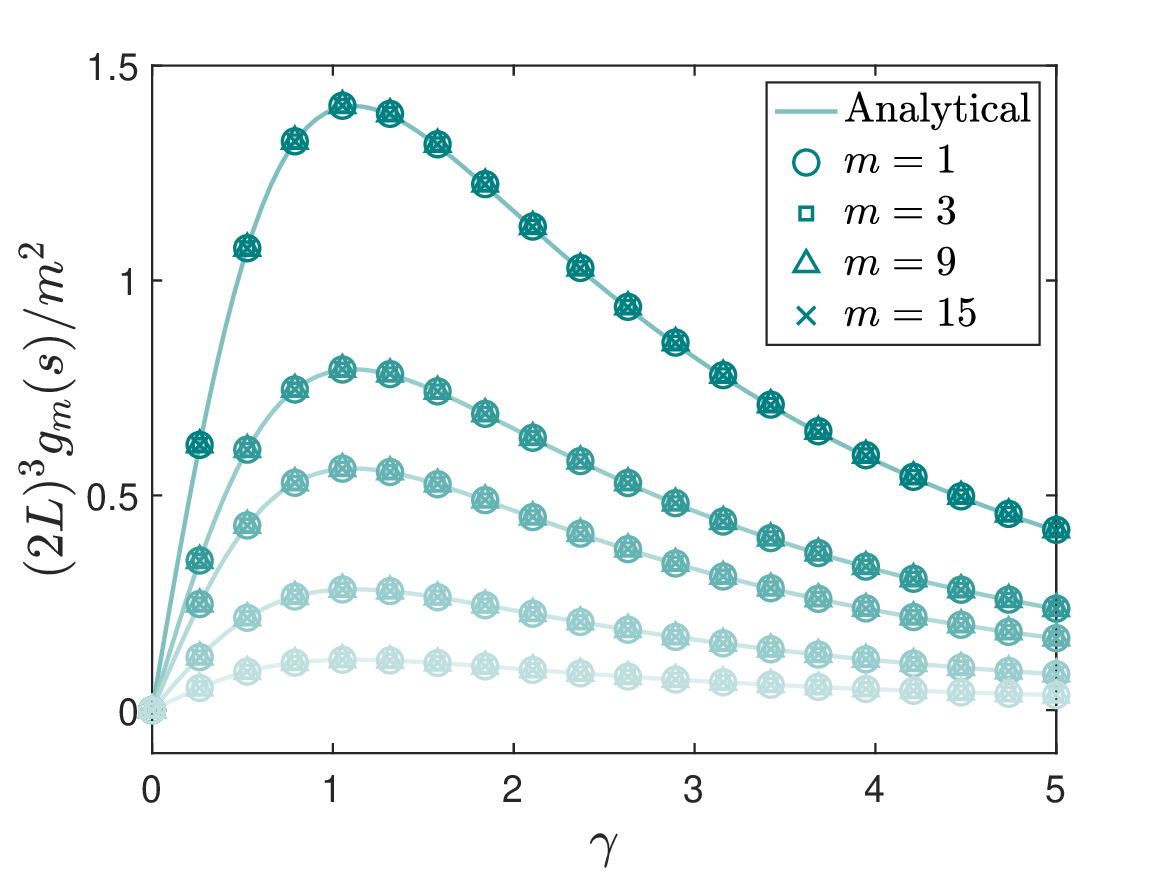}};
            \node[inner sep=0pt] (russell) at (-125pt, -85pt)
    {\includegraphics[width=0.25\textwidth]{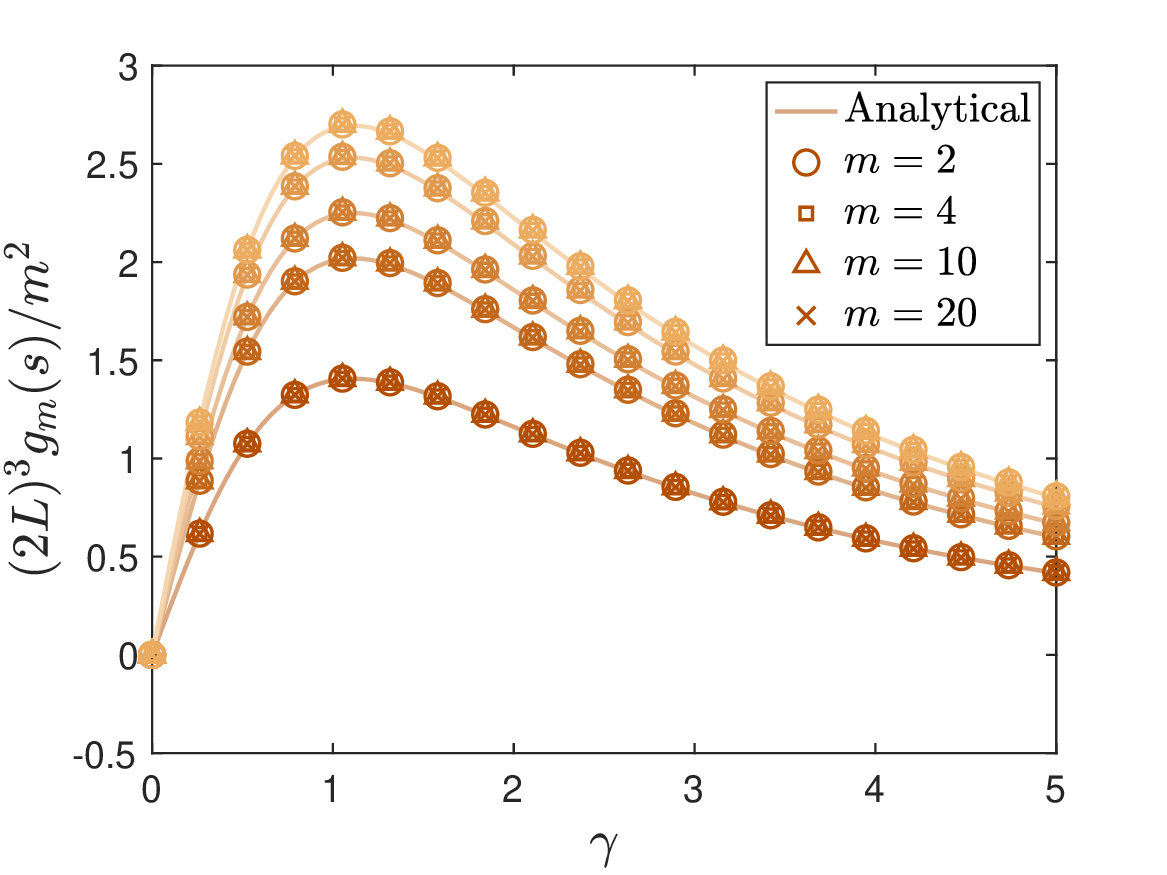}}; 
             \node[inner sep=0pt] (russell) at (0pt, -85pt)
    {\includegraphics[width=0.25\textwidth]{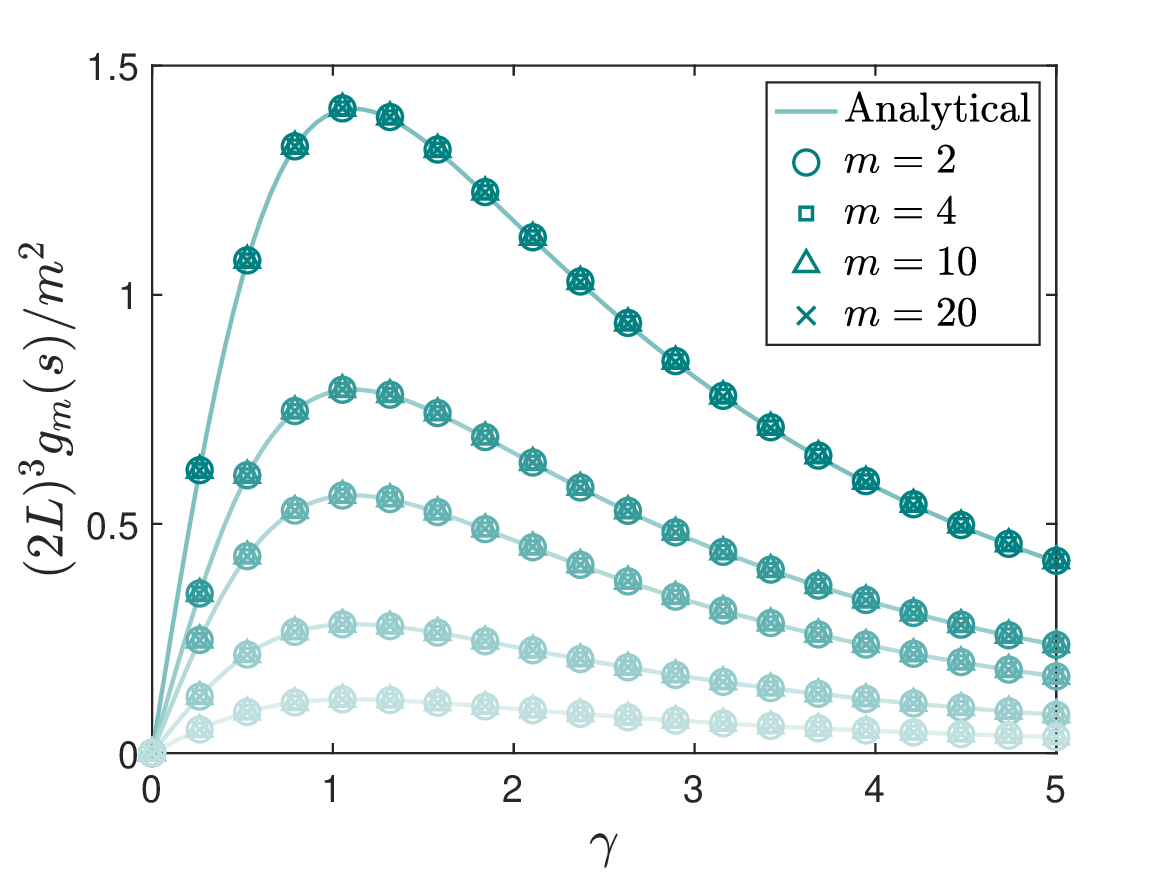}};
             \node[inner sep=0pt] (russell) at (125pt, -85pt)
    {\includegraphics[width=0.25\textwidth]{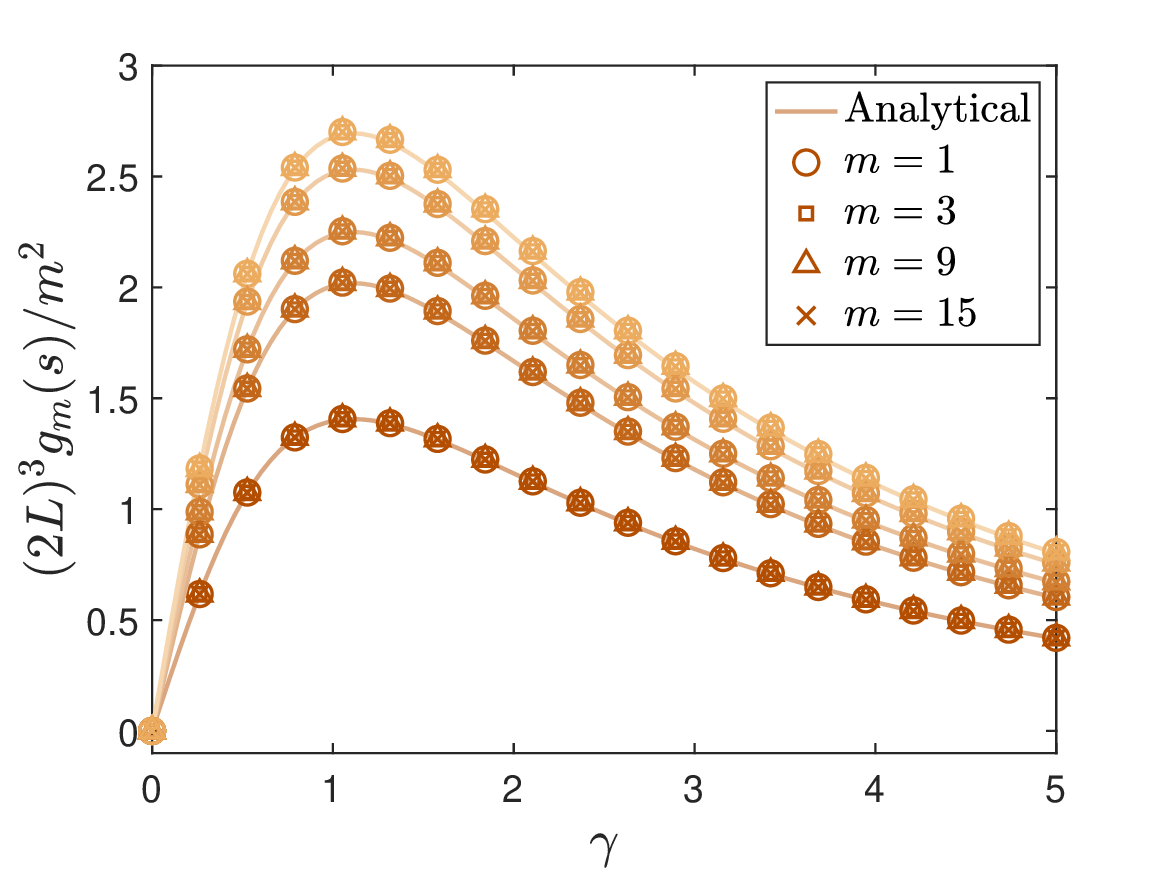}};

               \node[inner sep=0pt] (russell) at (-250pt, -185pt)
    {\includegraphics[width=0.25\textwidth]{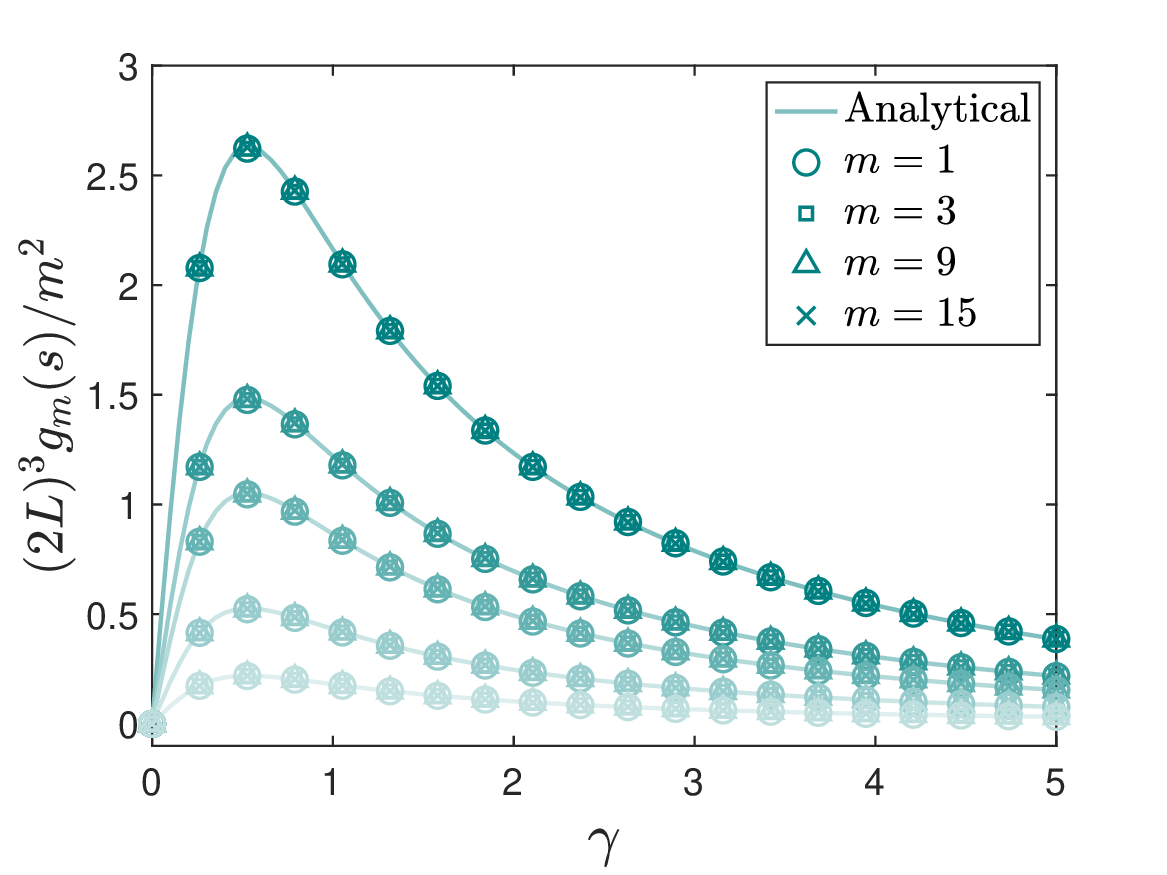}};
            \node[inner sep=0pt] (russell) at (-125pt, -185pt)
    {\includegraphics[width=0.25\textwidth]{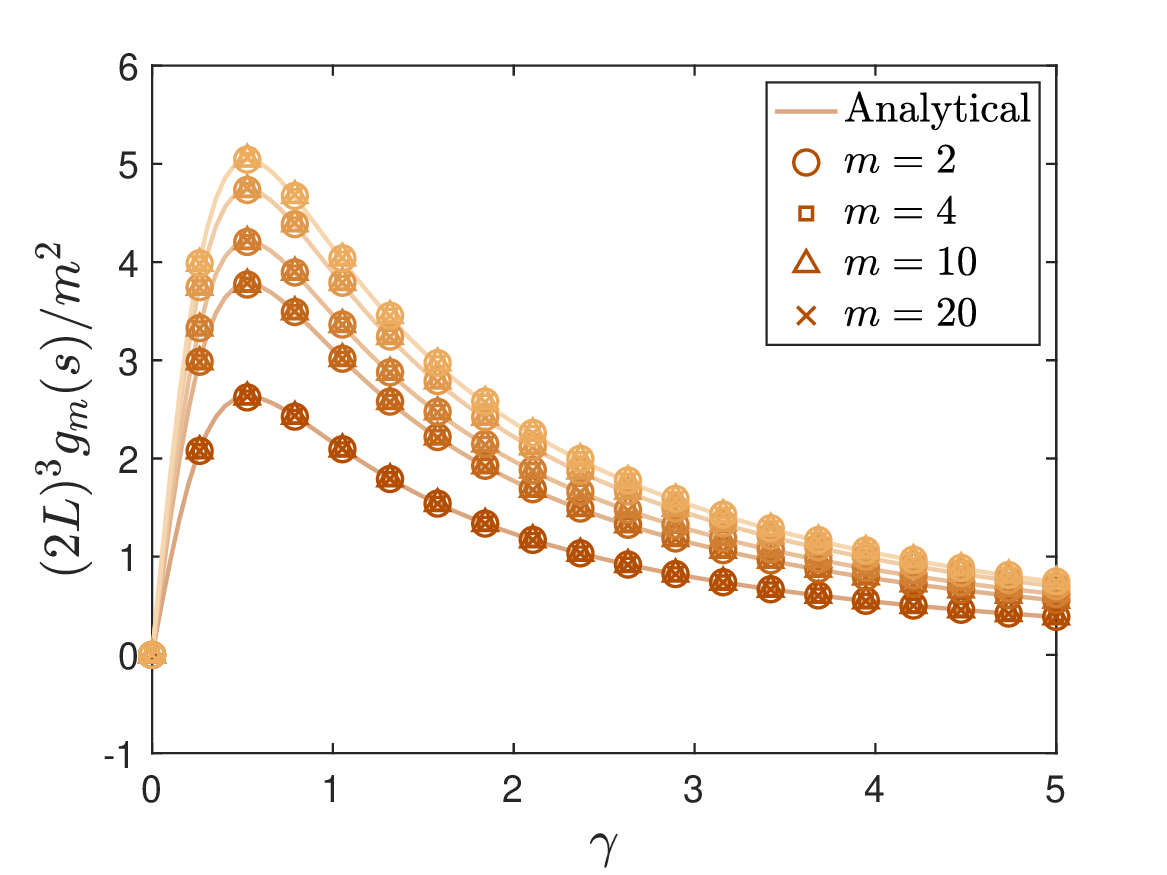}}; 
             \node[inner sep=0pt] (russell) at (0pt, -185pt)
    {\includegraphics[width=0.25\textwidth]{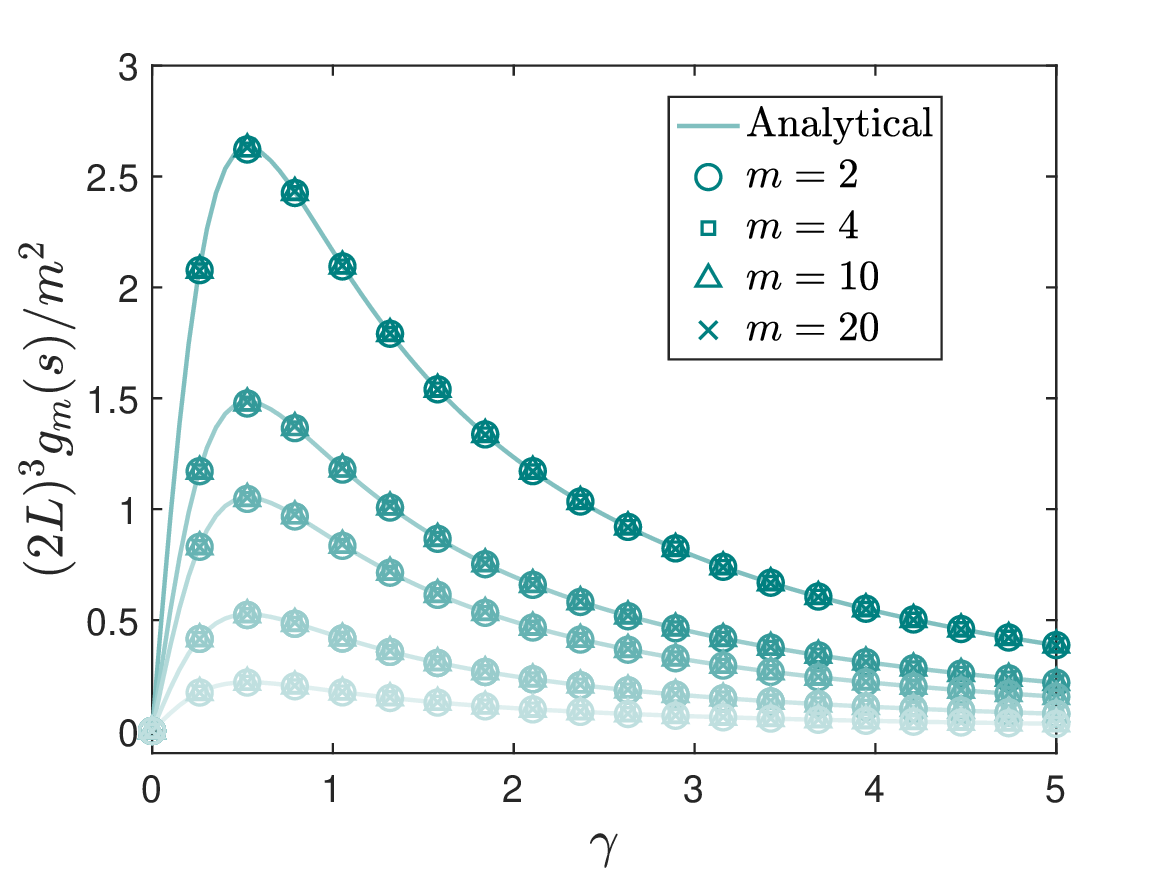}};
             \node[inner sep=0pt] (russell) at (125pt, -185pt)
    {\includegraphics[width=0.25\textwidth]{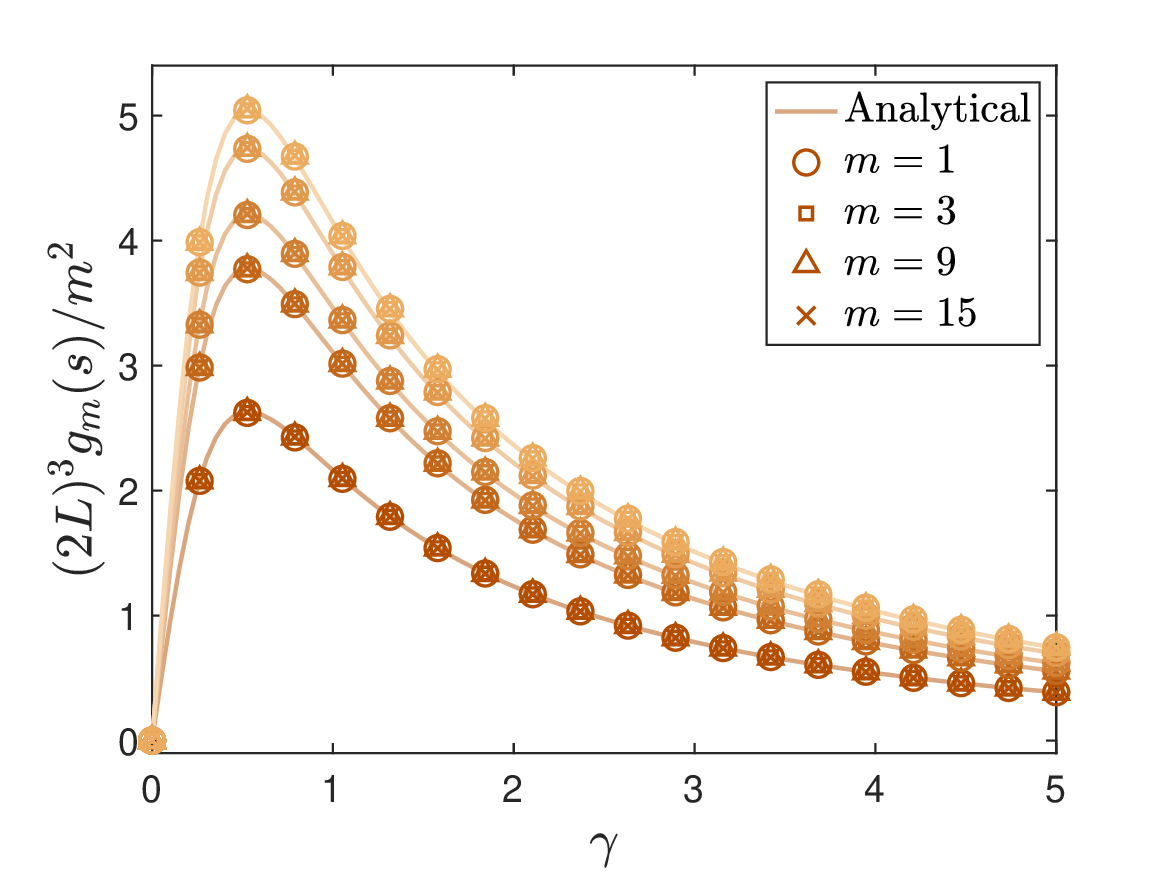}};

    \node at (-9,-2.5) {\bf (a)};
    \node at (-4.2,-2.5) {\bf (b)};
    \node at (0,-2.5) {\bf (c)};
    \node at (4.2,-2.5) {\bf (d)};
        \node at (-9,-6) {\bf (e)};
    \node at (-4.2,-6) {\bf (f)};
    \node at (0,-6) {\bf (g)};
    \node at (4.2,-6) {\bf (h)};

    \end{tikzpicture}
\caption{Harmonic chain: Analytical and numerical results of the rescaled relaxation rate  $(2L)^{3}g_{m}(s)/m^{2}$, plotted as a function of the dissipation strength $\gamma$. Panels (a)--(d) correspond to free boundary conditions, while panels (e)--(h) correspond to fixed boundary conditions. In each case, local dissipation is introduced at the left boundary, $n_d=1$, for the first two panels, and at the right boundary, $n_d=2L$, for the next two panels. The color gradient indicates decreasing values of the interface parameter $s$, from darker to lighter shades, with $s=1, 0.9, 0.8, 0.6, 0.4$. For each fixed $s$, different markers denote different slowly decaying modes $m$; their collapse shows that the scaled relaxation rate is independent of $m$ for slowly decaying modes. There is an excellent agreement between the numerical results (markers) and the exact analytical result (lines). For all plots, the total system size is $2L = 400$. For notational simplicity, we suppress the explicit dependence on the dissipation site $n_d$ in the figures and throughout the remainder of the paper, writing $g_m(s)$ instead of $g_{m,n_d}(s)$ whenever $n_d$ is fixed and clear from the context.
}
\label{fig: finite gamma boson}
\end{figure*}
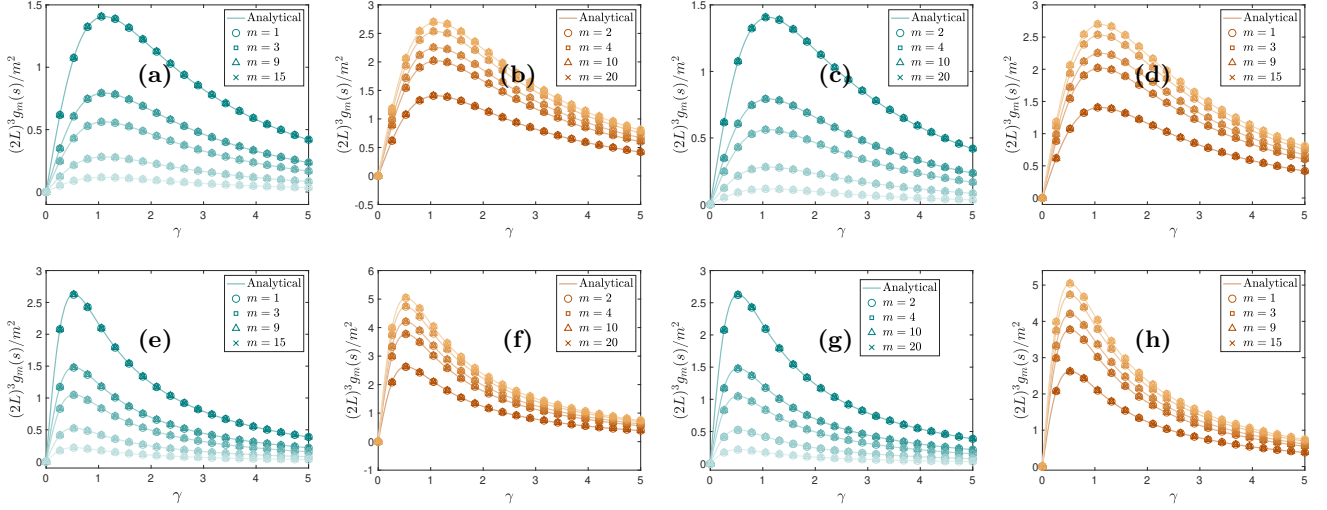

\subsubsection{Free boundary condition}
In this subsection, we consider the harmonic chain with a conformal interface subject to free boundary conditions and determine the relaxation rates as a function of the system parameters at a finite value of $\gamma$ for dissipation applied at the left and right boundaries. For free boundary conditions, we set $V_0 =0$ which reduces $\tilde{\gamma}_{j}$ in \eqref{eq: tilde gamma} to 
\begin{align}
   \tilde{\gamma}_{j} = -1 + \delta_{n_d,j}\left(\dfrac{\gamma^{2}}{4} - iE_{m}\gamma\right). 
\end{align}
First, we consider dissipation applied at the left boundary, $n_d =1$, for which $\tilde{\gamma}_{2L} = -1$. The quantization condition in Eq.\eqref{eq:quantization} then becomes 
\begin{align}
\label{eq:quantization for left bdry free}
\begin{split}
&\text{sin}\left(2L+1\right)\theta^{\rm free}_{m} +  \left(\tilde{\gamma}_{1}-1\right)\text{sin}\left(2L\right)\theta^{\rm free}_{m}\\ 
&-\tilde{\gamma}_{1}\text{sin}\left(2L-1\right)\theta^{\rm free}_{m}+ \left(\tilde\gamma_{1}+1\right)\sqrt{1-s^2} \,\,\text{sin}\theta^{\rm free}_{m}=0.
\end{split}
\end{align}
Note by setting $\gamma=0$, the above equation reduces to 
\begin{align}
\text{sin}[\left(2L+1\right)\theta^{\rm free}_{m}]+ \text{sin}[\left(2L-1\right)\theta^{\rm free}_{m}] = 2\,\text{sin}[\left(2L\right)\theta^{\rm free}_{m}], 
\end{align}
such that we recover the closed-system solution 
\begin{align}
    \theta^{\rm free}_{m'} =  \dfrac{m'\pi}{2L}. 
\end{align}
For finite dissipation strength $\gamma$, it is difficult to obtain a closed-form solution for $\theta^{\rm free}_m$ from the quantization condition in Eq.\eqref{eq:quantization for left bdry free}. We therefore expand $\theta^{\rm free}_m$ around the corresponding dissipation free solution in the large $L$ limit as
\begin{align}
\label{eq: ansatz for theta}
    \theta^{\rm free}_{m} \approx \dfrac{(2L-m)\pi}{2L} + \dfrac{\Theta_{R}+i\Theta_{I}}{L^{2}}.
\end{align}
After substituting the ansatz in Eq.\eqref{eq: ansatz for theta} into the quantization condition in Eq.\eqref{eq:quantization for left bdry free}, we obtain the following solutions for $\Theta_{R}$ and $\Theta_{I}$:
\begin{align}
\label{eq: Theta free left}
    &\Theta_{I} = -\dfrac{8m\pi\gamma\left(1+(-1)^{m}\sqrt{1-s^2}\right) }{\gamma^4+48\gamma^2+64} + \mathcal{O}(L^{-1}) \\ \nonumber
    &\Theta_{R} = -\dfrac{m\pi\left(1+(-1)^{m}\sqrt{1-s^2}\right)\gamma^{2}\left(56+\gamma^{2}\right) }{8\left(\gamma^4+48\gamma^2+64\right)} + \mathcal{O}(L^{-1})
\end{align}
The eigenvalue corresponding to this mode is
\begin{align}
&E_{m} = 2\,\text{sin}\left(\dfrac{\theta^{\rm free}_{m}}{2}\right).
\end{align}
Substituting the solutions for $\Theta_{I}$ and $\Theta_{R}$ from Eq.\eqref{eq: Theta free left} into the expression for  $\theta^{\rm free}_{m}$ in Eq.\eqref{eq: ansatz for theta}, we obtain
\begin{align}
&E_{m} \approx 2\, \left[1 + i\,\text{sin}\left(\dfrac{\pi}{4L}\right)\text{sin}\left(\dfrac{\Theta_{I}}{2L^{2}}\right) \right] \\ \nonumber
&\quad\quad\quad\approx 2\, \left[1 - i\dfrac{1}{L^3}\dfrac{m^{2}\pi^2\gamma\left(1+(-1)^{m}\sqrt{1-s^2}\right) }{\gamma^4+48\gamma^2+64} \right].
\end{align}
Thus, for the harmonic chain with free boundary conditions and left boundary dissipation, $n_d =1$, the relaxation rates of modes $m\ll L$ at finite $\gamma$ are given by
\begin{align}
    &g^{\rm free}_{m,1}(s) = - \text{Im}\left[E_{m}\right] \approx \dfrac{m^{2}\pi^2\gamma\left(1+(-1)^{m}\sqrt{1-s^2}\right) }{8L^3\left(\dfrac{\gamma^4}{16}+3\gamma^2+4\right)}, 
\end{align}
which reduces to the perturbative result in Eq.\eqref{eq:relax rate for weak boundary dissipation left} in the weak dissipation limit, $\gamma\ll 1$. \\

Similarly, keeping the same boundary conditions, we consider dissipation applied at the right boundary, $n_d =2L$. In this case, $\tilde\gamma_{1}=-1$, and quantization condition becomes
\begin{align}
\label{eq:quantization for right bdry free}
\begin{split}
&\text{sin}[\left(2L+1\right)\theta^{\rm free}_{m}] +  \left(\tilde{\gamma}_{2L}-1\right)\text{sin}[\left(2L\right)\theta^{\rm free}_{m}]\\ 
&-\tilde{\gamma}_{2L}\text{sin}[\left(2L-1\right)\theta^{\rm free}_{m}]- \left(\tilde\gamma_{2L}+1\right)\sqrt{1-s^2} \,\,\text{sin}[\,\theta^{\rm free}_{m}]=0.
\end{split}
\end{align}
For this case, the relaxation rates of modes with $m\ll L$ are given by
\begin{align}
    &g^{\rm free}_{m,2L}(s) \approx \dfrac{m^{2}\pi^2\gamma\left(1-(-1)^{m}\sqrt{1-s^2}\right) }{8L^3\left(\dfrac{\gamma^4}{16}+3\gamma^2+4\right)}, 
\end{align}
which recovers the perturbative result in Eq.\eqref{eq:relax rate for weak boundary dissipation right} for $\gamma\ll 1$.

\subsubsection{Fixed boundary condition}
In this subsection, we consider fixed boundary conditions at both ends of the harmonic chain. Setting $V_{0} = 1$ in Eq.\eqref{eq: boundary conditions}, $\tilde{\gamma}_{j}$ becomes
\begin{align}
       \tilde{\gamma}_{j} = \delta_{n_d,j}\left(\dfrac{\gamma^{2}}{4} - iE_{m}\gamma\right).
\end{align}
When dissipation is applied at either boundary, $n_d\in\{1,2L\}$, the quantization condition in Eq.\eqref{eq:quantization} reduces to 
\begin{align}
\label{eq: quantization fixed}
\begin{split}
&\text{sin}[\left(2L+1\right)\theta^{\rm fixed}_{m}]+\left(\tilde{\gamma}_{1}\delta_{n_d,1}+ \tilde{\gamma}_{2L}\delta_{n_d,2L}\right)\text{sin}[\left(2L\right)\theta^{\rm fixed}_{m}] \\
&+\left(\tilde{\gamma}_{1}\delta_{n_d,1}- \tilde{\gamma}_{2L}\delta_{n_d,2L}\right)\tilde{\gamma}_{n_d}\sqrt{1-s^2}\text{sin}[\,\theta^{\rm fixed}_{m}]= 0.
\end{split}
\end{align}
Similarly, setting $\gamma =0$ in the above quantization condition recovers the following closed system solution for $\theta^{\rm fixed}_{m'}$:
\begin{align}
    \theta^{\rm fixed}_{m'} = \dfrac{m'\pi}{2L+1}. 
\end{align}
For dissipation applied at the left boundary, $n_d =1$, the large $L$ expansion of $\theta^{\rm fixed}_{m}$ is given by
\begin{align}
        \theta^{\rm fixed}_{m} \approx \pi -\dfrac{m\pi}{2L+1} + \dfrac{\Theta_{R}+i\Theta_{I}}{(2L+1)^{2}}.
\end{align}
Substituting the above expansion for $\theta^{\rm fixed}_{m}$ into Eq.\eqref{eq: quantization fixed}, the imaginary part $\Theta_{I}$ is obtained as 
\begin{align}
    \Theta_{I} = -\dfrac{8m\pi\gamma\left(1+(-1)^{m}\sqrt{1-s^2}\right) }{\gamma^4+56\gamma^2+16} + \mathcal{O}(L^{-1}).
\end{align}
Using the above result for $\Theta_{I}$, the relaxation rate of the slowly decaying modes with $m\ll L$ for the harmonic chain with fixed boundary conditions and left boundary dissipation at finite $\gamma$ is given by
\begin{align}
    &g^{\rm fixed}_{m,1}(s) \approx \dfrac{m^{2}\pi^2\gamma\left(1+(-1)^{m}\sqrt{1-s^2}\right) }{(2L+1)^3\left(\dfrac{\gamma^4}{16}+\dfrac{7}{2}\gamma^2+1\right)}
\end{align}
For the right boundary $n_d = 2L$, substituting the solution for $\theta^{\rm fixed}_m$ into Eq.\eqref{eq: quantization fixed} gives 
\begin{align}
    \Theta_{I} = -\dfrac{8m\pi\gamma\left(1-(-1)^{m}\sqrt{1-s^2}\right) }{\gamma^4+56\gamma^2+16} + \mathcal{O}(L^{-1}),
\end{align}
and
\begin{align}
    &g^{\rm fixed}_{m,2L}(s) \approx   \dfrac{m^{2}\pi^2\gamma\left(1-(-1)^{m}\sqrt{1-s^2}\right) }{(2L+1)^3\left(\dfrac{\gamma^4}{16}+\dfrac{7}{2}\gamma^2+1\right)}.
\end{align}

As shown in Fig.\ref{fig: finite gamma boson}, we compare the finite $\gamma$ analytical results derived above with numerical results for the harmonic chain, obtained by diagonalizing the non-Hermitian matrix $\mathbf{X}$. We perform the comparison for both free and fixed boundary conditions, with dissipation applied at either the left or right boundary. The rescaled relaxation rates, $(2L)^3 g_{m}(s)/m^2$, of the slowly relaxing modes with $m\ll L$ agree very well with the analytical results and recover the perturbative results derived in the previous section in the weak dissipation limit. Moreover, for large system sizes, the relaxation rates of modes localized in the non-dissipative chain at $s=0$ gets suppressed by $1-\sqrt{1-s^2}$ as the interface transmission is decreased from $s=1$ to $s=0$. This suppression persists for large system sizes even at finite $\gamma$.

\begin{figure*}
\centering
    \begin{tikzpicture}
            \node[inner sep=0pt] (russell) at (-250pt, -85pt)
    {\includegraphics[width=0.25\textwidth]{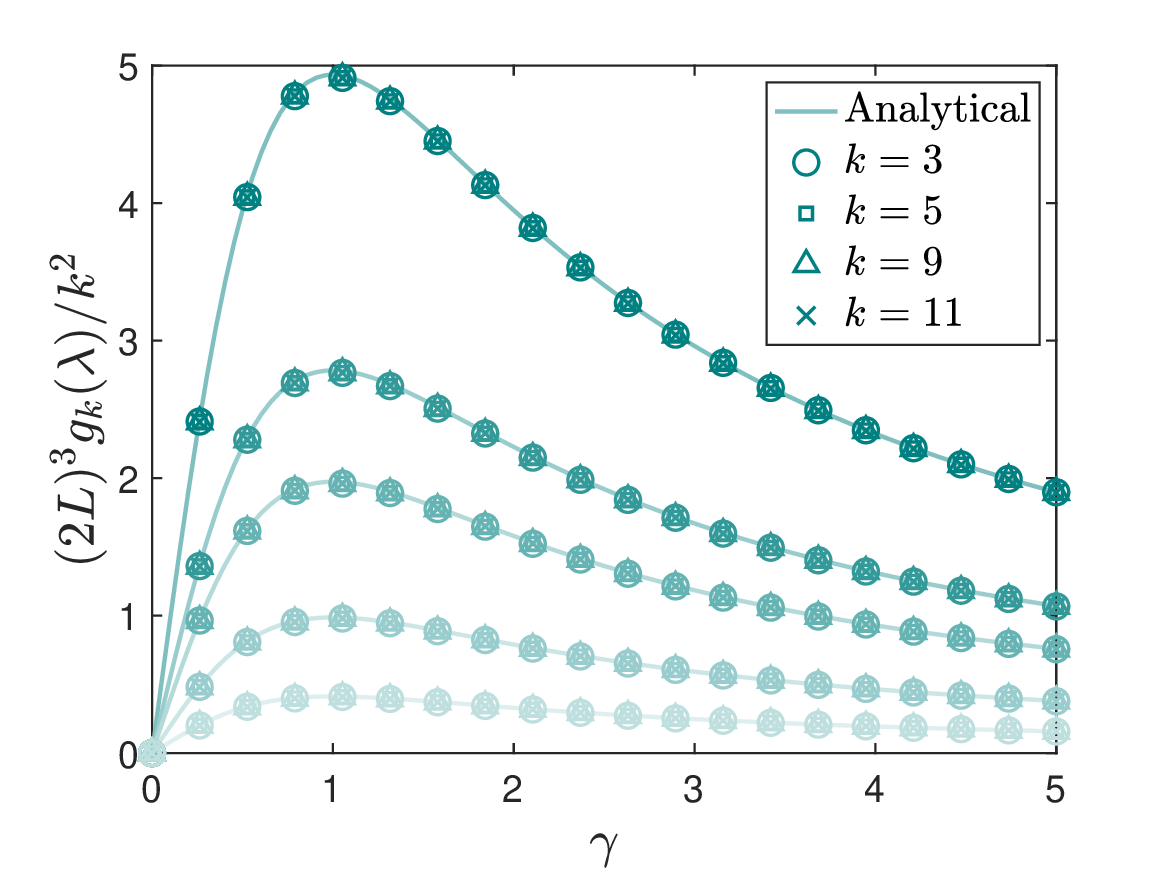}};
            \node[inner sep=0pt] (russell) at (-125pt, -85pt)
    {\includegraphics[width=0.25\textwidth]{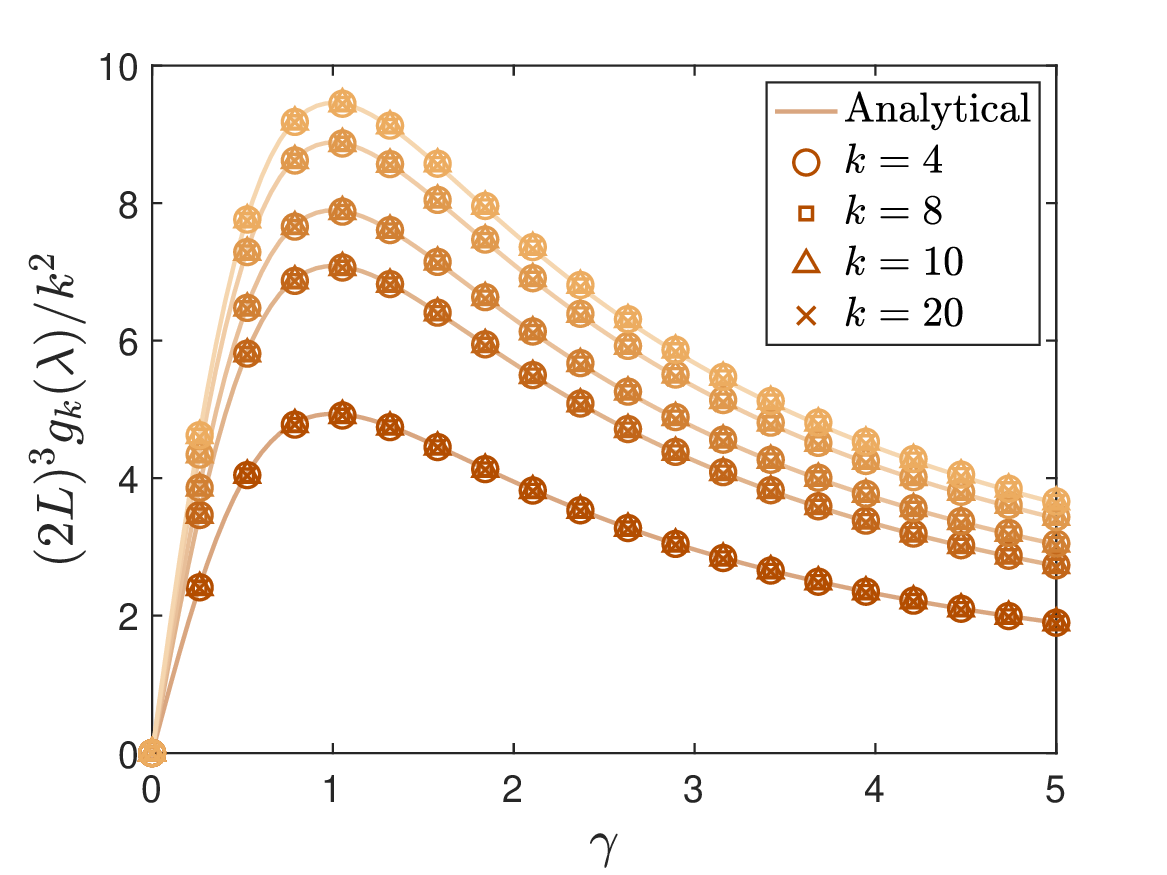}}; 
             \node[inner sep=0pt] (russell) at (0pt, -85pt)
    {\includegraphics[width=0.25\textwidth]{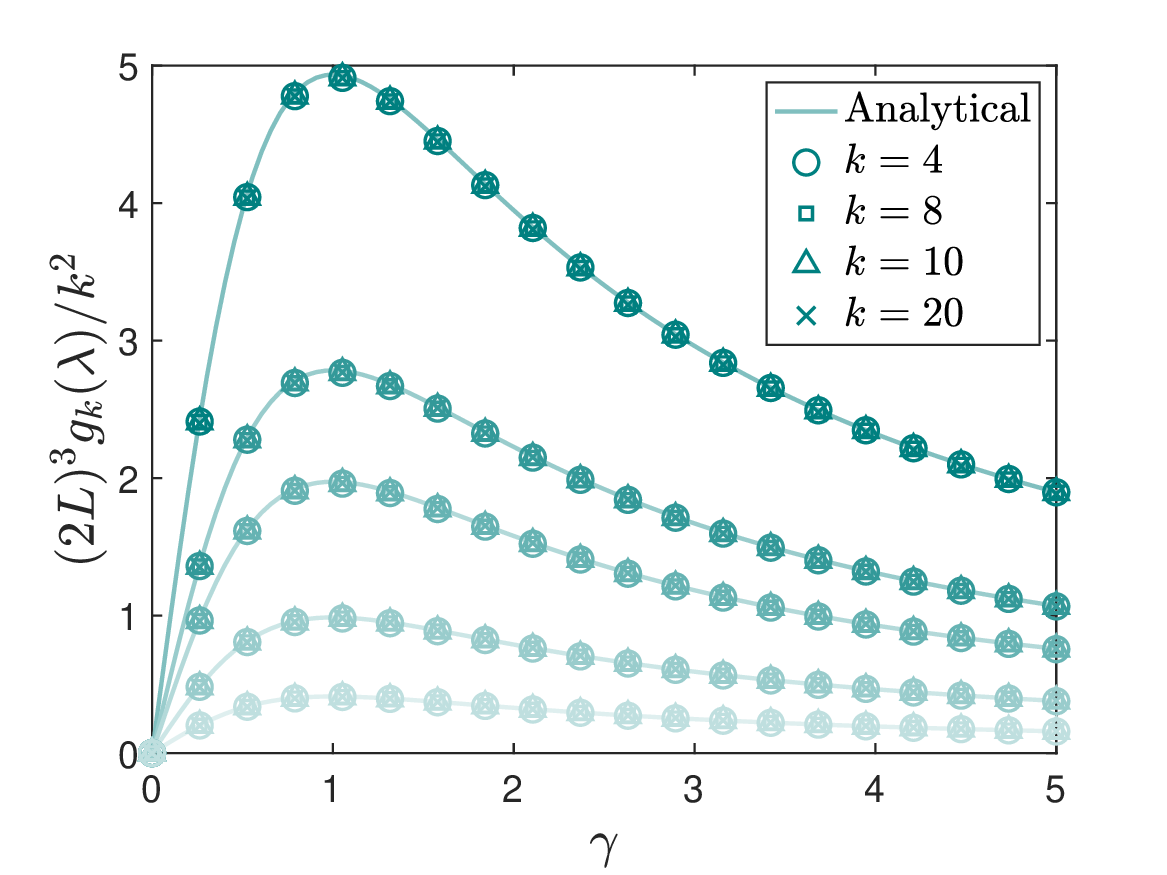}};
             \node[inner sep=0pt] (russell) at (125pt, -85pt)
    {\includegraphics[width=0.25\textwidth]{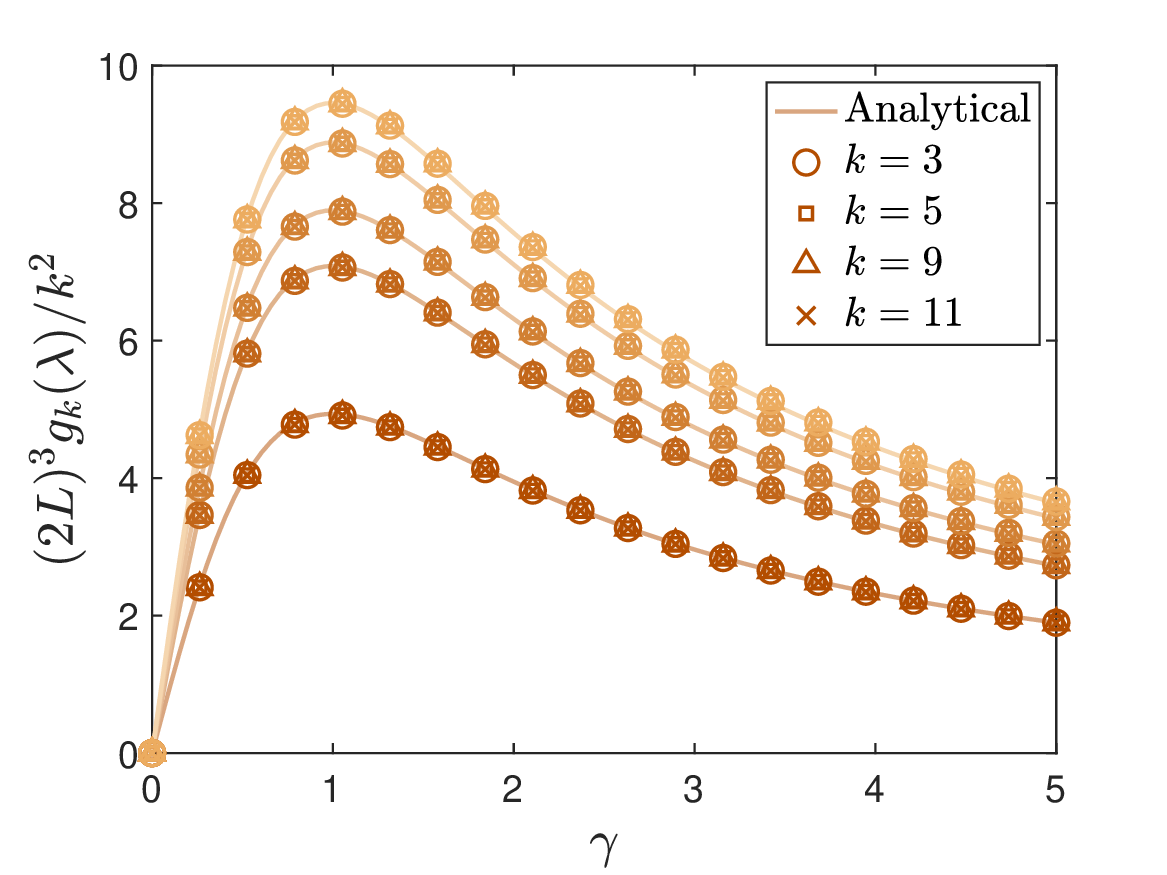}};

               \node[inner sep=0pt] (russell) at (-250pt, -185pt)
    {\includegraphics[width=0.25\textwidth]{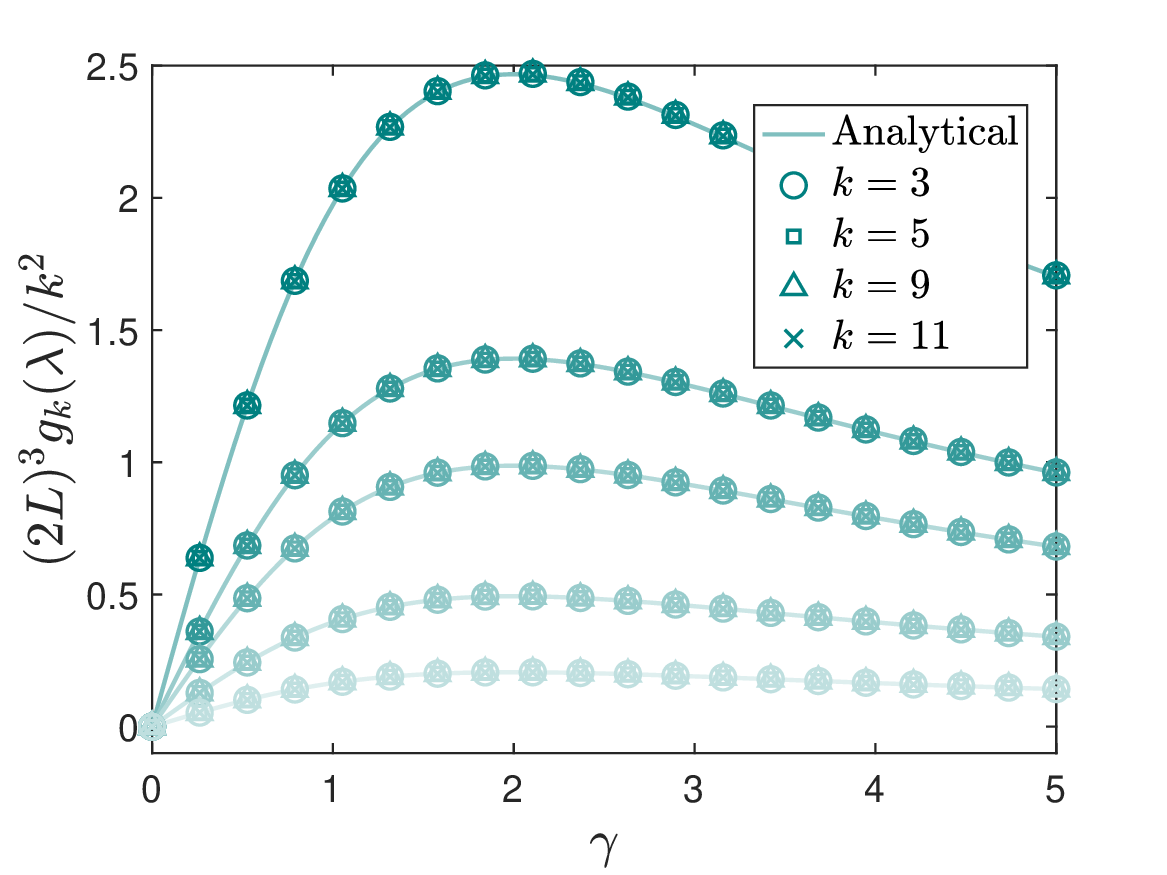}};
            \node[inner sep=0pt] (russell) at (-125pt, -185pt)
    {\includegraphics[width=0.25\textwidth]{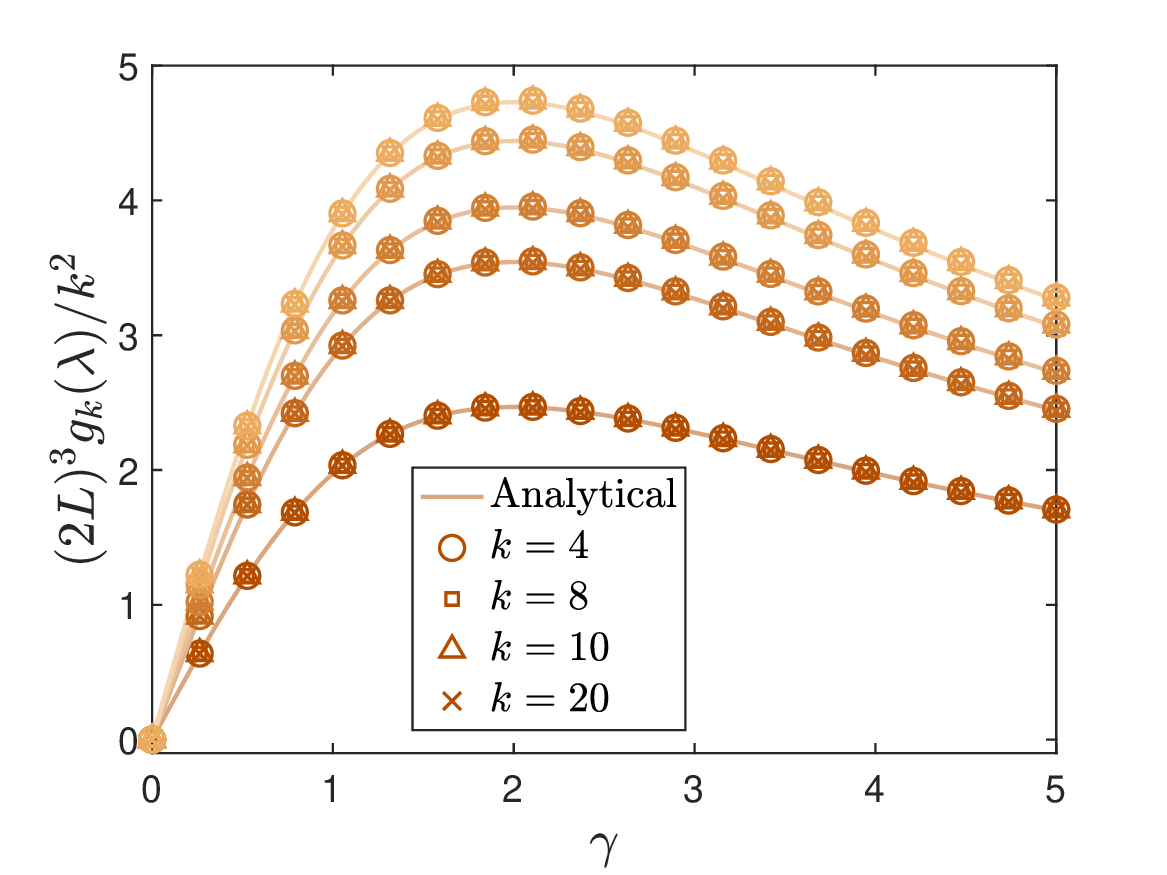}}; 
             \node[inner sep=0pt] (russell) at (0pt, -185pt)
    {\includegraphics[width=0.25\textwidth]{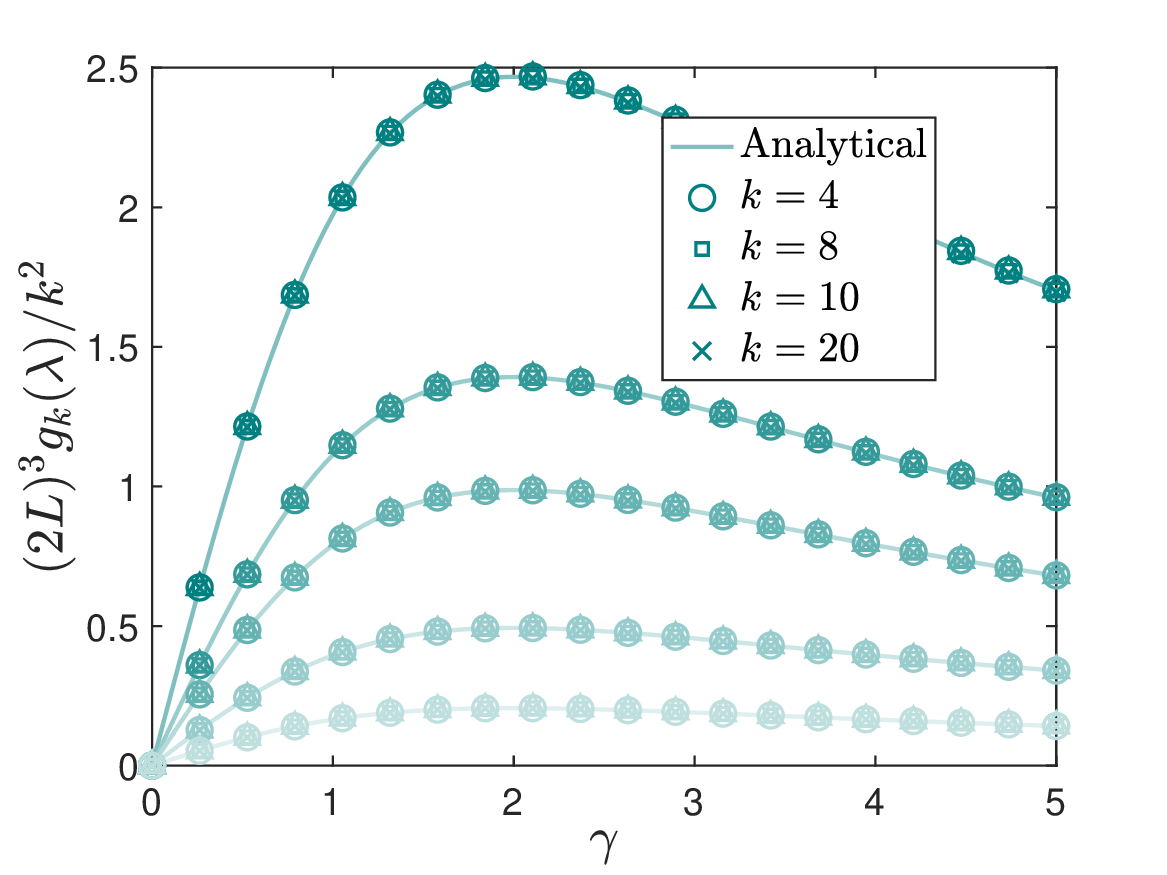}};
             \node[inner sep=0pt] (russell) at (125pt, -185pt)
    {\includegraphics[width=0.25\textwidth]{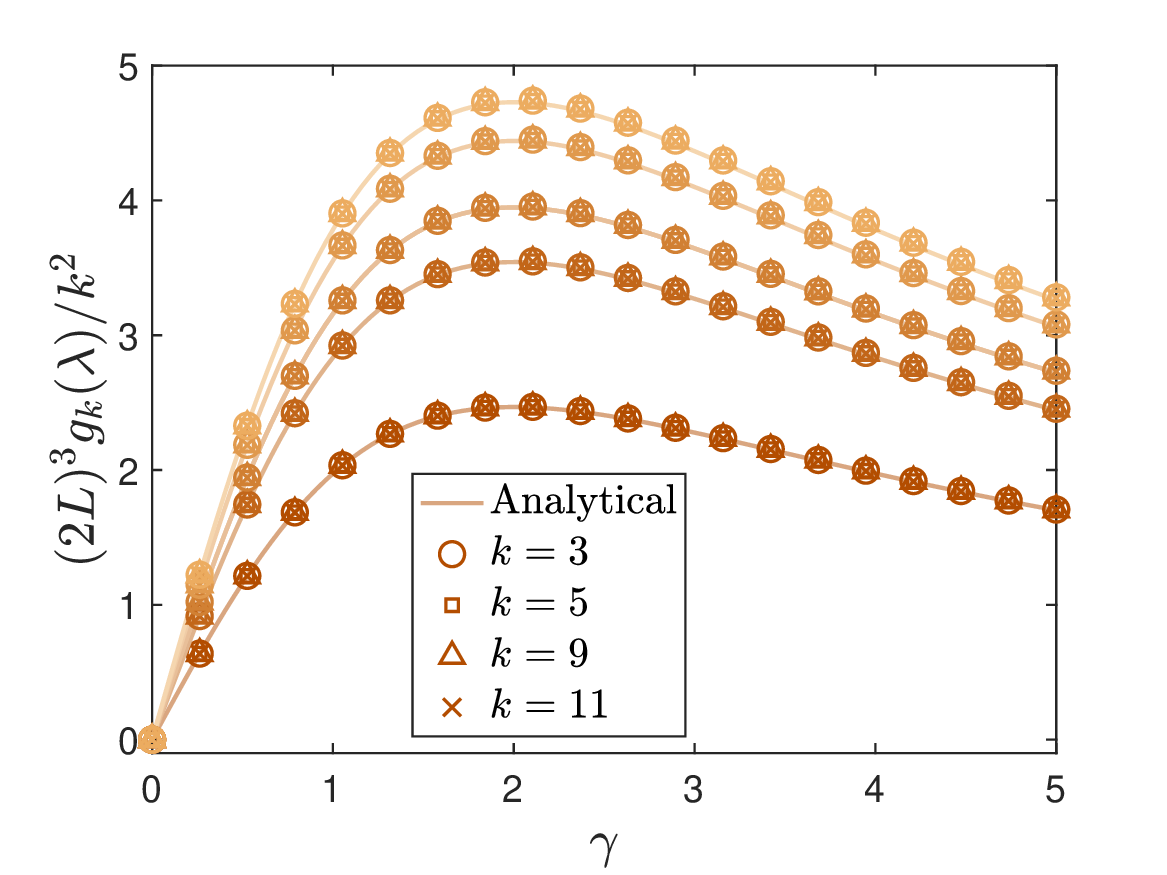}};

    \node at (-9,-2.5) {\bf (a)};
    \node at (-4.2,-2.5) {\bf (b)};
    \node at (0,-2.5) {\bf (c)};
    \node at (4.2,-2.5) {\bf (d)};
        \node at (-9,-6) {\bf (e)};
    \node at (-4.2,-6) {\bf (f)};
    \node at (0,-6) {\bf (g)};
    \node at (4.2,-6) {\bf (h)};

    \end{tikzpicture}
\caption{Free fermion chain: Analytical and numerical results of the rescaled relaxation rate  $(2L)^{3}g_{k}(\lambda)/k^{2}$, plotted as a function of the dissipation strength $\gamma$. Panels (a)--(d) correspond to free boundary conditions, while panels (e)--(h) correspond to fixed boundary conditions. In each case, local dissipation is introduced at the left boundary, $n_d=1$, for the first two panels, and at the right boundary, $n_d=2L$, for the next two panels. The color gradient indicates decreasing values of the interface parameter $\lambda$, from darker to lighter shades, with $\lambda = 1, 0.9, 0.8, 0.6, 0.4$. For each fixed $\lambda$, different markers denote different slowly decaying modes $k$; their collapse shows that the scaled relaxation rate is independent of $k$ for slowly decaying modes. There is an excellent agreement between the numerical results (markers) and the exact analytical result (lines). For all plots, the total system size is $2L = 400$. For notational simplicity, we suppress the explicit dependence on the dissipation site $n_d$ in the figures and throughout the remainder of the paper, writing $g_k(\lambda)$ instead of $g_{k,n_d}(\lambda)$ whenever $n_d$ is fixed and clear from the context.
}
\label{fig: finite gamma fermion}
\end{figure*}
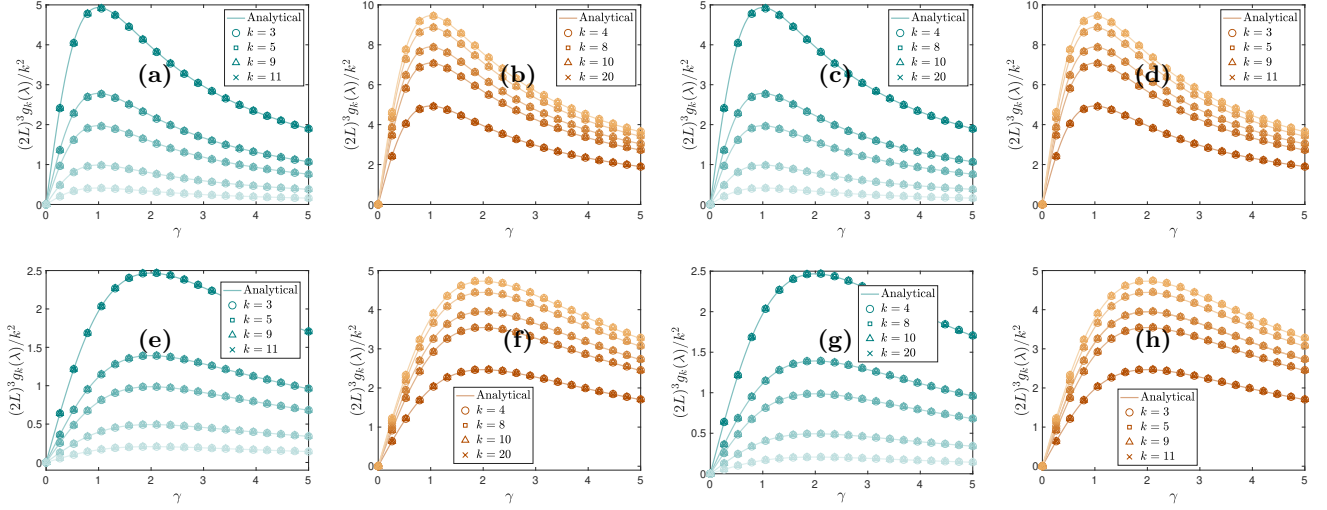

\subsection{Free fermion chain}
In this section, we derive an approximate analytical expression for the Liouvillian gap at finite dissipation strength $\gamma$ for critical free fermion chain with a conformal interface. We consider the eigenvalue problem  $\textbf{H}_{\text{eff}}(\lambda)|\boldsymbol{\psi}_k\rangle = E_k|\boldsymbol{\psi}_k\rangle$ under different boundary conditions. The corresponding single-particle eigenfunctions satisfy the following conditions:\\
(i) in the bulk, away from the boundaries and the conformal interface,
\begin{align}
    -\dfrac{1}{2}\left(\psi_{i-1} +\psi_{i+1}\right) = E_k\psi_{i},
\end{align}
(ii) at the two boundaries, they satisfy
\begin{align}
\label{eq:bdry condition free fermion}
\begin{split}
 &-\dfrac{1}{2}\left(\tilde{\gamma}_{1}\psi_{1} + \psi_{2}\right) = E_k\psi_{1} \\
&-\dfrac{1}{2}\left(\tilde{\gamma}_{2L}\psi_{2L} +\psi_{2L-1}\right)  = E_k\psi_{2L},   
\end{split}
\end{align}
with  $\tilde{\gamma}_{i} = \left(V_{0}+i\gamma\delta_{n_d,i}\right)$.
Thus, we introduce the plane-wave ansatz
\begin{align}
\psi_{j} = Az^{j}_k + Bz^{-j}_k, \quad z_k\in \mathbb{C}.
\end{align}
The bulk condition then gives the eigenvalue
\begin{align}
E = -\dfrac{1}{2}\left(z^{1}_k + z^{-1}_k\right).
\end{align}
In addition to the bulk and boundary conditions, the conformal interface at the center of the chain introduces two further matching conditions:
\begin{align}
    -\dfrac{1}{2}\left(\psi_{L-1} + \sqrt{1-\lambda^2}\psi_{L} + \dfrac{\lambda}{2}\psi_{L+1}\right) = E_k\psi_{L}, \\
    -\dfrac{1}{2}\left(\lambda\psi_{L} - \sqrt{1-\lambda^2}\psi_{L+1} + \psi_{L+2}\right) = E_k\psi_{L+1}.
\end{align}
Following the construction used for the harmonic chain and in Ref.~\cite{Barad_2025}, we introduce the following plane-wave ansatz for the system with a conformal interface:
\begin{align}
    \psi_{j}(k) =
   \begin{cases}
       \alpha_{k}(Az^{j}_k + Bz^{-j}_k) \qquad\quad 1\le j \le L \\
       \beta_{k}(Cz^{j}_k + Dz^{-j}_k) \quad L+1\le j \le 2L, 
   \end{cases} 
\end{align}
Here, $z_k=e^{i\theta_{k}}$. Substituting the ansatz into the boundary and interface conditions, we obtain the following equations:
\begin{align}
\begin{split}
&A(1-\tilde{\gamma}_{1} z_k) + B(1-\tilde{\gamma}_{1}z^{-1}_k) = 0\\
     \begin{split}       
     &A(z^{L+1}_k\alpha_{k} -z^{L}_k\alpha_{k}\mu) + B(z^{-L-1}_k\alpha_{k} -z^{-L}_k\alpha_{k}\mu)\\  &+  C(-z^{L+1}_k\beta_{k}\lambda) +D(-z^{-L-1}_k\beta_{k}\lambda) = 0
     \end{split}\\
     \begin{split}
         &A(-z^{L}_k\alpha_{k}\lambda) + B(-z^{-L}_k\alpha_{k}\lambda)\\ &+C(z^{L}_k\beta_{k} +z^{L+1}_k\beta_{k}\mu) +  D(z^{-L}_k\beta_{k} +z^{-L-1}_k\beta_{k}\mu) = 0 
         \end{split}\\
     &C(z^{2L+1}_k- \tilde{\gamma}_{2L}z^{2L}_k) + D(z^{-2L-1}_k- \tilde{\gamma}_{2L}z^{-2L}_k) = 0 
\end{split}
\end{align}
with $\mu = \sqrt{1-\lambda^2}$. Combining the boundary and interface conditions, we obtain the following quantization condition:
\begin{align}
\begin{split}
&\text{sin}[(2L+1)\theta_{k}] - (\tilde{\gamma}_{1}+\tilde{\gamma}_{2L})\text{sin}[(2L)\theta_{k}]\\
+&\tilde{\gamma}_{1}\tilde{\gamma}_{2L}\text{sin}[(2L-1)\theta_{k}]  -\mu\,(\tilde{\gamma}_{1}-\tilde{\gamma}_{2L})\text{sin}[\theta_{k}] = 0.
        \end{split}
\label{eq:bdry condition free fermion}
\end{align}
In the remainder of this subsection, we use the above quantization condition to derive analytical expressions for the relaxation rates at finite $\gamma$ for both free and fixed boundary conditions.

\subsubsection{Free boundary condition}

Here, we consider the critical free-fermion chain with a conformal interface subject to free boundary conditions and determine the relaxation rates at finite dissipation strength $\gamma$ for dissipation applied at either the left or right boundary. For free boundary conditions, we set $V_0 =0$ which reduces $\tilde{\gamma}_{j}$ in \eqref{eq:bdry condition free fermion} to 
\begin{align}
 \tilde{\gamma}_{i} = i\gamma\delta_{n_d,i},
\end{align}
and the quantization condition in Eq.\eqref{eq:bdry condition free fermion} reduces to 
\begin{align}
\label{eq:quantization for left bdry free fermion}
\begin{split}
&\text{sin}[\left(2L+1\right)\theta^{\rm free}_{k}] -  i\gamma\left(\text{sin}[\left(2L\right)\theta^{\rm free}_{k}]-(-1)^{n_d}\mu \,\,\text{sin}[\theta^{\rm free}_{k}]\right)\\&=0.
\end{split}
\end{align}
Note that, by setting $\gamma=0$, the above quantization condition reduces to 
\begin{align}
\text{sin}[\left(2L+1\right)\theta^{\rm free}_{k'}] = 0, \end{align}
from which we recover the closed-system solution
\begin{align}
    \theta^{\rm free}_{k'} =  \dfrac{k'\pi}{2L+1}. 
\end{align}
In the large $L$ limit, we therefore expand the solution for $\theta^{\rm free}_k$ around the corresponding closed-system value as
\begin{align}
\label{eq: ansatz for theta fermion}
    \theta^{\rm free}_{k} \approx \pi- \dfrac{k\pi}{2L+1} + \dfrac{\Theta_{R}+i\Theta_{I}}{L^{2}}.
\end{align}
After substituting the ansatz in Eq.\eqref{eq: ansatz for theta fermion} into the quantization condition in Eq.\eqref{eq:quantization for left bdry free fermion}, we obtain
\begin{align}
\label{eq: Theta free left fermion}
    &\Theta_{I} = -\dfrac{k\pi\gamma\left(1+(-1)^{k}\sqrt{1-\lambda^2}\right) }{4\left(\gamma^2+1\right)} + \mathcal{O}(L^{-1}), \\
    &\Theta_{R} = -\dfrac{k\pi\gamma^2\left(1+(-1)^{k}\sqrt{1-\lambda^2}\right)}{4\left(\gamma^2+1\right)} + \mathcal{O}(L^{-1}).
\end{align}
Thus, using the eigenvalue $E_{k} = -\,\text{cos}\left(\theta^{\rm free}_{k}\right)$, the relaxation rates of the modes with $k\ll L$ are given by $g^{\rm free}_{k,1}(\lambda)  = - 2 \Im[E_{k}]$ :
\begin{align}
    &g^{\rm free}_{k,1}(\lambda) \approx \dfrac{k^{2}\pi^2\gamma\left(1+(-1)^{k}\sqrt{1-\lambda^2}\right) }{4\left(\gamma^2+1\right)L^3}, 
\end{align}
which is valid at finite dissipation strength $\gamma$.\\

Similarly, for right boundary dissipation, we expand the solution for $\theta^{\rm free}_{k}$ in the large $L$ limit as
\begin{align}
        \theta^{\rm free}_{k} \approx \pi - \dfrac{k\pi}{2L+1} + \dfrac{\Theta_{R}+i\Theta_{I}}{L^{2}}.
\end{align}
Using this solution in the quantization condition, the finite $\gamma$ relaxation rates of the modes with $k \ll L$ are given by
\begin{align}
    &g^{\rm free}_{k,2L}(\lambda) \approx \dfrac{k^{2}\pi^2\gamma\left(1-(-1)^{k}\sqrt{1-\lambda^2}\right) }{4\left(\gamma^2+1\right)L^3}.
\end{align}

\subsubsection{Fixed boundary condition}
Next, we consider the critical free fermion chain with a conformal interface under fixed boundary conditions. For fixed boundary conditions, we set $V_0 = \pm 1$ which reduces $\tilde{\gamma}_{j}$ in Eq.\eqref{eq:bdry condition free fermion} to 
\begin{align}
 \tilde{\gamma}^{\pm}_{i} = \left(i\gamma\delta_{n_d,i}\pm1\right),
\end{align}
such that quantization condition Eq.\eqref{eq:bdry condition free fermion} becomes 
\begin{align}
\label{eq:quantization for right bdry free fermion}
\begin{split}
&\text{sin}[\left(2L+1\right)\theta^{\rm fixed}_{k}] - (i\gamma\pm 2)\text{sin}[\left(2L\right)\theta^{\rm fixed}_{k}] \\
&+(1\pm i\gamma)\, \text{sin}[\left(2L-1\right)\theta^{\rm fixed}_{k}]+i(-1)^{n_d} \mu\gamma \,\,\text{sin}[\theta^{\rm fixed}_{k}]=0.
\end{split}
\end{align}
Note that, by setting $\gamma=0$, the above equation reduces to 
\begin{align}
\text{sin}[\left(2L+1\right)\theta^{\rm fixed}_{k'}]+ \text{sin}[\left(2L-1\right)\theta^{\rm fixed}_{k'}] = \pm 2\,\text{sin}[\left(2L\right)\theta^{\rm fixed}_{k'}], \end{align}
such that we recover the same closed-system solution for both $V_0 = \pm 1$:
\begin{align}
    \theta^{\rm fixed}_{k'} =  \dfrac{k'\pi}{2L}. 
\end{align}
For the remainder of this section, we focus on $V_{0} =+1$ and derive analytical expressions for the relaxation rates of the slowly relaxing modes.\\

For left boundary dissipation, $n_d =1$, we expand the general solution for $\theta^{\rm fixed, +}_{k}$ as 
\begin{align}
        \theta^{\rm fixed, +}_{k} \approx \pi- \dfrac{k\pi}{2L} + \dfrac{\Theta_{R}+i\Theta_{I}}{L^{2}}.
\end{align}
After substituting this expansion into the quantization condition in Eq.\eqref{eq:quantization for right bdry free fermion}, we obtain
\begin{align}
\label{eq: Theta free left fermion}
    &\Theta_{I} = -\dfrac{k\pi\gamma\left(1+(-1)^{k}\sqrt{1-\lambda^2}\right) }{4\,\left(\gamma^2+4\right)} + \mathcal{O}(L^{-1}), \\
    &\Theta_{R} = -\dfrac{k\pi\gamma^2\left(1+(-1)^{k}\sqrt{1-\lambda^2}\right)}{8\,\left(\gamma^2+4\right)} + \mathcal{O}(L^{-1}).
\end{align}
Thus, from the eigenvalue corresponding to this mode 
$E_{k} = -\,\text{cos}\left(\theta^{\rm fixed, +}_{k}\right)$, the finite $\gamma$ relaxation rates of modes with with $k\ll L$ for $V_{0} = +1$ are given by
\begin{align}
    &g^{\rm fixed, +}_{k,1}(\lambda) \approx \dfrac{k^2\pi^2\gamma\left(1+(-1)^{k}\sqrt{1-\lambda^2}\right) }{8\left(\gamma^2+4\right)L^3}. 
\end{align}
Similarly, for right boundary dissipation, $n_d =2L$, we expand $\theta^{\rm fixed, +}_{k}$ around the corresponding closed-system solution in the large $L$ limit as
\begin{align}
        \theta^{\rm fixed, +}_{k} \approx \pi - \dfrac{k\pi}{2L} + \dfrac{\Theta_{R}+i\Theta_{I}}{L^{2}},
\end{align}
and substituting this solution into the quantization condition in Eq.\eqref{eq:quantization for right bdry free fermion}, we obtain
\begin{align}
    &\Theta_{I} = -\dfrac{k\pi\gamma\left(1-(-1)^{k}\sqrt{1-\lambda^2}\right) }{4\,\left(\gamma^2+4\right)} + \mathcal{O}(L^{-1}), \\ \nonumber
    &\Theta_{R} = -\dfrac{k\pi\gamma^2\left(1-(-1)^{k}\sqrt{1-\lambda^2}\right)}{8\,\left(\gamma^2+4\right)} + \mathcal{O}(L^{-1}).
\end{align}
Therefore, in this case, the finite $\gamma$ relaxation rates of the slowly decaying modes with $k\ll L$ are given by  
\begin{align}
    &g^{\rm fixed, +}_{k, 2L}(\lambda) = -  \text{Im}\left[E_{k}\right]\\
    &\quad\quad\approx  \dfrac{k^2\pi^2\gamma\left(1-(-1)^{k}\sqrt{1-\lambda^2}\right) }{8\left(\gamma^2+4\right)L^3}.
\end{align}

In Fig.\ref{fig: finite gamma fermion}, we compare the finite $\gamma$ analytical results derived in this subsection with  numerical results for free fermion chain, obtained by diagonalizing the non-Hermitian structure matrix $\mathbf{A}$ in Eq.(23) of \cite{Barad_2025}. We present the comparison for both free and fixed boundary conditions, considering dissipation at the left and right boundaries separately. In the large $L$ limit, the relaxation rates of slowly decaying modes localized in the non-dissipative chain in the perfectly reflective limit at $\lambda=0$ get suppressed by $1-\sqrt{1-\lambda^2}$ as the interface transmission is decreased continuously from $\lambda=1$ to $\lambda=0$, even at finite dissipation strength $\gamma$.

\section{Liouvillian Gap and Relaxation coefficients $c^{\text{global}}_{\rm relax}$ } 
\label{sec:6}

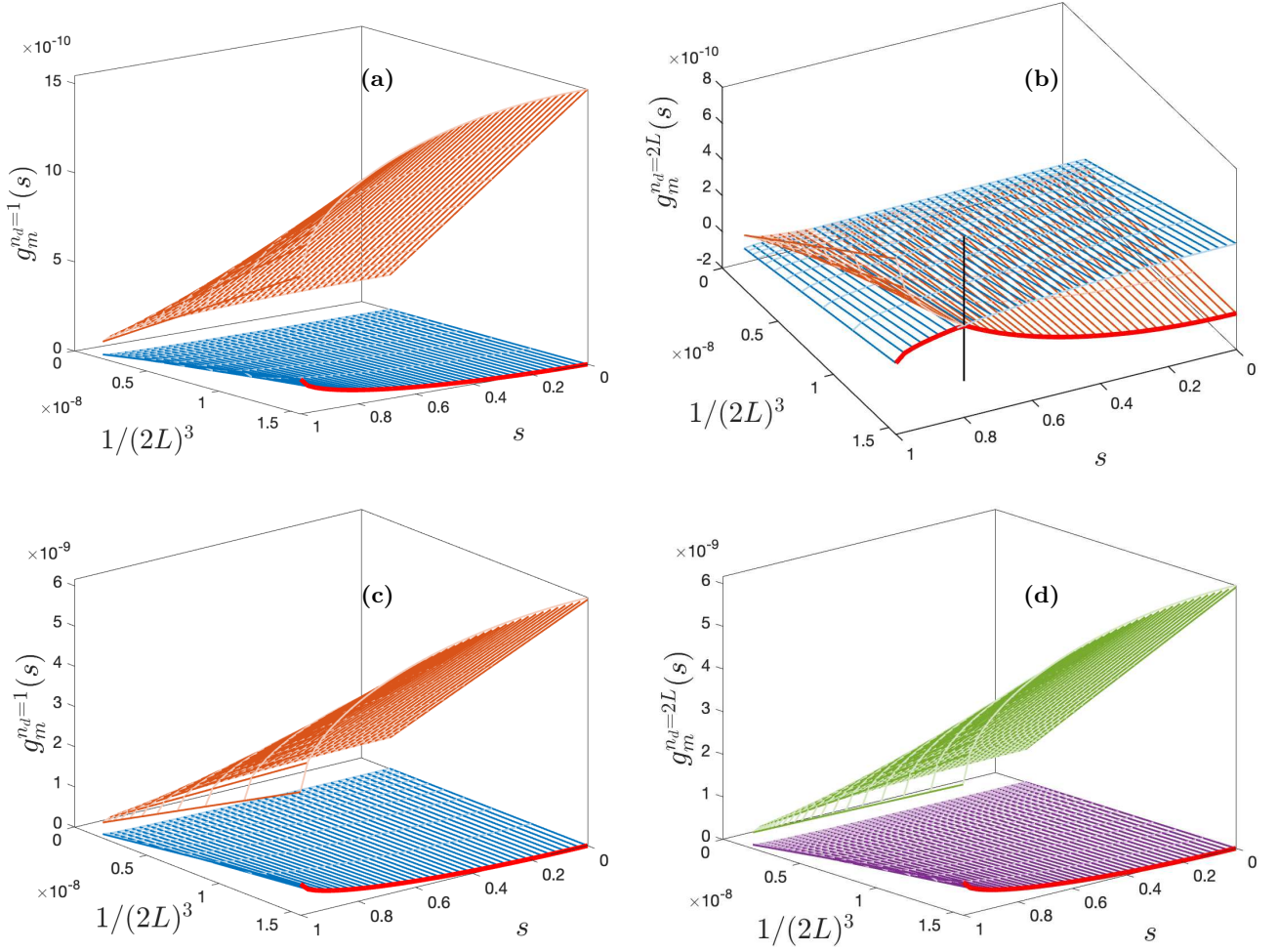
\begin{figure*}[t]
    \centering
    \begin{tikzpicture}
            \node[inner sep=0pt] (russell) at (-250pt, 0pt)
    {\includegraphics[width=0.5\textwidth]{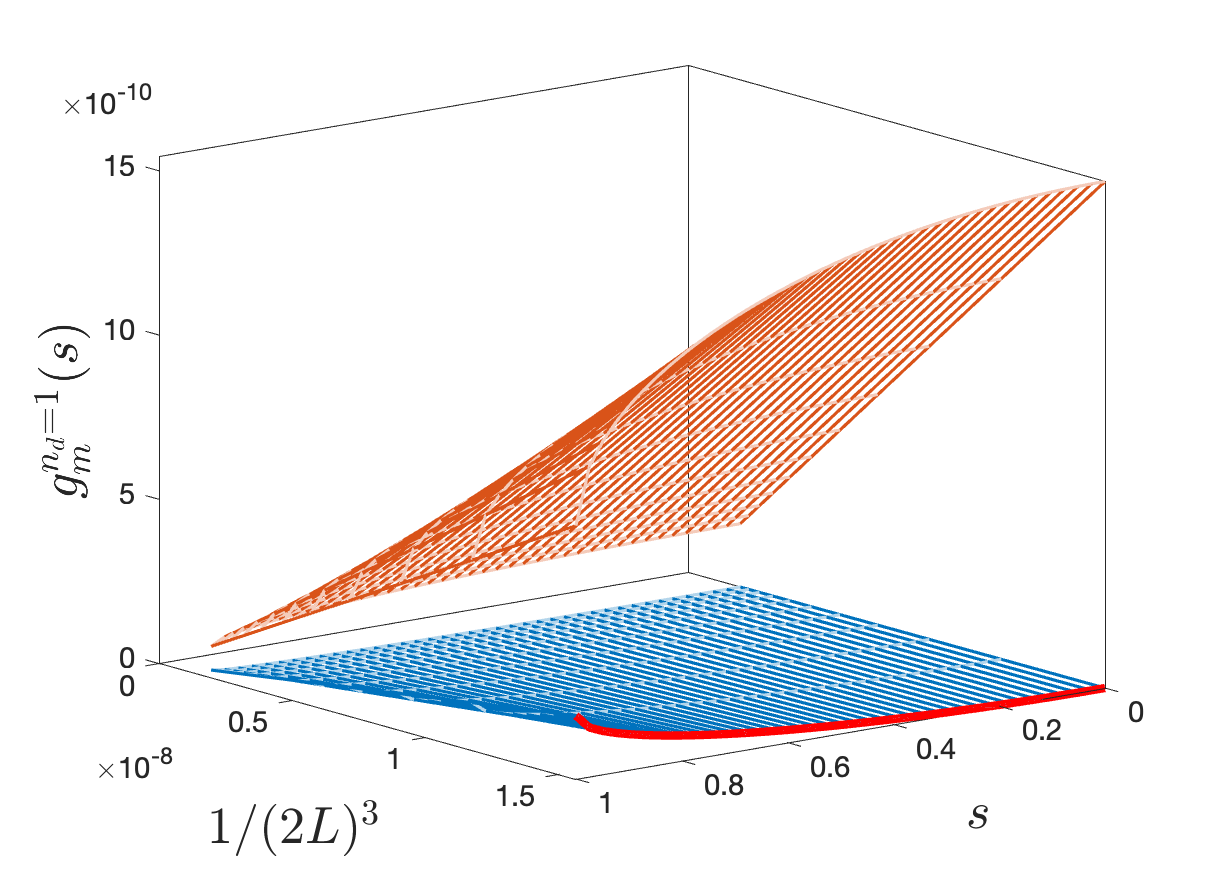}};
            \node[inner sep=0pt] (russell) at (0pt, 0pt)
    {\includegraphics[width=0.5\textwidth]{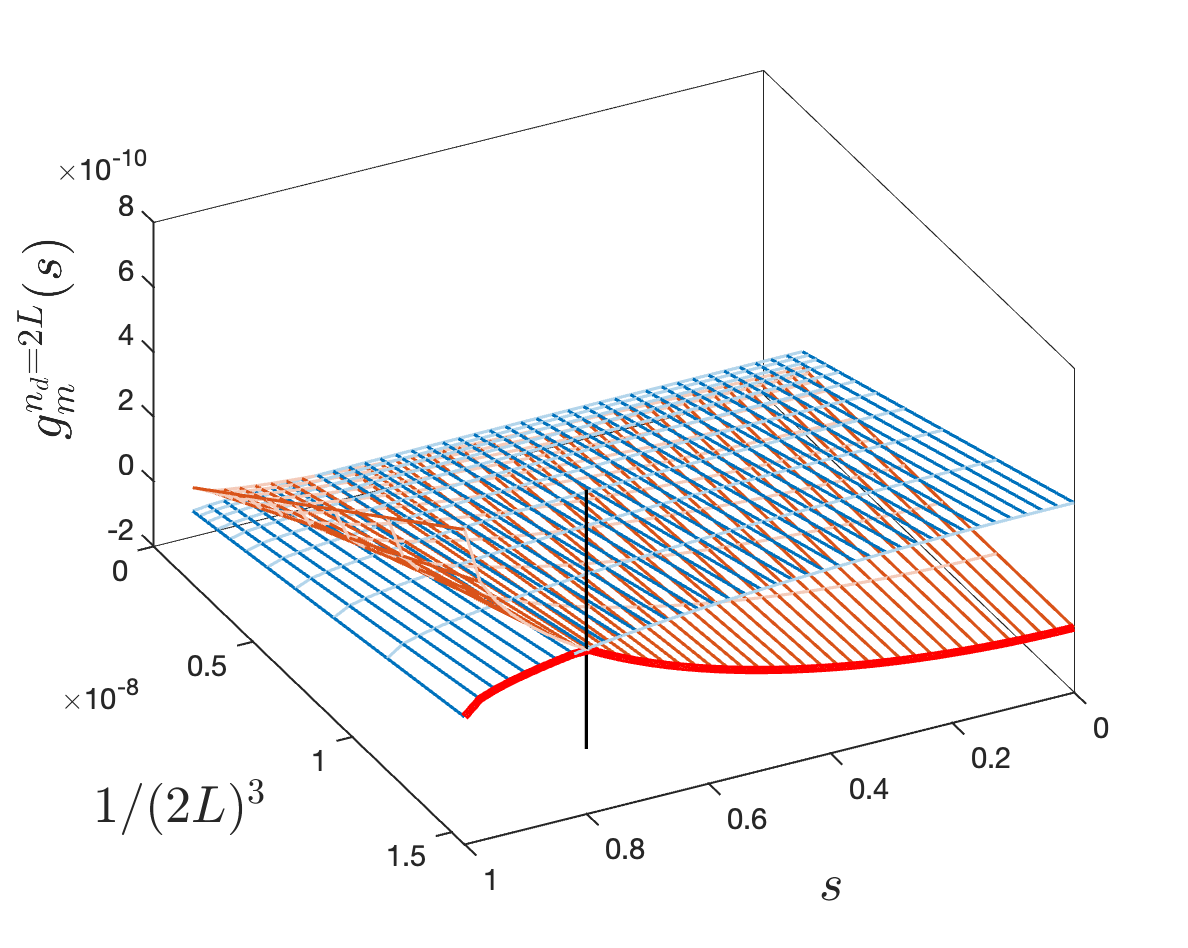}}; 
                         \node[inner sep=0pt] (russell) at (-250pt, -190pt)
    {\includegraphics[width=0.5\textwidth]{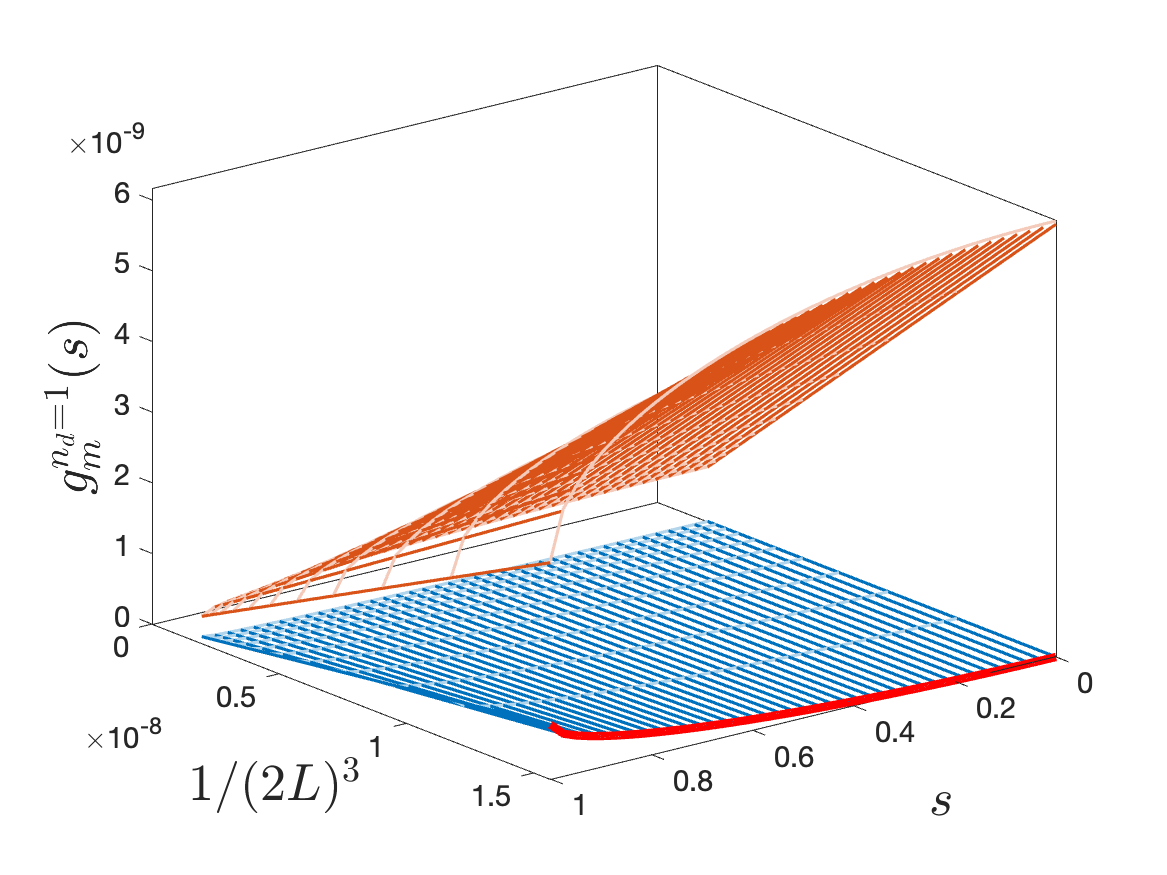}};
             \node[inner sep=0pt] (russell) at (0pt, -190pt)
    {\includegraphics[width=0.5\textwidth]{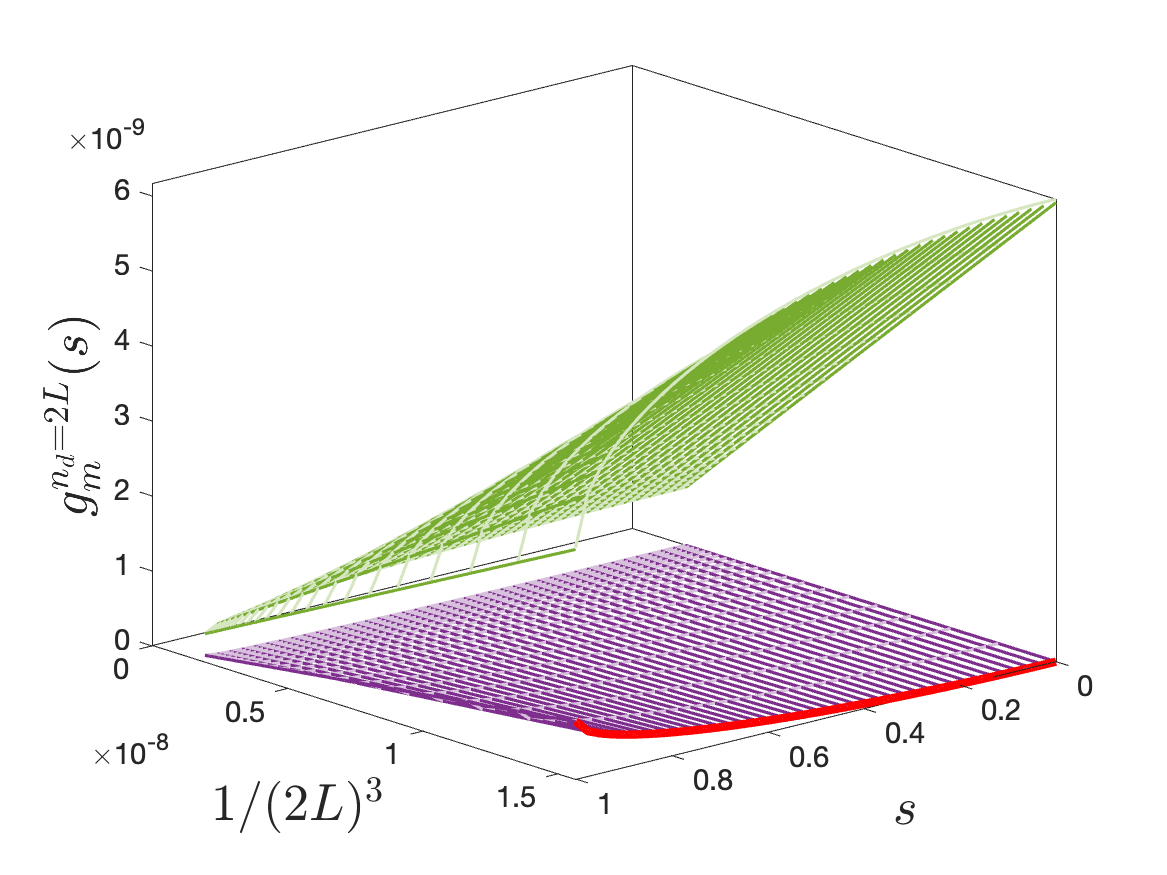}};    
    \node at (-8,2) {\textbf{(a)}};
    \node at (1,2) {\textbf{(b)}};
    \node at (-8,-5) {\textbf{(c)}};
    \node at (1,-5) {\textbf{(d)}};

    \end{tikzpicture}
    \caption{Numerical and analytical results for the relaxation rates of the two slowest-decaying modes as functions of inverse system size $L^{-3}$ and interface transmission $s$ in critical harmonic chain. Panels (a) and (b) correspond to free boundary conditions, while panels (c) and (d) correspond to fixed boundary conditions. For each boundary condition, the left plot corresponds to left boundary dissipation ($n_d=1$), while the right plot corresponds to right boundary dissipation ($n_d =2L$). In each plot, darker lines show the numerical results, while lighter lines show the analytical results. Red solid lines correspond to $c^{\rm global}_{\rm relax}$ (after an appropriate normalization) obtained using the previous definition Eq.\eqref{eq:old crelax formula} based on the Liouvillian gap.}
    \label{fig: Harmonic chain free left and right}
\end{figure*}

Following the discussion of the exact solution for the relaxation rates of the slowly decaying modes whose relaxation time scales as $\mathcal{O}(L^{3})$, we now comment on the Liouvillian gap, or equivalently the global relaxation rate, of the two chains connected by a conformal interface. In the previous section, we showed that the relaxation rates $g_m(t)$ of the slowly decaying modes exhibit a universal scaling i.e. $g_m(t) \propto 1/\left(m^{2}L^{3}\right)$ in both critical harmonic chain and critical free fermion chain with a conformal interface, for both fixed and free boundary conditions. Although the exact closed-form expressions depends on the dissipation site $n_d$, and the dissipation strength $\gamma$, the relaxation rates of the slowly decaying modes can still be obtained for any fixed choice of $n_d$. These modes are of particular interest because they control the long time relaxation dynamics and determine the Liouvillian gap of the full combined system. \\ 

In Fig.\ref{fig: Harmonic chain free left and right}(a) and Fig.\ref{fig: Harmonic chain free left and right}(b), we show the relaxation rates of the two slowest-relaxing modes for the harmonic chain with free boundary conditions. The Liouvillian gap of the full system is determined by the smaller of these two rates. The left plot, Fig.\ref{fig: Harmonic chain free left and right}(a), shows that when dissipation is applied to the left chain, the Liouvillian gap is determined throughout $t\in[0,1]$ by the mode that is localized in the non-dissipative chain at $t=0$. However, for dissipation applied to the right chain, as shown in Fig.~\ref{fig: Harmonic chain free left and right}(b), Liouvillian gap is determined by a competition between the two slowest-relaxing modes, which at $t=0$ are localized in the non-dissipative and dissipative chains, respectively. This shows that the Liouvillian gap based relaxation coefficient $c^{\rm global}_{\rm relax}$, defined as the ratio between the Liouvillian gaps at partial and full transmission, does not necessarily track the same mode as the interface transmission is varied. Instead, Liouvillian gaps at $0\le t<1$ and $t=1$ may be determined by different modes. This explains the behavior discussed in the Sec.\ref{sec:1}; see Fig.\ref{fig:relaxation coefficient}. The previous definition $c^{\rm global}_{\rm relax}$ follows the Liouvillian gap, but the mode determining the gap can change as the interface transmission is varied. As a result, it may compare the relaxation rates of different modes rather than tracking the same mode throughout.\\  

In Fig.\ref{fig: Harmonic chain free left and right}(c) and Fig.\ref{fig: Harmonic chain free left and right}(d), we similarly show the relaxation rates of the two slowest-relaxing modes for the harmonic chain with fixed boundary conditions. Here, Fig.\ref{fig: Harmonic chain free left and right}(c) and Fig.\ref{fig: Harmonic chain free left and right}(d) show the cases of left and right boundary dissipation, respectively. Again, the Liouvillian gap is obtained by minimizing over these relaxation rates. In this case, the gap is always determined by a mode that is initially localized in the non-dissipative chain at $t=0$, although the specific mode index can change as the interface transmission is varied. For example, in Fig.\ref{fig: Harmonic chain free left and right}(c), using
\begin{align}
    \theta^{\rm fixed}_m = \pi - \dfrac{m\pi}{2L+1}
\end{align}
the Liouvillian gap is determined by the mode with index $m=1$ throughout $0\le t<1$. In contrast, in Fig.\ref{fig: Harmonic chain free left and right}(d), the Liouvillian gap is determined by a different mode, with index $m=2L$ throughout $0\le t<1$. However, in the fully transmissive limit, $t=1$, the modes with indices $m=1$ and $m=2L$ become degenerate in their relaxation rates and give the same Liouvillian gap. Thus, $c^{\rm global}_{\rm relax}$ may involve the ratio of Liouvillian gaps determined by different modes at $0\le t <1$ and $t=1$, rather than tracking the relaxation rate of the same mode. This motivates defining a mode-resolved ratio that tracks the same mode as the interface transmission is varied. Specifically, we compare its relaxation rate at partial transmission $t$ with that in the fully transmissive limit, $t=1$.

In the next section, we build on the above discussion and classify the modes according to their behavior in the perfectly reflective limit, $t=0$, as illustrated in Fig.\ref{fig:physical picture}. The modes separate into two sets: those localized in the dissipative chain and those localized in the non-dissipative chain. This classification allows us to introduce a universal mode-resolved relaxation coefficient that characterizes how the conformal interface modifies the relaxation of individual modes.

\section{Mode-Resolved Relaxation coefficient $c_{\rm relax}$} 
\label{sec:7}
As shown in Fig.\ref{fig:physical picture}, in the perfectly reflective limit, $t=0$, the conformal interface separates the system into two halves: a dissipative chain and a non-dissipative chain, with the corresponding modes localized in each half. Figure \ref{fig: Mode resolved rapidities} then shows how the relaxation rates of these modes evolve as the interface transmission is varied.  Taking the fully transmissive case, $t=1$, as the reference point\footnote{We choose the fully transmissive limit, $t=1$, as the reference point because the full system relaxes to a time-independent steady state in this limit. In contrast, at $t=0$ the two chains are decoupled, and the non-dissipative chain does not relax, so the full system need not approach a time-independent steady state.}, decreasing the interface transmission has opposite effects on the two classes of modes. The relaxation rates of modes that are purely oscillating and localized in the non-dissipative chain at $t=0$ are suppressed, whereas those of modes localized in the dissipative chain at $t=0$ are enhanced.

This distinction motivates a universal mode-resolved relaxation coefficient. To characterize how the conformal interface modifies the relaxation of individual modes as the transmission is varied, we make two choices. First, we compare relaxation rates mode by mode, rather than only through the Liouvillian gap. Second, for each mode, we measure the change in its relaxation rate relative to the perfectly reflective limit, $t=0$, and normalize this change by its value in the fully transmissive limit, $t=1$. In other words, we define $c_{\rm relax}$ as a \textit{mode-resolved} relaxation coefficient that measures how the relaxation rate of a given mode evolves between the perfectly reflective and fully transmissive limits. This sets up naturally the definition
\begin{align}
\label{eq: relaxation coeff}
   \dfrac{c_{\rm relax}}{c} := \dfrac{g_{m}(t)-g_{m}(0)} {g_m(1)-g_{m}(0)}.
\end{align}
Here, $g_m(t)$ is the relaxation rate of the $m$-th mode, and $t$ denotes the interface transmission. The quantities $g_m(0)$ and $g_m(1)$ correspond to the relaxation rates in the perfectly reflective and fully transmissive limits, respectively. We distinguish two classes of modes according to their behavior at $t=0$: purely oscillating modes with $g_m(0)=0$, and relaxing modes with $g_m(0)>0$. We now examine the mode-resolved relaxation coefficient for these two classes in the weak-dissipation regime using Eq.\eqref{eq: relaxation coeff}. For a purely oscillating mode with $g_m(0)=0$, we find, for any mode index $m$ and any location $n_d$ of the local dissipation, and either choice of boundary conditions,
\begin{align}
\dfrac{c_{\rm relax}}{c} \approx 1-\sqrt{1-t^{2}}. 
\end{align}
For a relaxing mode with $g_m(0)>0$, the same expression is obtained independently of $\gamma$, $n_d$, $m$, and the choice of boundary conditions:
\begin{align}
\dfrac{c_{\rm relax}}{c} \approx \dfrac{1+\sqrt{1-t^{2}}-2}{1-2} \approx1-\sqrt{1-t^{2}}.
\end{align}
Thus, redefining $c_{\rm relax}$ as \textit{mode-resolved} relaxation coefficient relative to the perfectly reflective limit provides a unified characterization of how the conformal interface modifies the relaxation rates.
\begin{figure*}[t]
\centering
    \begin{tikzpicture}
            \node[inner sep=0pt] (russell) at (-250pt, -85pt)
    {\includegraphics[width=0.25\textwidth]{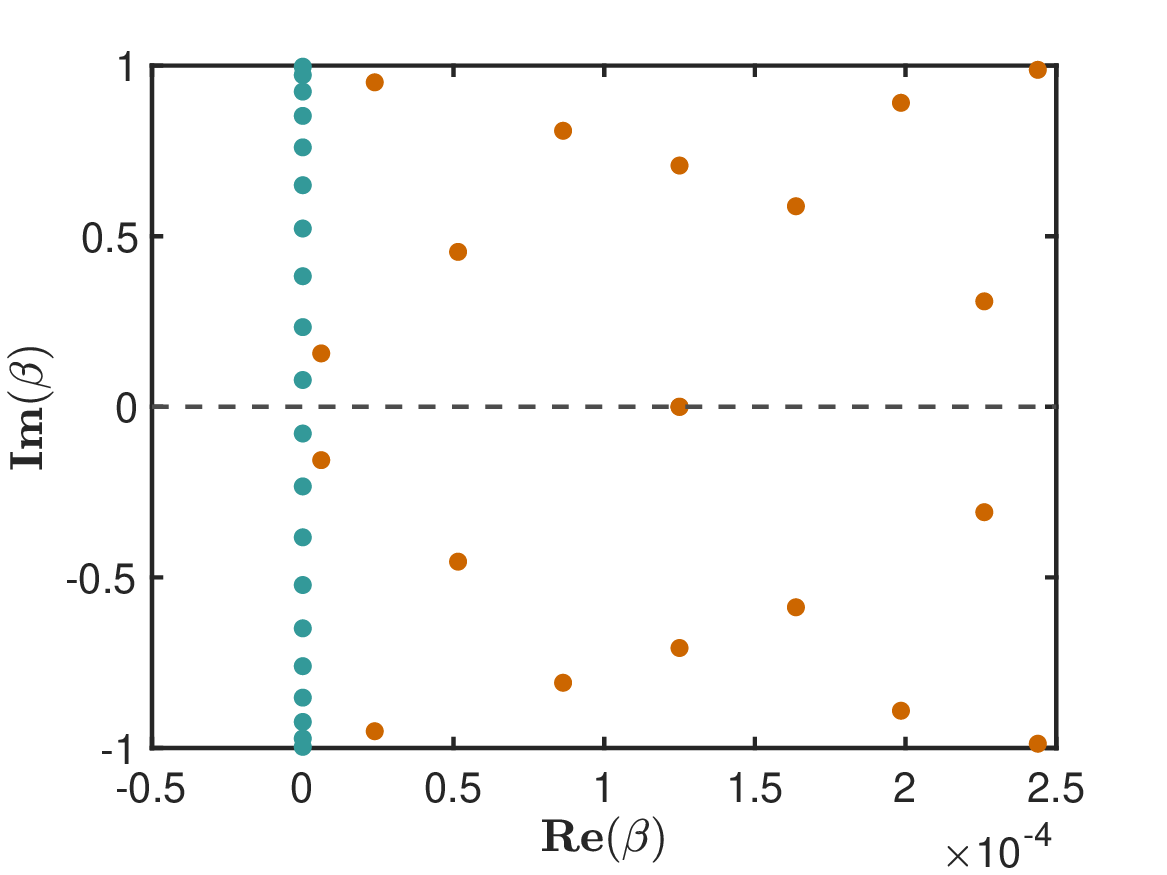}};
            \node[inner sep=0pt] (russell) at (-125pt, -85pt)
    {\includegraphics[width=0.25\textwidth]{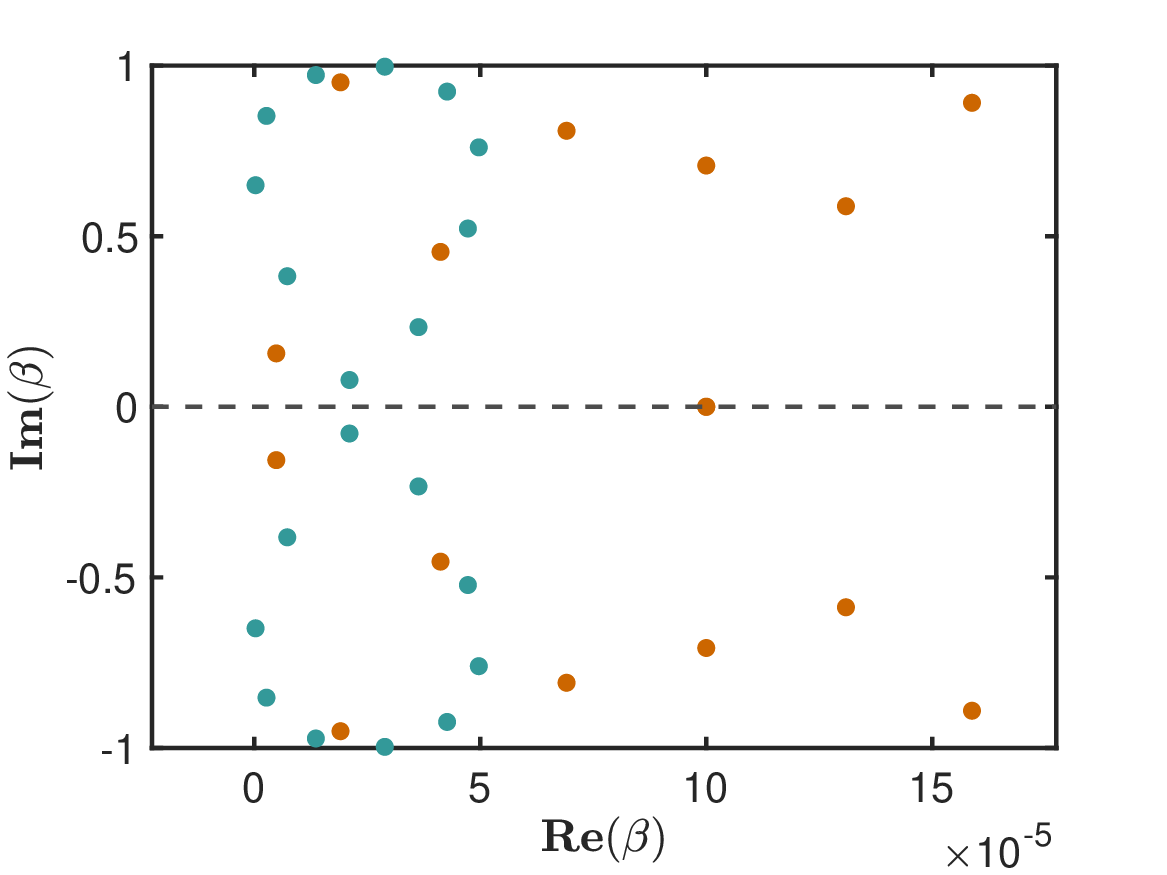}}; 
             \node[inner sep=0pt] (russell) at (0pt, -85pt)
    {\includegraphics[width=0.25\textwidth]{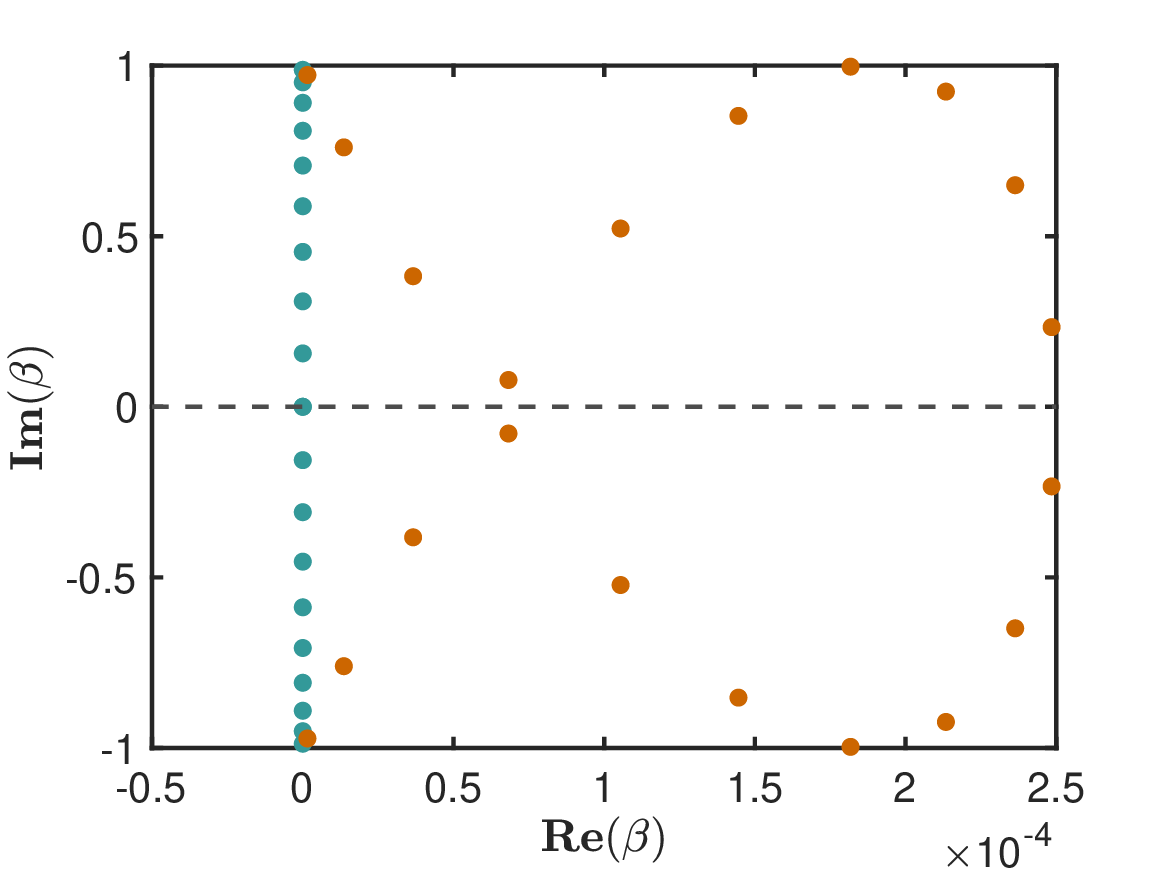}};
             \node[inner sep=0pt] (russell) at (125pt, -85pt)
    {\includegraphics[width=0.25\textwidth]{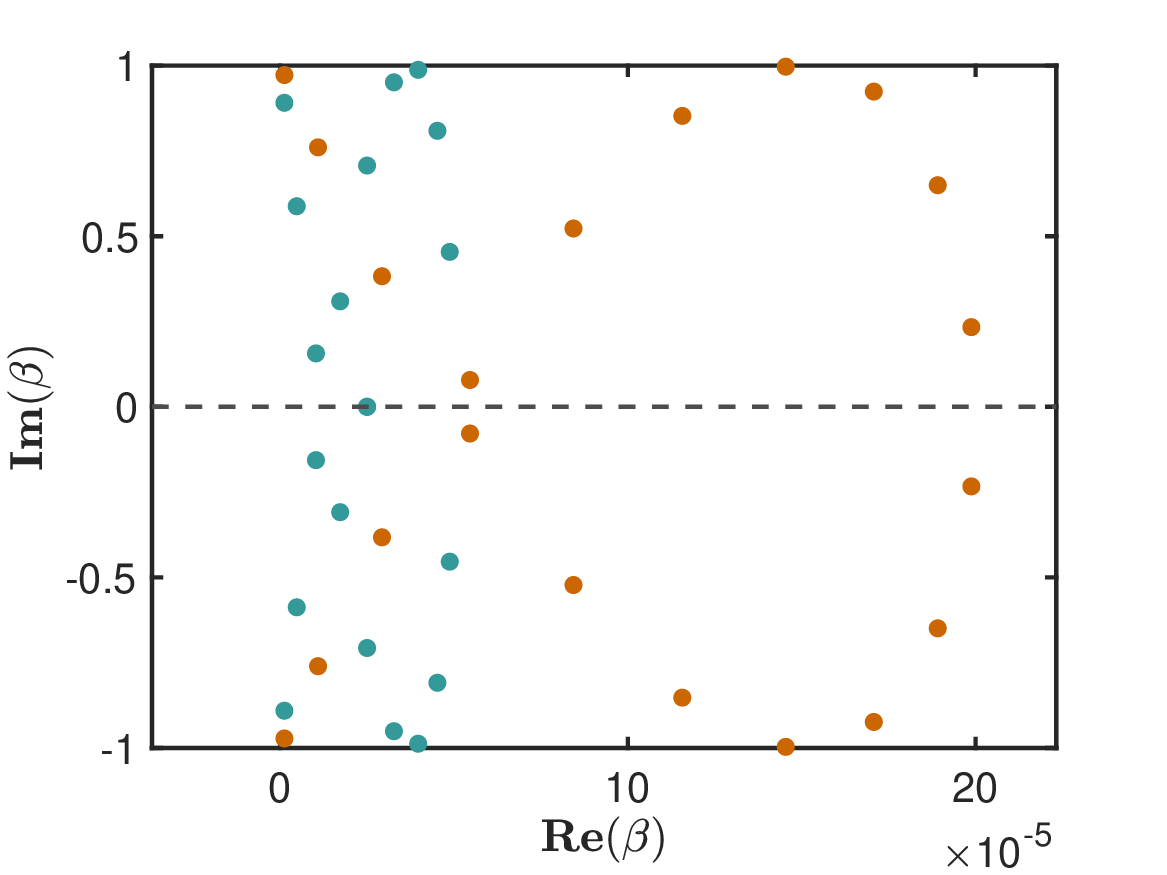}};

    \node at (-9,-2.5) {\bf (a)};
    \node at (-4.2,-2.5) {\bf (b)};
    \node at (0,-2.5) {\bf (c)};
    \node at (4.2,-2.5) {\bf (d)};
    \end{tikzpicture}
\caption{(a$-$d) Rapidity spectrum $\{\beta_{i}\}$ for harmonic chain  of length $N = 2L = 20$. The local dissipation is introduced on the left chain at $n_d =L-4$ in (a,\,b) and on the right chain at $n_d =L+4$ in (c,\,d) with interface parameter $s$ and dissipation strengths ($\gamma_l$ and $\gamma_g$, are set as:  $\gamma_l=0.015$ and $\gamma_g=0.01$. We consider free boundary condition in (a$-$d). In (a,\,\,c) we tune the conformal interface parameter $s=0$ such that green dots show purely oscillating modes that do not decay or localized in non-dissipative chain. In (b,\,\,d), we tune $s =0.8$ such that both modes have finite relaxation rate.}
\label{fig:Rapidity spectrum app}
\end{figure*}
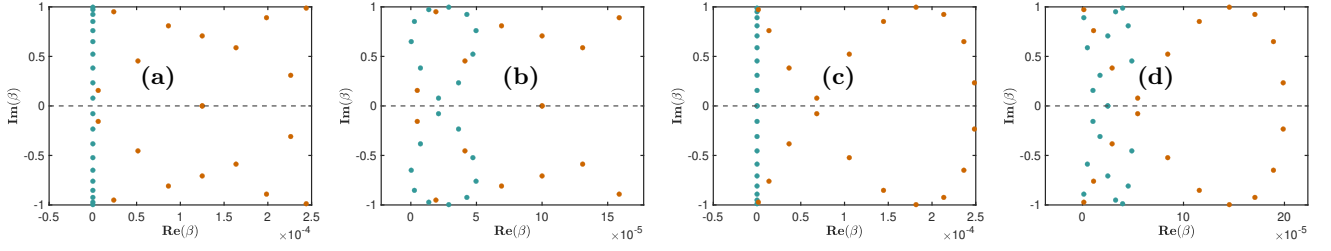

\section{Relaxation rate for arbitary choice of location $n_d$ of local dissipation}\label{sec:8}

In this section, we discuss the relaxation rates for an arbitrary choice of the local dissipation site $n_d$ in the weak-dissipation regime using perturbation theory. For concreteness, we present the analysis for the critical harmonic chain with a conformal interface characterized by the transmission parameter $s$, under free boundary conditions. The extension to the other setups discussed in the main text is straightforward. As shown in the main text, the general expression for the relaxation rate in the weak-dissipation regime is given by Eq.~\eqref{eq: g for free} as
\begin{align}
\label{eq: g for arbitary nd}
    g^{\rm free}_{m, n_d}(s) \approx \dfrac{\gamma}{2} \frac{\eta^{\rm free}_{n_d}(m)}{L}\cos^{2}\left(\frac{\left(2n_d-1\right)\theta_m}{2}\right),
\end{align}
with $\eta^{\rm free}_{n_d}(m)$ as follows:
\begin{align}
    \eta^{\rm free}_{n_d}(m) = \begin{cases}
         \alpha^{2}_{m}, \quad  n_d \in [1,L]\\
         \beta^{2}_{m}, \quad   n_d \in [L+1, 2L].
    \end{cases}
\end{align}

\begin{figure*}[t]
    \centering
    \begin{tikzpicture}
            \node[inner sep=0pt] (russell) at (-250pt, 0pt)
    {\includegraphics[width=0.5\textwidth]{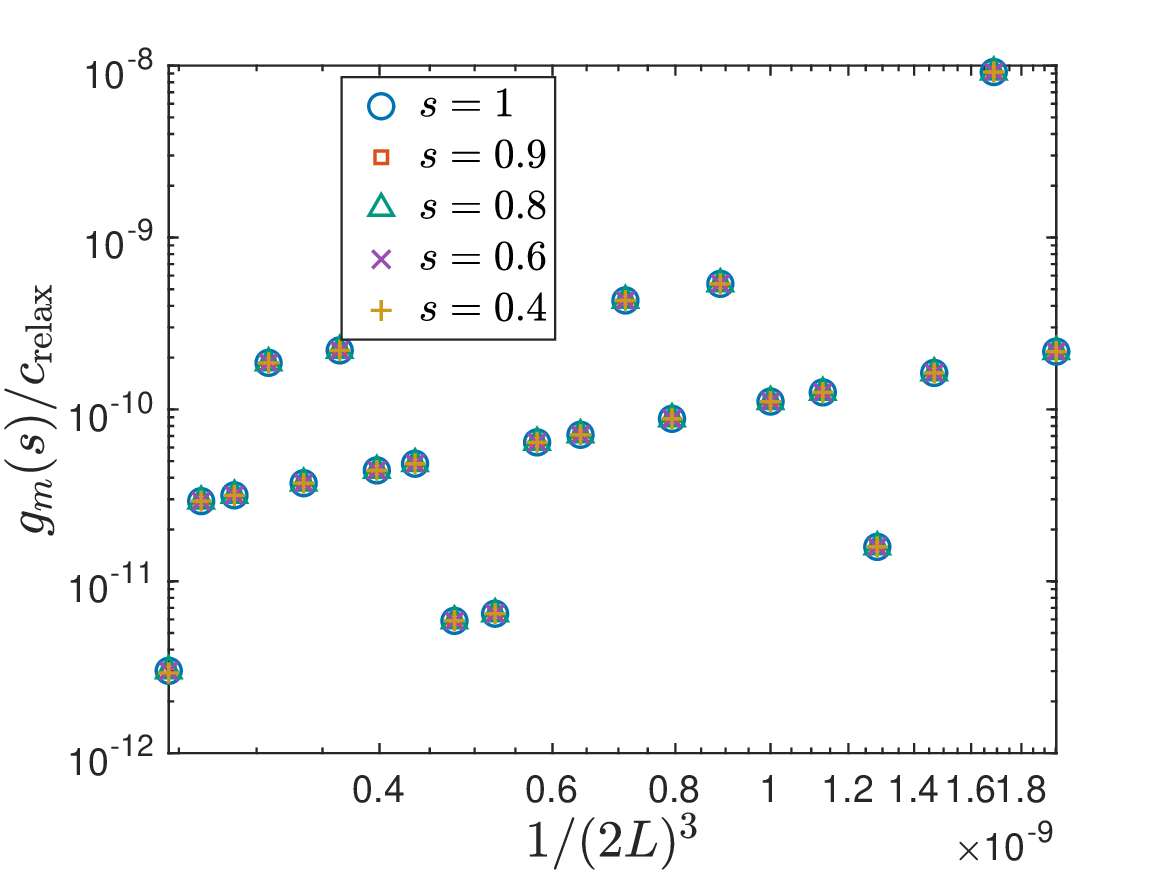}};
            \node[inner sep=0pt] (russell) at (0pt, 0pt)
    {\includegraphics[width=0.5\textwidth]{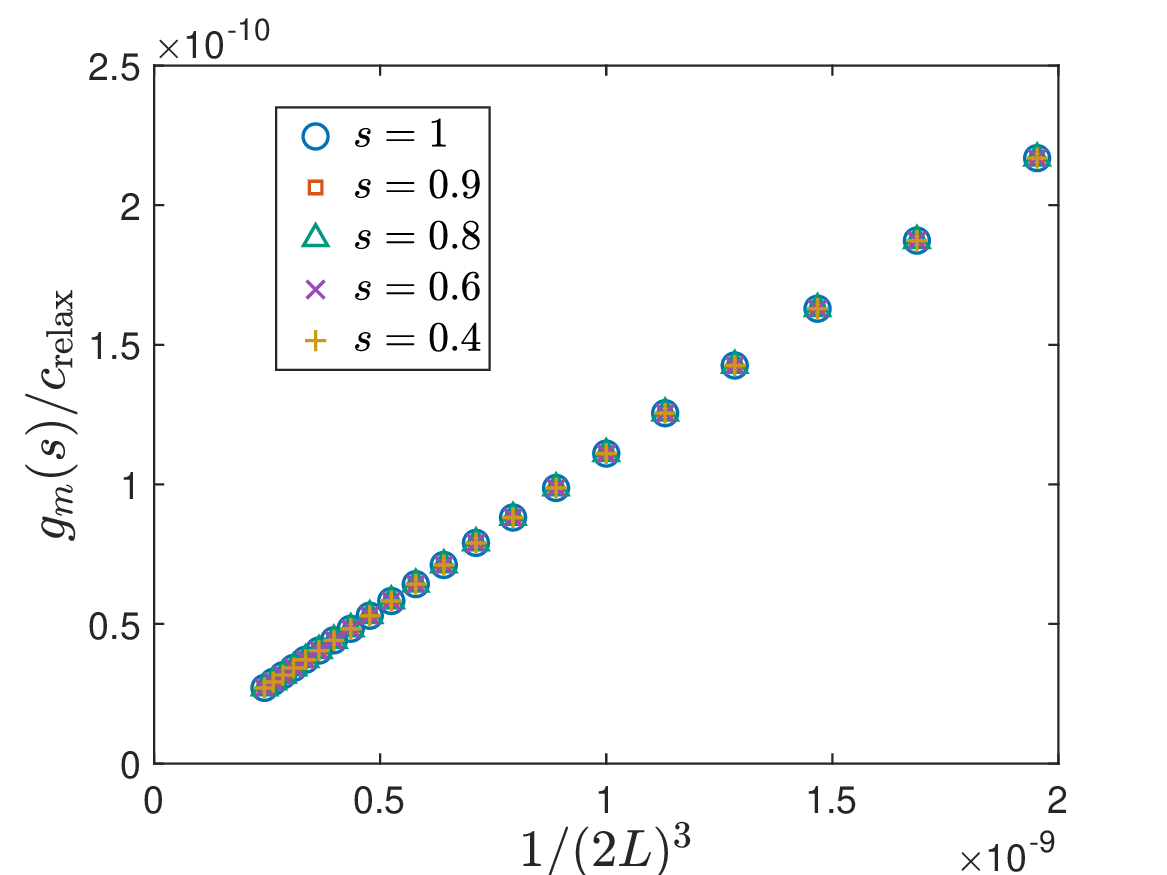}}; 
                         \node[inner sep=0pt] (russell) at (-250pt, -190pt)
    {\includegraphics[width=0.5\textwidth]{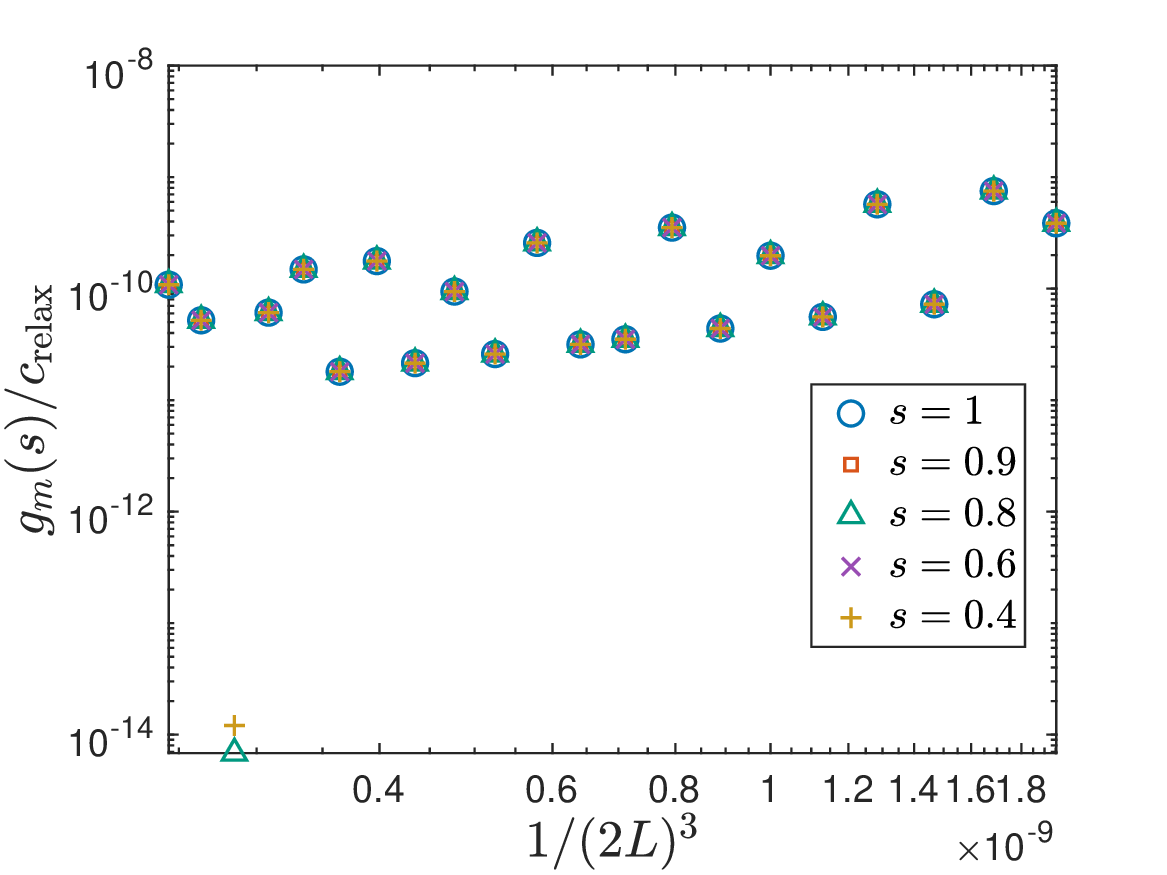}};
             \node[inner sep=0pt] (russell) at (0pt, -190pt)
    {\includegraphics[width=0.5\textwidth]{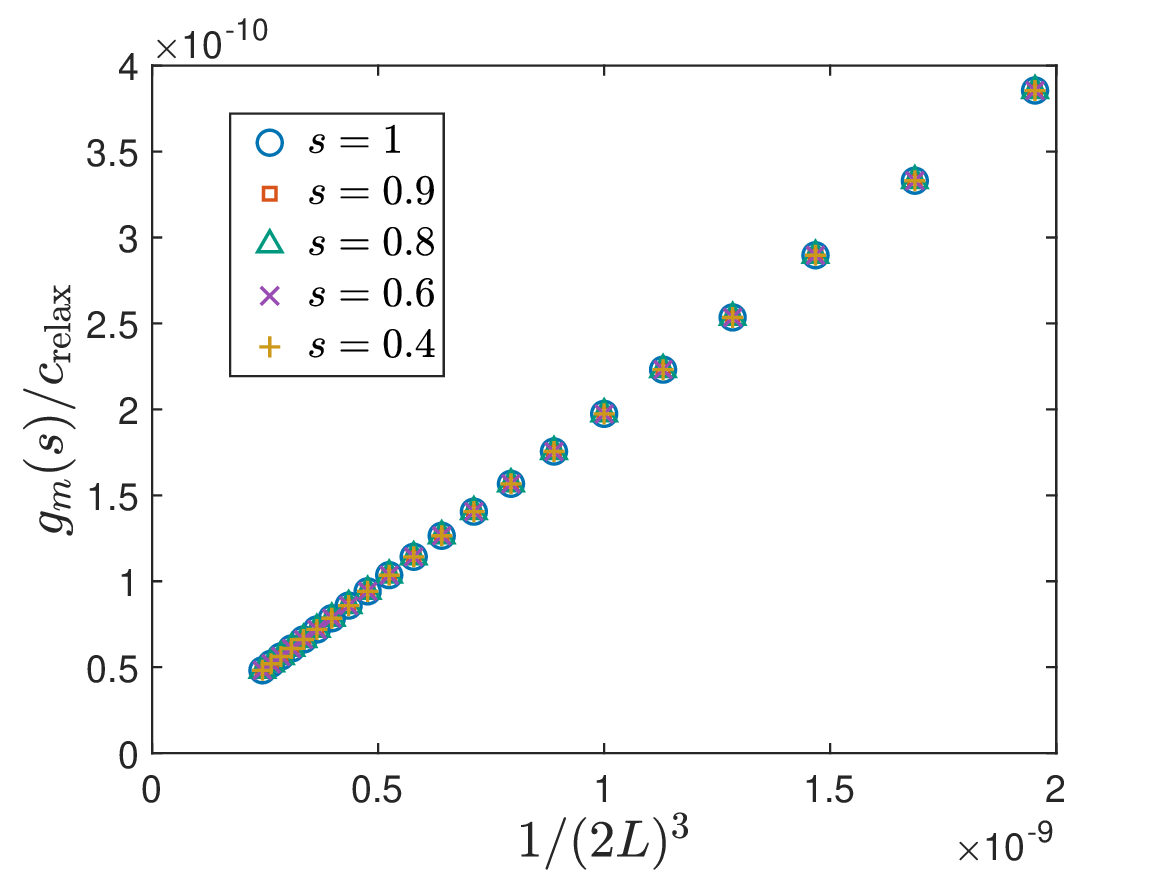}};    
    \node at (-8,2) {\textbf{(a)}};
    \node at (1,2) {\textbf{(b)}};
    \node at (-8,-4.6) {\textbf{(c)}};
    \node at (1,-5) {\textbf{(d)}};

    \end{tikzpicture}
    \caption{Numerical results for the relaxation rates rescaled by $c_{\rm relax}$ of modes initially localized in non-dissipative chain as functions of inverse system size $L^{-3}$ and interface transmission $s$ in critical harmonic chain with free boundary conditions. We consider four choices of the local dissipation site $n_d$
: two in the left chain, shown in panels (a) and (b), and two in the right chain, shown in panels (c) and (d). Specifically, panel (a) corresponds to $n_d = L-10$,  panel (b) to $n_d = L/2$, panel (c) to $n_d = L+10$,  and panel (d) to $n_d = L+L/2$. For each choice of the dissipation site $n_d$, the numerical relaxation rates collapse onto the same curve after rescaling by $c_{\rm relax}$.}
    \label{fig: arbitary n_d relaxation rate}
\end{figure*}
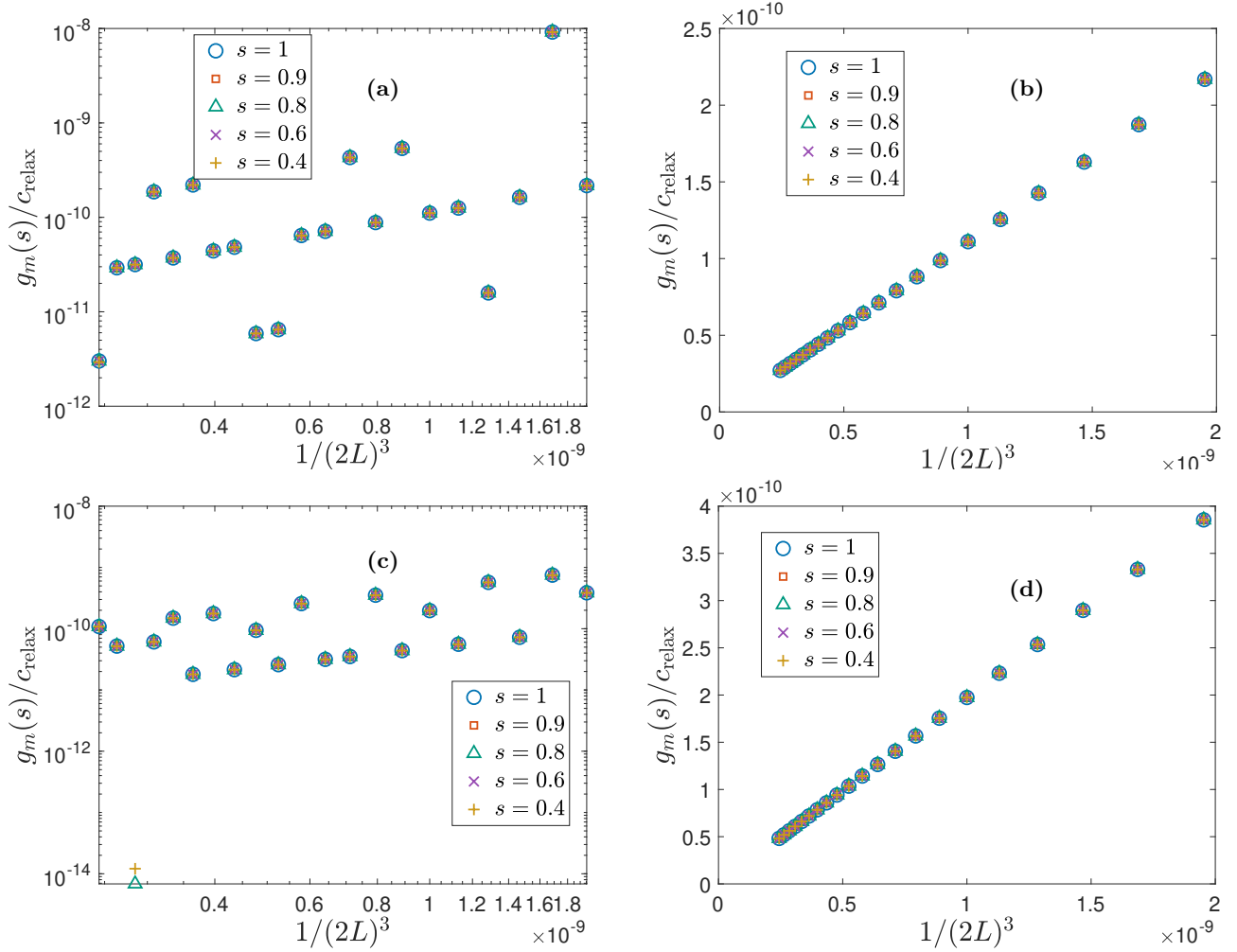

First, at $s=0$, we separate the modes into two sets : purely oscillating modes localized in the non-dissipative chain and relaxing modes localized in the dissipative chain. When local dissipation is applied to the left chain, the purely oscillating modes correspond to odd mode indices, $m=1,3,5,\dots 2L-1$, while the relaxing modes correspond to even mode indices, $m=0,2,4,\dots 2L-2$. This separation is also reflected in the rapidity spectrum shown in Fig.\ref{fig:Rapidity spectrum app}. Using Eq.\eqref{eq: relaxation coeff} and Eq.\eqref{eq: g for arbitary nd}, the mode-resolved relaxation coefficient for a mode with index $m$ is then given by
\begin{align}
\label{eq: crelax result}
    \dfrac{c_{\rm relax}}{c} \approx 1-\sqrt{1-s^{2}}. 
\end{align}
Similarly, when local dissipation is applied to the right chain, the parity assignment of the modes is reversed. The purely oscillating modes correspond to even mode indices, $m=0, 2,4,\dots 2L-2$, while the relaxing modes correspond to odd mode indices. After this relabeling, the mode-resolved relaxation coefficient reduces to the same expression as in the left chain dissipation case, independent of $n_d$. This agreement is also confirmed by the numerical results in Fig.\ref{fig: arbitary n_d relaxation rate}, where the relaxation rates of different modes are plotted as functions of system size after rescaling by $c_{\rm relax}$. For different values of the interface parameter $s$, the relaxation rates collapse onto the same curve after rescaling by $c_{\rm relax}$. We consider four choices of the local dissipation site $n_d$, with two sites in the left chain and two in the right chain. Thus, for modes that are purely oscillating at $s=0$, the relaxation rate is given by
\begin{align}
        g_{m, n_d}(s) \approx \frac{\gamma \left(1-\sqrt{1-s^2}\right)}{2L} \left\{\cos^{2}\left(\frac{\left(2p\pm1\right)m\pi}{4L}\right)\right\},
\end{align}
where we choose $n_d = L\mp p$, with the minus and plus signs corresponding to dissipation sites in the left and right chains, respectively.

\section{Discussion and conclusion}
\label{sec:9}

In this work, the central result is that the suppression of relaxation by a conformal interface admits a universal characterization once the dissipative dynamics are resolved mode by mode. The non-universality of the previously introduced $c^{\rm global}_{\rm relax}$ in Ref.\cite{Barad_2025} originates from the fact that the Liouvillian gap does not necessarily track the same mode as the interface transmission is varied: for certain boundary conditions, different modes can successively determine the slowest decay rate. By instead continuously following the relaxation rate of a fixed Liouvillian rapidity mode, this mode-switching ambiguity is avoided. This mode-resolved viewpoint also leads to a simple physical picture: Since the dissipation is local, the relaxation rate of a given mode is controlled by its weight in the dissipative region. The conformal interface modifies the spatial profile of the mode and therefore controls how strongly it overlaps with the local dissipation. In particular, decreasing the interface transmission suppresses the penetration of modes initially localized on the nondissipative side into the dissipative region, thereby reducing their relaxation rates.

Quantitatively, for both the critical harmonic chain and the critical free-fermion chain, and for the different boundary conditions considered in this work, the mode-resolved relaxation coefficient takes the universal form
\begin{align}
\frac{c_{\rm relax}}{c}\approx 1-\sqrt{1-t^2}
\end{align}
in the weak-dissipation regime. Remarkably, this result is independent of the boundary conditions, the dissipation strength, the location of the local dissipation, and the particular mode being tracked. The same expression applies to both modes that are purely oscillatory at $t=0$ and modes that already have a finite relaxation rate in the perfectly reflecting limit. For boundary dissipation, this universal dependence on the interface transmission persists even at finite dissipation strength. Moreover, for both solvable models, the universal quantities continue to satisfy the hierarchy proposed in Ref.\cite{Barad_2025}
\begin{align}
0\leq c_{\rm relax}\leq c_{\rm LR}\leq c_{\rm eff}\leq c,
\end{align}
placing the suppression of relaxation alongside energy and information transmission as a universal property controlled by the conformal interface.

\bigskip

There are several interesting future directions for extending the study of conformal interfaces in open quantum critical systems.

Frist, the mode-resolved framework developed here also suggests a natural generalization beyond quadratic Lindblad dynamics. In the present work, the linear jump operators and quadratic Hamiltonians allow the relaxation modes to be described in terms of effective single-particle rapidities. For more general dissipative processes, such as local dephasing in a free fermion chain\cite{2025_Alba_dephasing}, or for interacting CFTs \cite{2026_Tang_Barad_Wen}, this single-particle description is no longer available in general. Nevertheless, the same idea can be extended to a many-body mode-resolved picture, in which one continuously tracks the relaxation of the same many-body Liouvillian mode as the interface transmission is varied. This raises the question of whether a universal $c_{\rm relax}$ can be defined beyond the quasi-free setting.

Another important direction is to develop a continuum field-theory description of the relaxation dynamics studied here \cite{2016_Sieberer,2023_Kamenev,2025_RMP}. In the present lattice models, the suppression of relaxation can be understood in terms of how the conformal interface modifies the spatial weight of a mode in the dissipative region. In a CFT description, 
one may introduce dissipation through jump operators constructed by smearing local CFT operators over a finite spatial region.
In the weak-dissipation regime, the relaxation rate of a given CFT excitation is controlled by matrix elements of the smeared jump operators between many-body CFT states in the presence of the conformal interface.
This suggests that $c_{\rm relax}$ may admit a direct formulation in terms of universal interface-CFT correlation functions, providing a continuum analog of the lattice mode-weight picture developed in this work.

\section{Acknowledgments}

We thank Andreas Karch, Ainesh Sanyal, and Mianqi Wang for helpful discussions. They pointed out to us that $c^{\rm global}_{\rm relax}$ is not universal for certain choices of boundary conditions in lattice systems, which motivated us to identify the universal quantity $c_{\rm relax}$ presented in this work.

\appendix
\section{Eigenvalues and Eigenfunctions}\label{section:Eigenvalues and Eigenfunctions}

In this appendix, we present the single-particle eigenspectrum of the critical harmonic chain and the critical free fermion chain. For both models, we consider the two boundary conditions discussed in the main text: free boundary conditions at both ends and fixed boundary conditions at both ends.
\subsection{Harmonic chain}
In this part of the appendix, we consider the harmonic chain Hamiltonian defined in Eq.\eqref{eq:initial Hamiltonian}, with the coupling constants and mass terms specified in Eq.\eqref{eq:conformal parameters}. Following the coordinate rescaling introduced in Ref.\cite{Eisler_2012},
\begin{equation}
u_n = \sqrt{m_n}x_n, \quad\quad\pi_n = \frac{p_n}{\sqrt{m_n}},
\end{equation}
the Hamiltonian in Eq.\eqref{eq:initial Hamiltonian} becomes 
\begin{align}
\label{eq: initial Hamiltonian  after rescaling}
 H = \dfrac{1}{2}\sum_{n}^{2L} \left(-\dfrac{\partial^{2}}{\partial u^{2}_{n}} + \Omega^{2}_{0}u^{2}_{n}\right) + \dfrac{1}{2} \sum_{n\neq L}\left(u_{n}-u_{n+1}\right)^{2} \\ \nonumber
    + \dfrac{1}{2}\left( \dfrac{K_0}{K_1}u^{2}_{L} + \dfrac{K_0}{K_2}u^{2}_{L+1} - 2\dfrac{K_0}{\sqrt{K_1K_2}}u_{L}u_{L+1}\right).
\end{align}
Equivalently, this can be rewritten as
\begin{align}
    H = \dfrac{1}{2}\sum_{i=1}^{2L}\pi^{2}_{i} + \dfrac{1}{2}\sum_{i,j=1}^{2L}u_i\left(\mathbf{K}(s)\right)_{ij}u_j,
\end{align}
where $\mathbf{K}(s)$ denotes the stiffness matrix defined in Eq.\eqref{eq:K matrix}. The corresponding normal-mode frequencies and eigenfunctions are obtained from the eigenvalue equation 
\begin{align}
    \mathbf{K}(s)|\Phi_{m}\rangle = \Omega^{2}_m|\Phi_{m}\rangle
\end{align}
where $\langle j|\Phi_{m}\rangle = \phi_m(j)$ denotes the $j$-th component of the eigenvector $|\Phi_{m}\rangle$ corresponding to the normal-mode frequency $\Omega_m$.

\subsubsection{Free boundary condition}
For free boundary conditions at both ends of the harmonic chain, $V_{0}=0$ in Eq.\eqref{eq:K matrix}. In the homogeneous limit, $s=1$, 
\begin{align}
    \mathbf{K}(1)|\Phi^{\rm free}_{m}\rangle = \left(\Omega^{\rm free}_m\right)^{2}|\Phi^{\rm free}_{m}\rangle.
\end{align}
The corresponding normal-mode frequencies are  
\begin{equation}
\Omega^{\rm free}_m = \sqrt{\Omega_0^2 + 2\left(1-\cos\frac{m\pi}{2L}\right)},\label{eq:eigenfrenquency Omega_m}
\end{equation}
with quantized wave number
\begin{align}
  \theta^{\rm free}_m = \frac{m\pi}{2L},
\end{align}
where $m=0,1,\dots,2L-1$ is the mode index. The eigenfunctions in the homogeneous case are
\begin{equation}
\phi^{\rm free}_m(j) = \sqrt{\frac{1}{L}}\cos\left(\frac{\left(j-\frac{1}{2}\right)m\pi}{2L}\right),\quad\phi^{\rm free}_0(j)=\frac{1}{\sqrt{2L}},\label{eq:homogeneous eigenfunctions free}
\end{equation}
with site index $j = 1, 2, \dots, 2L$. In the inhomogeneous case, 
\begin{align}
    \mathbf{K}(s)|\tilde{\Phi}^{\rm free}_{m}\rangle = \left(\Omega^{\rm free}_m\right)^{2}|\tilde{\Phi}^{\rm free}_{m}\rangle,
\end{align}
here $|\tilde{\Phi}^{\rm free}_{m}\rangle$ denote the modified eigenvectors, which leave the eigenvalues unchanged \cite{Eisler_2012}. The corresponding eigenfunctions take the following form:
\begin{equation}
\tilde{\phi}^{\rm free}_m(j) = \begin{cases}\alpha_m\phi^{\rm free}_m(j),\quad 1 \le j \le L,\\\beta_m\phi^{\rm free}_m(j),\quad L+1 \le j \le 2L,\end{cases}\label{eq:modified single-particle eigenfunction free}
\end{equation}
with coefficients $\alpha_m$ and $\beta_m$ characterizing the two sides of the interface, respectively, and satisfying
\begin{equation}
\alpha^2_m = 1 + (-1)^m\sqrt{1-s^2},\quad \beta^2_m = 1 - (-1)^m\sqrt{1-s^2}.\label{eq:alpha and beta in terms of s}
\end{equation} 
At the critical point $\Omega_{0} = 0$, the spectrum contains a zero mode in the case of free boundary conditions. 

\subsubsection{Fixed boundary condition}
 For fixed boundary conditions at both ends of the harmonic chain, $V_{0}=1$ in Eq.\eqref{eq:K matrix}. In the homogeneous limit, $s=1$, the eigenvalue problem for $\mathbf{K}(1)$ is 
 \begin{align}
    \mathbf{K}(1)|\Phi^{\rm fixed}_{m}\rangle = \left(\Omega^{\rm fixed}_m\right)^{2}|\Phi^{\rm fixed}_{m}\rangle.
\end{align}
The corresponding normal-mode frequencies are 
\begin{equation}
\Omega^{\rm fixed}_m = \sqrt{\Omega_0^2 + 2\left(1-\cos\frac{m\pi}{2L+1}\right)},\label{eq:eigenfrenquency Omega_m fixed}
\end{equation}
with quantized wave number
\begin{align}
  q^{\rm fixed}_m = \frac{m\pi}{2L+1}
\end{align}
where $m=1,2,\dots,2L$ is the mode index. The eigenfunctions in the homogeneous case are
\begin{equation}
\phi^{\rm fixed}_m(j) = \sqrt{\frac{2}{2L+1}}\sin\left(\frac{\pi mj}{2L+1}\right).
\label{eq:homogeneous eigenfunctions fixed}
\end{equation}
In the inhomogeneous case, eigenfunctions take a form analogous to the free-boundary-condition case, with the coefficients on the two sides of the interface interchanged i.e.,
\begin{equation}
\tilde{\phi}^{\rm fixed}_m(j) = \begin{cases}\beta_m\phi^{\rm fixed}_m(j),\quad 1 \le j \le L,\\\alpha_m\phi^{\rm fixed}_m(j),\quad L+1 \le j \le 2L.\end{cases}\label{eq:modified single-particle eigenfunction fixed}
\end{equation}

\subsection{Free fermionic chain}
\label{Appendix:fermion_BC}
\subsubsection{Free boundary condition}
Next, we consider the critical free fermion chain with free boundary conditions, corresponding to ordinary open boundary conditions with no additional boundary potentials at either end, $V_{0}=0$ in Eq.\eqref{eq: boundary free fermion}. The Hamiltonian with a conformal interface is
\be
\widetilde{\mathbf{H}}^{\rm free}(\lambda) = \sum_{i, j=1}^{2L} \widetilde{H}^{\rm free}_{i,j}(\lambda) c_i^\dagger c_j,
\ee
where the nonzero elements of $\widetilde{H}^{\rm free}(\lambda)$ are given in Eq.\eqref{eq:H2}. The corresponding single-particle eigenvalue problem is
\be
\widetilde{H}^{\rm free}(\lambda) \, |\widetilde{\phi}^{\rm free}_k\rangle = E^{\rm free}_k \, |\widetilde{\phi}^{\rm free}_k\rangle.
\ee
The eigenvalues are independent of the interface transmission and coincide with those of the homogeneous limit, $\lambda=1$:
\be
E^{\rm free}_k = - \cos \theta^{\rm free}_k, \quad \theta^{\rm free}_k = \frac{k \pi}{2L+1} , \quad k=1,\cdots,2L,
\ee
The interface modifies only the eigenfunctions, which take the form
\begin{equation}
	\widetilde{\phi}^{\rm \,free}_k(j) = \begin{cases} 
		\alpha_k \phi^{{\rm free}}_k(j), 
		& \quad 1 \le j \le L, \\ 
		\beta_k \phi^{{\rm free}}_k(j), 
		& \quad L+1 \le j \le 2L. \\ 
	\end{cases}
\end{equation}
Here the coefficients satisfies
\begin{equation}
	\alpha_k^2 = 1 - (-1)^k \sqrt{1-\lambda^2} , \quad \beta_k^2 = 1 + (-1)^k \sqrt{1-\lambda^2},
\end{equation}
and $\phi^{\rm free}_k(m)$ denotes the eigenfunction of the corresponding homogeneous open chain:
\begin{equation}
\phi^{{\rm free}}_k(j) = \sqrt{\frac{2}{2L+1}} \sin \left(j\theta_k^{\rm free}\right). 
\end{equation}
We introduce two virtual sites at $j=0$ and $j=2L+1$, located just outside the left and right boundaries of the chain. The single-particle wavefunction vanishes at these sites,
\begin{align}
   \phi^{\rm free}(0)=  \phi^{\rm free}(2L+1) =0, 
\end{align}
 corresponding to Dirichlet-type boundary conditions for the single-particle wavefunction. 

\subsubsection{Fixed boundary condition}

For fixed boundary conditions, additional boundary potentials are applied at the two ends of the open chain. In particular, we consider
\be
\widetilde{\mathbf{H}}^{\rm fixed}(V_{0}, \lambda) = \widetilde{\mathbf{H}}^{\rm free}(\lambda) - \dfrac{V_{0}}{2} \left( c_1^\dagger c_1 + c_{2L}^\dagger c_{2L} \right),
\ee
which can be solved when $V_{0}=\pm1$. For both choices, the single-particle eigenvalues are
\be
E^{{\rm fixed}}_k = - \cos \theta^{\rm fixed}_k\quad\text{and}\quad\theta^{{\rm fixed}}_k = \frac{k \pi}{2L}
\ee
for both $V_{0}=+1$ and $V_{0}=-1$. The corresponding eigenfunctions in the presence of the conformal interface take the form
\begin{equation}
	\widetilde{\phi}^{\rm \,fixed{,\pm}}_k(j) = \begin{cases} 
		\beta_k \phi^{\rm \,fixed{,\pm}}_k(j), 
		& \quad 1 \le j \le L, \\ 
		\alpha_k \phi^{\rm \,fixed{,\pm}}_k(j), 
		& \quad L+1 \le j \le 2L, \\ 
	\end{cases}
\end{equation}
where the single-particle wavefunction for $V_{0}=+1$ is given by
\begin{align}
\label{eq: phi fix plus}
\phi^{\rm \,fixed{,+}}_k(j) = \begin{cases}\displaystyle
    \sqrt{\frac{2}{2L}} \sin\left[ \frac{\pi k}{2L} (j-\frac12) \right], & k\in[1,2L-1], \\\\\displaystyle
    \sqrt{\frac{1}{2L}} \sin\left[ \pi (j-\frac12) \right], & k=2L,
\end{cases}
\end{align}
and for $V_{0}=-1$:
\be
\phi^{\rm \,fixed{,-}}_k(j) = \begin{cases}\displaystyle
    \sqrt{\frac{2}{2L}} \cos\left[ \frac{\pi k}{2L} (j-\frac12) \right], & k\in[1,2L-1], \\\\\displaystyle
    \sqrt{\frac{1}{2L}} \cos\left[ \pi (j-\frac12) \right], & k=2L.
\end{cases}
\ee
To characterize the boundary conditions, we again introduce two virtual sites at $j=0$ and $j=2L+1$. The eigenfunctions satisfy
\be
\begin{aligned}
    \phi_k^{\rm fixed{,\pm}}(0) &= \mp \phi_k^{\rm fixed{,\pm}}(1), \\
    \phi_k^{\rm fixed{,\pm}}(2L) &= \mp \phi_k^{\rm fixed{,\pm}}(2L+1).
\end{aligned}
\ee
These relations characterize the two fixed-boundary cases corresponding to $V_0 = +1$ and $V_0 = -1$, respectively.

\section{Equation of motion for the covariance matrix}\label{appendix:Equation of motion for the covariance matrix}
In this appendix, we discuss the time evolution of two-point correlation functions in the harmonic chain governed by the Lindblad master equation. We begin by expressing the Hamiltonian in Eq.\eqref{eq: initial Hamitonian in simplified form} in terms of the local bosonic operators
\begin{align}
    a_{i} = \dfrac{u_i + i\pi_i}{\sqrt{2}},\quad a^{\dagger}_i = \dfrac{u_i-i\pi_i}{\sqrt{2}},
\end{align}
which satisfy the canonical commutation relations
\begin{align}
 \left[a_i,a_j^\dagger\right] = \delta_{ij}, \quad \left[a_i,a_j\right] = \left[a^\dagger_i,a^\dagger_j\right] = 0.    
\end{align}
Using these operators, the Hamiltonian takes the quadratic form
\begin{equation}
H = \sum_{i,j = 1}^{2L}A_{ij}a_i^{\dagger}a_j + \frac{1}{2}\left(B_{ij}a_i^{\dagger}a_j^{\dagger} + B_{ij}^{\ast}a_ia_j\right)+\text{const},\label{eq:bosonic quadratic Hamiltonian}
\end{equation}
where the matrix elements $A_{ij}$ and $B_{ij}$ are expressed in terms of the stiffness matrix elements $K_{ij}$ defined in Eq.\eqref{eq:K matrix} as
\begin{align} 
\label{eq:A and B}
A_{ij} = \dfrac{1}{2}\left(K_{ij} +\delta_{ij}\right), \quad\quad B_{ij} = \dfrac{1}{2}\left(K_{ij} -\delta_{ij}\right). 
\end{align}

We now provide a detailed derivation of the effective non-Hermitian Hamiltonian from the equations of motion for the covariance matrix. We begin with the following two-point correlation functions: 
\begin{align}
    \mathbb{C}_{ij}(t) = \text{Tr}\left[\rho(t)a^{\dagger}_i a_j\right], \quad   \mathbb{G}_{ij}(t) = \text{Tr}\left[\rho(t)a_i a_j\right],
\end{align}
in order to derive the equation of motion for the covariance matrix $\Phi(t)$, defined as
\begin{align}
    \Phi(t) :=  \left\langle \begin{pmatrix} a^{\dagger} \\ a \end{pmatrix} \begin{pmatrix} a & a^{\dagger} \end{pmatrix}\right\rangle_{t}
    =  \begin{pmatrix}  \mathbb{C} &&  \mathbb{G}^{\dagger} \\ \mathbb{G} && \mathbb{I} +\mathbb{C}^{T}
    \end{pmatrix}_{t}.
\end{align}
We start from the Lindblad master equation. For any observable $O$, the corresponding Heisenberg-picture equation of motion is
\begin{align}
    \dfrac{dO}{dt} = i[H, O] + \sum_{\mu}\left(L^{\dagger}_{\mu}OL_{\mu} - \dfrac{1}{2}\left\{L^{\dagger}_{\mu}L_{\mu}, O \right\}\right).
    \label{eq:operator evolution}
\end{align}
In this work, we consider local particle loss and gain at site $n_d$, described by the Lindblad jump operators
\begin{align}
    L_{1} = \sqrt{\gamma_l}a_{n_d} \quad\quad L_{2} = \sqrt{\gamma_g}a^{\dagger}_{n_d}.
\end{align}
Substituting $O = a^{\dagger}_ia_j$ into Eq.\eqref{eq:operator evolution}, we obtain the equation of motion for the correlation matrix $\mathbb{C}$: 
\begin{align}
    \dfrac{d\mathbb{C}}{dt} = i[\mathbf{A},\mathbb{C}] + i \mathbf{B}\mathbb{G} - i  \mathbb{G}^{\dagger}\mathbf{B} -\dfrac{1}{2}\left\{\mathbf{\Gamma},\mathbb{C} \right\} + \mathbf{\Gamma}_+,
    \label{eq:two-point correlation}
\end{align}
where $A$ and $B$ are the matrices defined in Eq.\eqref{eq:A and B}. Similarly, substituting $O = a_ia_j$ into Eq.\eqref{eq:operator evolution} gives the equation of motion for the anomalous correlation matrix $\mathbb{G}$:
\begin{align}
    \dfrac{d\mathbb{G}}{dt} = -i\left\{\mathbf{A},\mathbb{G}\right\} - i \mathbf{B}\mathbb{C} - i(\mathbb{C}^{T}+\mathbb{I})\mathbf{B} -\dfrac{1}{2}\left\{\mathbf{\Gamma},\mathbb{G} \right\}.
    \label{eq:two-point anomalous correlation}
\end{align}
From Eq.\eqref{eq:two-point correlation}) and Eq.\eqref{eq:two-point anomalous correlation}, we obtain
\begin{align}
    \dfrac{d\Phi}{dt} = \mathbf{X}\Phi +\Phi \mathbf{X}^{\dagger} +\mathbf{\Lambda},
\end{align}
where 
\begin{align}
    \mathbf{X} := \begin{pmatrix}
        i\mathbf{A}-\dfrac{\mathbf{\Gamma}}{2} && i\mathbf{B} \\
        -i\mathbf{B} && -i\mathbf{A}-\dfrac{\mathbf{\Gamma}}{2}
    \end{pmatrix},
\end{align}
and
\begin{align}
    \mathbf{\Gamma} = \mathbf{\Gamma_-}-\mathbf{\Gamma_+}, \quad \mathbf{\Lambda} = 
        \mathbf{\Gamma_+} \oplus \mathbf{\Gamma_-}.
\end{align}
Therefore, the relaxation rates can be obtained from the non-Hermitian effective Hamiltonian
\begin{align}
    \mathbf{H}_{\rm eff} = i\mathbf{X^{\ast}}\label{eq:bosonic effective Hamiltonian}.
\end{align}

\section{Finite dissipation}\label{appendix:Finite dissipation case}
In this appendix, we provide a detailed derivation of the eigenspectrum of the harmonic chain at finite dissipation strength $\gamma$. We begin with the effective non-Hermitian Hamiltonian, whose eigenvalue equation is 
\begin{align}
\mathbf{H}_{\text{eff}}|\Phi_m\rangle = E_m|\Phi_m\rangle.    
\end{align}
Here, 
\begin{align}
    \mathbf{H}_{\text{eff}} = \begin{pmatrix}
        \mathbf{A} - \dfrac{i}{2}\mathbf{\Gamma} && \mathbf{B}\\
        -\mathbf{B}    && -\mathbf{A} - \dfrac{i}{2}\mathbf{\Gamma}
    \end{pmatrix}, \quad|\Phi_m\rangle := \begin{pmatrix}
        \boldsymbol{|\phi^{+}_m\rangle} \\ \boldsymbol{|\phi^{-}_m\rangle}
    \end{pmatrix}.
\end{align}
For $\gamma=0$, the effective Hamiltonian reduces to
\begin{align}
    \mathbf{H}_{\text{eff}} = \begin{pmatrix}
        \mathbf{A} && \mathbf{B}\\
        -\mathbf{B} && -\mathbf{A}
    \end{pmatrix},
\end{align}
where $\mathbf{A}$ and $\mathbf{B}$ are defined in Eq.\eqref{eq:A and B}. For finite dissipation strength $\gamma \neq0$, the eigenvalue equation leads to the coupled equations
\begin{subequations}
\begin{align}
    &\left(\mathbf{A} - \dfrac{i}{2}\mathbf{\Gamma}\right)|\boldsymbol\phi^{+}_m\rangle + \mathbf{B}|\boldsymbol\phi^{-}_m\rangle = E_m|\boldsymbol\phi^{+}_m\rangle, 
    \label{eq:setup1}\\
    &-\mathbf{B}|\boldsymbol\phi^{+}_m\rangle -\left(\mathbf{A} + \dfrac{i}{2}\mathbf{\Gamma}\right)|\boldsymbol\phi^{-}_m\rangle = E_m|\boldsymbol\phi^{-}_m\rangle.
    \label{eq:setup2}
\end{align}
\end{subequations}
Defining
\begin{align}
|\boldsymbol\phi^{+}_m\rangle +|\boldsymbol\phi^{-}_m\rangle = |\boldsymbol\psi_m\rangle \quad\quad  |\boldsymbol\phi^{+}_m\rangle -|\boldsymbol\phi^{-}_m\rangle = |\boldsymbol\chi_m\rangle
\end{align}
the coupled equations become
\begin{subequations}
\begin{align}
    &\left(\mathbf{A} -\mathbf{B}\right)|\boldsymbol\chi_m\rangle - \dfrac{i}{2}\mathbf{\Gamma}|\boldsymbol\psi_m\rangle = E_m|\boldsymbol\psi_m\rangle, \\
    &\left(\mathbf{A} +\mathbf{B}\right)|\boldsymbol\psi_m\rangle - \dfrac{i}{2}\mathbf{\Gamma}|\boldsymbol\chi_m\rangle = E_m|\boldsymbol\chi_m\rangle.
\end{align}
\end{subequations}
Using $\mathbf{A} -\mathbf{B} = \mathbb{I}$ and $\mathbf{A} + \mathbf{B} = \mathbf{K}$, we obtain
\begin{align}
    \mathbf{K}|\boldsymbol\psi_m\rangle - \dfrac{i}{2}\mathbf{\Gamma}\left(\dfrac{i}{2}\mathbf{\Gamma} + E_m\mathbb{I}\right)|\boldsymbol\psi_m\rangle = E_m\left(\dfrac{i}{2}\mathbf{\Gamma} + E_m\mathbb{I}\right)|\boldsymbol\psi_m\rangle,
\end{align}
which gives 
\begin{equation}
\left(\mathbf{K} + \dfrac{\mathbf{\Gamma}^{2}}{4} -i E_m\mathbf{\Gamma} - E^{2}_m\mathbb{I}\right)\boldsymbol|\boldsymbol\psi_m\rangle = 0.\label{eq:eigenvalue equation in Gamma matrix form}
\end{equation}

\section{Lattice calculation of $c_{\rm LR}$ and $c_{\rm eff}$ for harmonic chain}\label{appendix:cLR and ceff Harmonic chain}
\begin{figure}[!htbp]
    \centering
\includegraphics[width=1\linewidth]{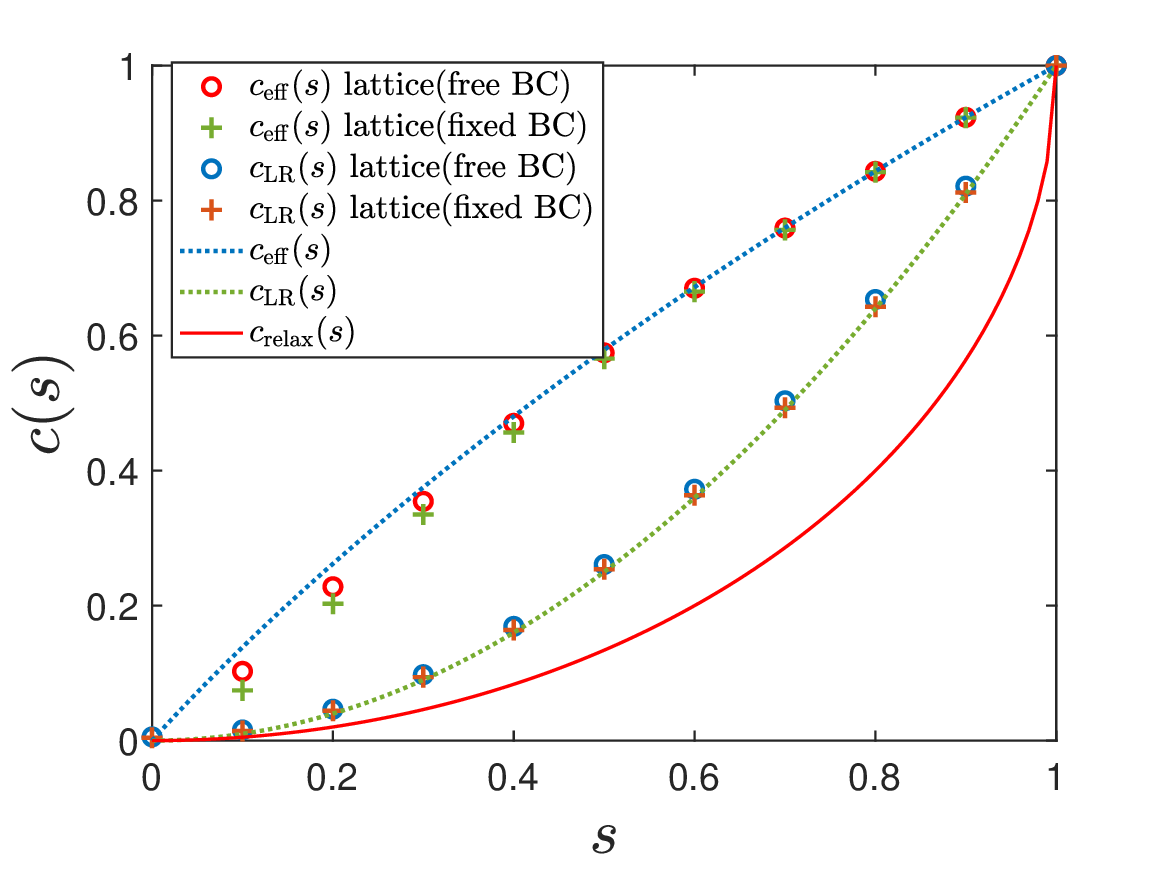}
    \caption{Comparison of $c_{\rm eff}$ and $c_{\rm LR}$ obtained from analytical results in Eq.\eqref{eq: c_LR for boson} and Eq.\eqref{eq:definition of c_eff} (dashed lines) and lattice calculations (circles and plus symbols) for the harmonic chain with free and fixed boundary conditions.  To extract $c_{\rm LR}$ from the lattice simulations, the two initially decoupled chains are defined on the intervals $[-400,100]$ and $[100,400]$, respectively, and are joined at $x=100$, while the conformal interface is located at $x=0$. We take $\Omega_0=10^{-2}$ to avoid singularities due to zero mode in free boundary conditions. For the extraction of $c_{\rm eff}$, we use a system size $2L=1600$ and $\Omega_0=10^{-4}$. For comparison, we also show the analytical result for $c_{\rm relax}$ in the perturbative regime derived in Eq.\eqref{eq: crelax result}.}
\label{fig:cLR and ceff lattice}
\end{figure}
\begin{figure}[!htbp]
    \centering
    \begin{tikzpicture}
            \node[inner sep=0pt] (russell) at (-250pt, 0pt)
    {\includegraphics[width=0.45\textwidth]{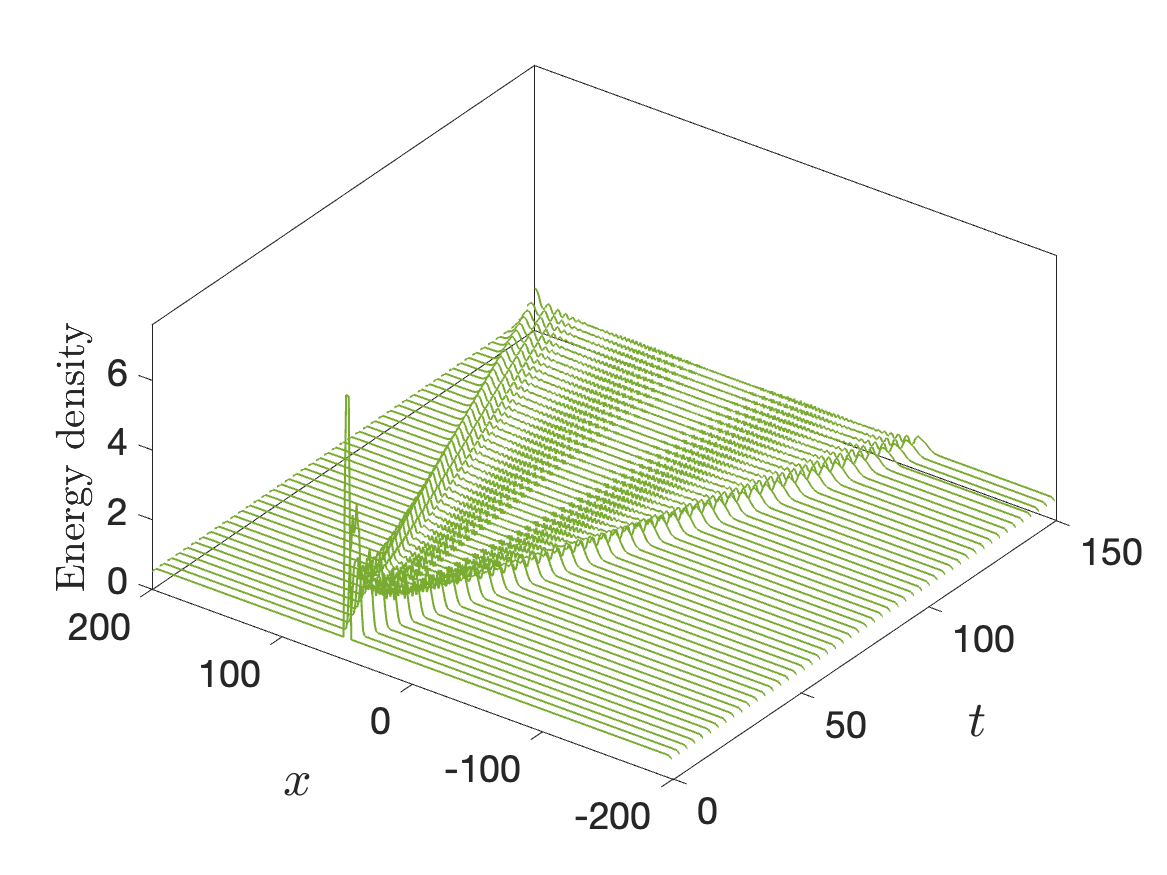}};

    \node[inner sep=0pt] (russell) at (-250pt, -165pt)
    {\includegraphics[width=0.45\textwidth]{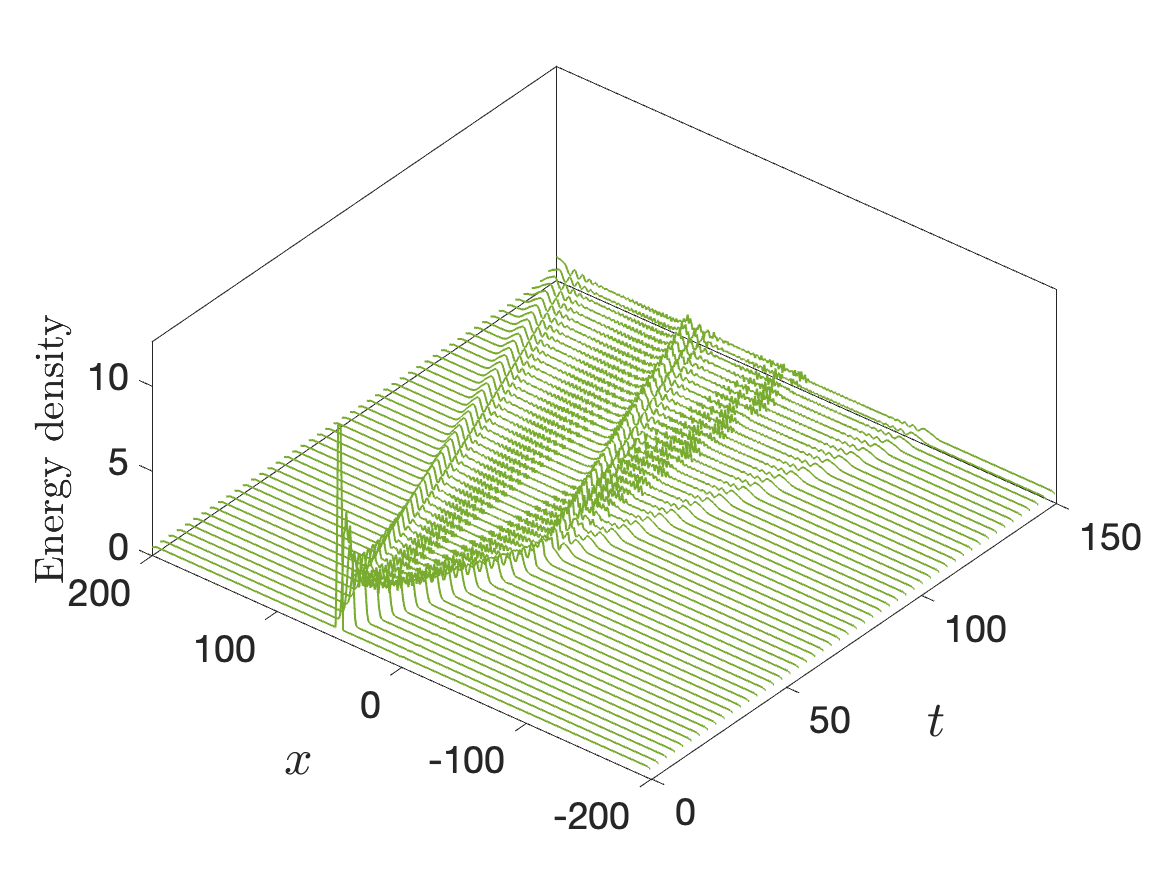}};
                       \node[inner sep=0pt] (russell) at (-250pt, -330pt)
    {\includegraphics[width=0.45\textwidth]{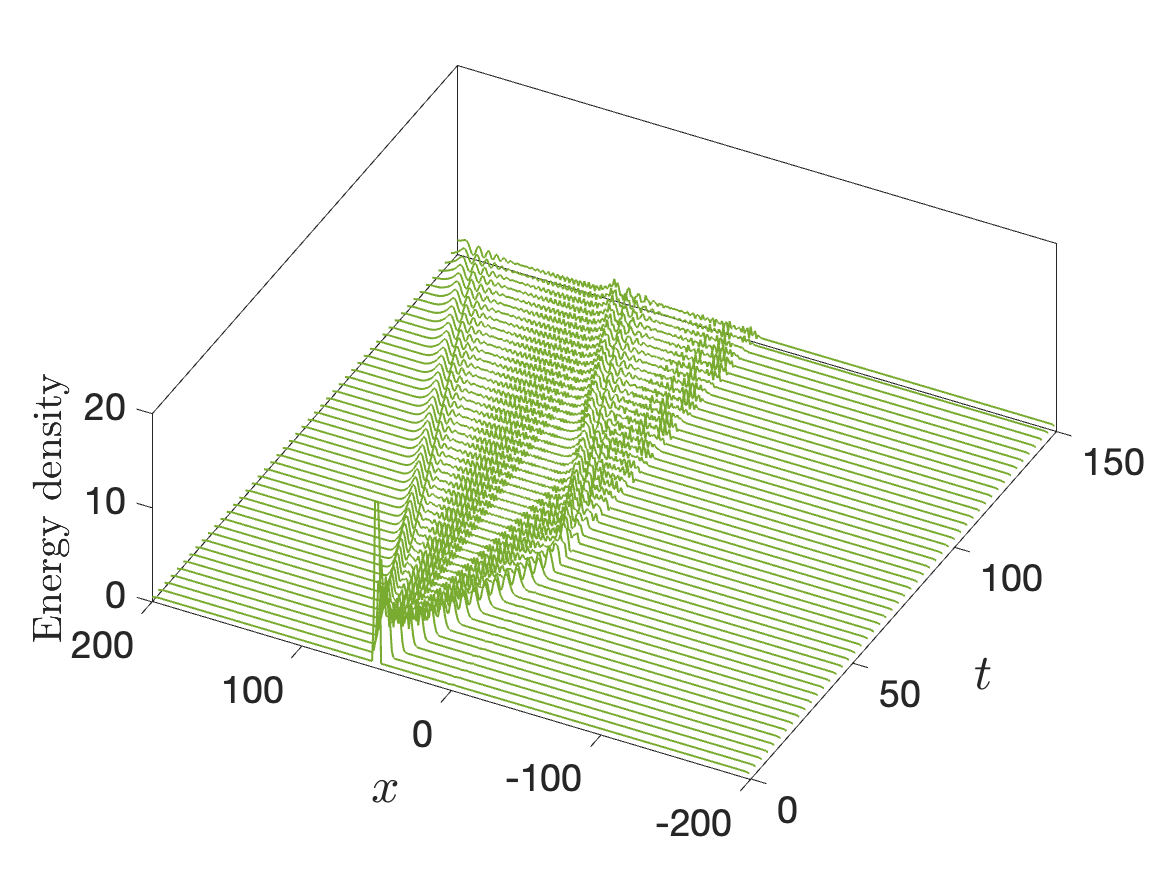}};
    \node at (-10,2) {\textbf{(a)}};
    \node at (-10,-4) {\textbf{(b)}};
    \node at (-10,-10) {\textbf{(c)}};
    \node at (-8,2) {\text{Total transmission}};
    \node at (-8,-4) {\text{Partial transmission}};
    \node at (-8,-10) {\text{Total reflection}};
    \end{tikzpicture}
    \caption{Energy scattering across a conformal interface following a local quantum quench in the harmonic chain with free boundary conditions. The representative plots correspond to total transmission ($s=1$), partial transmission and reflection ($s=0.7$), and total reflection ($s=0$). For the lattice simulations, the two initially decoupled chains are defined on the intervals $[-200,50]$ and $[50,200]$, are joined at $x=50$, while the conformal interface is located at $x=0$.}
    \label{fig: cLR lattice}
\end{figure}
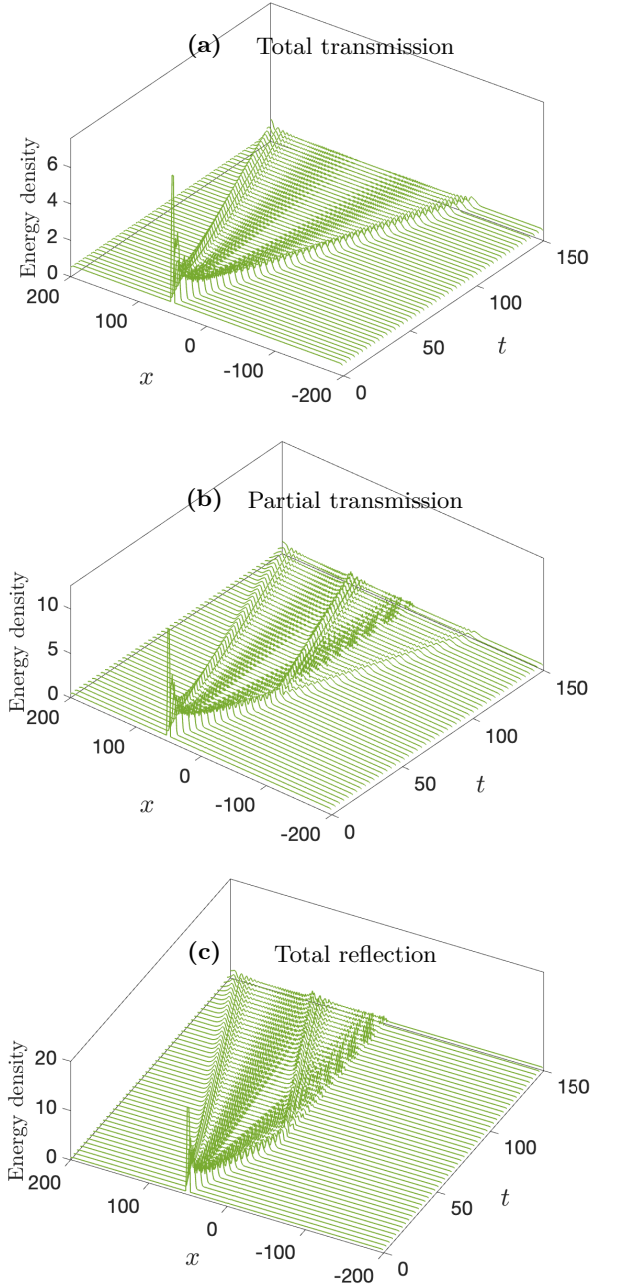
For completeness, in this appendix we explain how the effective central charges $c_{\rm LR}$and  $c_{\rm eff}$ are extracted from the lattice models. Here, $c_{\rm LR}$ characterizes energy transmission, while $c_{\rm eff}$ characterizes information transmission across the conformal interface. The corresponding analysis for the critical free fermion chain was presented in our recent work \cite{Barad_2025}, to which we refer the reader. Here, we present the analysis for the harmonic chain with both free and fixed boundary conditions.

\subsection{Energy transmission coefficient : $c_{\rm LR}$}
Here, we extract $c_{\rm LR}$ from the energy scattering experiment, following Refs.~\cite{Sakai_2008, Barad_2025}. The energy transmission coefficient is defined as the ratio of the transmitted energy pulse to the incident energy pulse across the conformal interface, namely
\begin{align}
&\mathcal{T} = \frac{\text{transmitted energy}}{\text{incident energy}},\,\,
&\mathcal{R} = \frac{\text{reflected energy}}{\text{incident energy}}.\label{eq:original definition of energy transmission and reflection coefficients}
\end{align}
For a conformal interface, these quantities are independent of the incoming excitation, which makes them universal. One can therefore define an effective central charge $c_{\rm LR}$ associated with energy transmission as follows:
\begin{align}
    \mathcal{T}_{L} = \frac{c_{\rm LR}}{c_L} \quad\text{and}\quad \mathcal{T}_{R} = \frac{c_{\rm LR}}{c_R}.\label{eq:energy-transmission coefficients T_L and T_R}
\end{align}
Here, the subscripts indicate the direction and origin of the incoming excitation relative to the interface. We refer the reader to Appendix~A of Ref.\cite{Barad_2025} for details of the lattice energy scattering experiment. Briefly, an energy pulse is generated by a local quench and propagated toward the conformal interface, where it is partially transmitted and reflected. The corresponding transmission coefficient is extracted from the ratio of the transmitted to incident energy. Here, we present the resulting analysis for the harmonic chain in Fig.~\ref{fig: cLR lattice}. For $c_{\rm LR}$, the analytical dependence on the interface transmission was derived in Ref.\cite{Sakai_2008, Eisler_2012} and is given by
\begin{align}
\label{eq: c_LR for boson}
  c^{\rm boson}_{\rm LR}= s^2.
\end{align}
We compare this analytical prediction with our lattice results.

\subsection{Effective central charge: $c_{\rm eff}$}
Next, we compute the effective central charge $c_{\rm eff}$, which measures the amount of entanglement transmitted through the conformal interface \cite{Sakai_2008,Karch_2023}. Here, we follow a method similar to that used in previous work to extract $c_{\rm eff}$ for the lattice harmonic chain. For completeness, we briefly recall the method used to extract $c_{\rm eff}$. Following Ref. \cite{Wen_2018}, we compute the entanglement entropy of a subsystem containing the conformal interface.
\begin{figure}[!htbp]
    \centering
\includegraphics[width=1\linewidth]{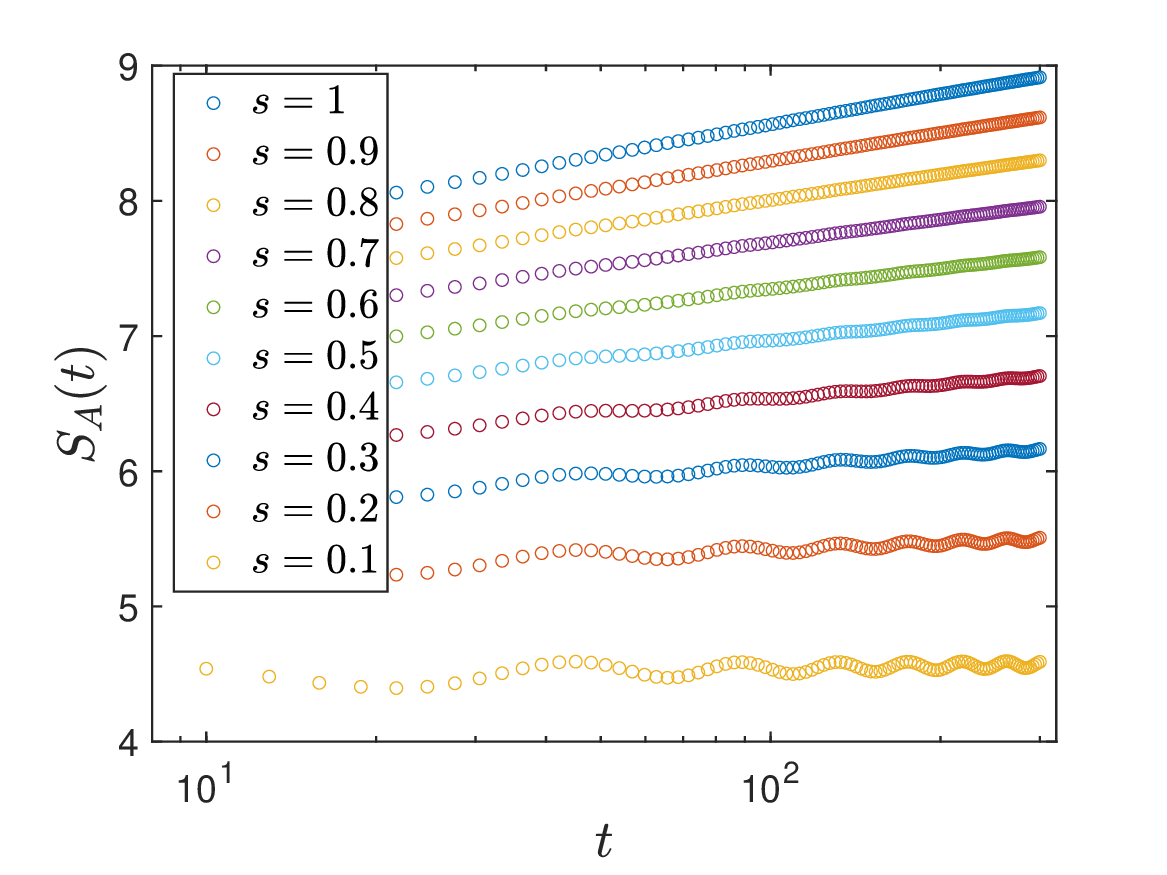}
    \caption{Time evolution of the entanglement entropy for $A = [0, L]$ based on a harmonic chain (free boundary conditions) calculations with different strengths of conformal interfaces. The interface is inserted at $x = L = 800$, the total length of the chain is chosen as $2L = 1600$, and $\Omega_0$ is taken as $10^{-6}$.}
    \label{fig:ceff lattice}
\end{figure}
We start from the ground state of two decoupled harmonic chains and keep  $\Omega_{0} \sim (10^{-4}-10^{-6})$ finite but small to regularize the zero mode. At $t=0$, we connect the initially decoupled chains through the conformal interface. We then compute the time evolved entanglement entropy of a subsystem across the conformal interface such that coefficient of the logarithmic growth defines the effective central charge (see Fig.\ref{fig:ceff lattice}),
\begin{align}
    S_{A}(t)- S_{A}(0) \simeq \dfrac{c_{\rm eff}}{3} \text{log}(t). 
\end{align}
The extraction is performed for both free and fixed boundary conditions, as shown in Fig.\ref{fig:cLR and ceff lattice}. For $c_{\rm eff}$, the analytical dependence on the interface transmission has also been derived previously and is given by
\begin{equation}
\label{eq:definition of c_eff}
    \begin{split}
c^{\rm Boson}_{\rm eff} = \frac{3s}{2} + \frac{3}{\pi^2}\sum_{\kappa = \pm}\Big[(1+\kappa s){\rm Li}_2(-\kappa s) \\ 
+ (1+\kappa s)\ln(1+ \kappa s)\ln s \Big]. 
\end{split}
\end{equation}

\bibliography{defect_boson_new}
\end{document}